\documentclass[useAMS,usegraphicx,times]{mn2e}

\def\kms{$\rm km\, s^{-1}$}
\def\cm3{$\rm cm^{-3}$}
\def\Ts{$\rm T_{*}$}
\def\Vs{$\rm V_{s}$}
\def\n0{$\rm n_{0}$}
\def\B0{$\rm B_{0}$}

\def\Fh{$\rm F_{H}$}
\def\erg{$\rm erg\, cm^{-2}\, s^{-1}$}
\def\Hb{H$\beta$}

\def\mm{$\mu$m}

\title[Dust-to-gas ratios in the starburst regions of LIGs]{Dust-to-gas ratios in the starburst regions of luminous infrared galaxies}
\author[M. Contini \& T. Contini]{Marcella Contini$^{1}$\thanks{E-mail:
contini@ccsg.tau.ac.il} \& Thierry Contini$^2$ \\
$^{1}$School of Physics and Astronomy, Tel Aviv University, Tel Aviv 69978, Israel\\
$^{2}$Observatoire Midi--Pyr\'en\'ees, Laboratoire d'Astrophysique (UMR 5572), 14 avenue E. Belin, F-31400 Toulouse, France\\}
\begin{document}

\date{Accepted . Received ; in original form }

\pagerange{\pageref{firstpage}--\pageref{lastpage}} \pubyear{2002}

\maketitle

\label{firstpage}

\begin{abstract}

We investigate the properties of dust and dust-to-gas ratios in different  starburst
regions of luminous infrared galaxies (LIGs). We refer to the sample of seven LIGs
recently observed in the mid-infrared by Soifer et al. (2001) using the Keck
telescopes with spatial resolution approaching the diffraction limit. These seven
objects are among the closest LIGs and have been classified as starburst galaxies
from optical spectroscopy. Our goal consists in modelling the continuum spectral energy
distribution (SED) of each galaxy, particularly in the infrared range. Models are further
constrained by observed emission-line ratios in the optical range.
The multi-cloud models
consistently account for the coupled effect of shock, photoionization by hot stars,
and diffuse secondary radiation from the shock-heated gas.
Emission from clouds in the neighbourhood of evolved starbursts and with high
shock velocities ($\sim$ 500 \kms) explains both the bremsstrahlung and
reradiation from dust in the mid-infrared. Clouds with lower velocity (100 \kms) and
corresponding to younger starbursts also contribute to both line and continuum spectra.
Both low- and high-velocity clouds are thus present in nearly all the sample galaxies.
For all the galaxies, an old stellar population is revealed by black body emission in
the optical-NIR range. Dust-to-gas ratios vary in different regions of individual galaxies.

\end{abstract}


\section{Introduction}

IRAS surveys have revealed that a significant fraction of extragalactic objects 
produce the bulk of their radiation at far-infrared (FIR) wavelengths (Soifer 1986; 
Sanders \& Mirabel 1996 for a recent review). 
Imaging surveys of luminous infrared galaxies (LIGs; $L_{\rm IR} > 10^{11}L_{\odot}$) 
showed that the vast majority of these objects  is morphologically disturbed, 
with spatial structures indicative of galactic mergers (e.g. Sanders et al. 1988; 
Clements et al. 1996). 

The interest on the nature of LIGs grew  during the last few years as 
they may represent an important step in the global process of galaxy formation 
and evolution. Indeed, strong evolution in the space density of LIGs has been 
detected in the deep mid-infrared (MIR; Taniguchi et al. 1997; Aussel et al. 1999) 
and FIR surveys (Kawara et al. 1998; Puget et al. 1999) with ISO satellite. 
Reports from the first deep sub-millimeter (sub-mm) surveys with SCUBA 
(Smail et al. 1997; Hughes et al. 1998; Barger et al. 1998; Eales et al. 1999) 
showed that the space density of LIGs at high redshift ($z \geq 1$) appears 
sufficient to account for nearly all of the FIR/sub-mm background radiation 
(e.g. Barger, Cowie \& Sanders 1999). It seems clear that detailed studies of 
nearby LIGs are a required first step in order to better understand their 
high-redshift counterparts, and in particular to reveal the nature of the dominant 
energy source responsible for their enormous FIR luminosity (Lutz \& Tacconi 1999). 
Results from recent studies at MIR (e.g. Genzel et al. 1998) 
and radio wavelengths suggest that the majority of LIGs are 
powered predominantly by young and hot stars rather than by active galactic 
nuclei (AGNs). However, there is growing evidence from optical/near-infrared (NIR) 
spectroscopy that the frequency of AGN occurrence among LIGs increases with 
increasing FIR luminosity (e.g. Veilleux et al. 1995; Kim, Veilleux \& Sanders 1998; 
Veilleux, Sanders \& Kim 1999).  

In this paper, we refer to the sample of seven LIGs recently observed in the MIR 
(from 8 to 18 \mm) by Soifer et al. (2001, hereafter S01) using the Keck telescopes 
with spatial resolution approaching the diffraction limit. 
These seven objects are among the closest LIGs and have been classified as 
starburst galaxies from optical spectroscopy (Armus et al. 1989; Veilleux et al. 1995). 

S01  give some important results by the analysis of the
MIR data (S01, Table 3) about, sizes, temperatures, and luminosities in the
different sources. We would like, however, to explain the results by  modelling the
emitting clouds or clumps in dusty regions surrounding the starbursts.
Dust plays a major role in IR emission. We will thus investigate the distribution
of dust in the different starburst regions  within the galaxies, and dust-to-gas 
properties on small local scales. For sake of consistency, we will focus 
on both the continuum radiations and  the
optical emission lines from each galaxy of the sample. When the spectra are integrated, 
we will try to distinguish the contributions of clouds with different physical 
conditions, photoionized by starbursts of different ages.

The models adopted up to now for the calculation of dust emission in the IR
(e.g. Siebenmorgen et al 1999) accounted for heating of dust by  star radiation 
in starbursts, and for power-law radiation in  AGN.
The grains are destroyed by photo-dissociation.
However, in  the turbulent regime created by the starburst, shocks are playing
a major role  by  heating  gas and dust and
by  leading  to sputtering of  the grains.
We will focus on the clouds in the very starburst regions.
Therefore, in our models we account for the coupled effect
of radiation from an external source and for the shocks.

Gas and dust  mutually heat each other by collisions 
in  shock turbulent  regimes. Regarding the spectral energy distribution 
(SED) of the continuum, recall that the  frequency corresponding to the dust reradiation 
peak depends on the  shock velocity (e.g. Viegas \& Contini 1994). 
At high shock velocities, the grains will 
 reach high temperatures in the immediate post shock region, before being destroyed 
by sputtering, so that dust reradiation will  peak in the MIR, whereas, 
at low velocities,  the peak appears in the FIR.

Actually, our goal consists in  modelling the continuum SED of each galaxy in the 
S01 sample, particularly in the IR range, adopting the data in the MIR presented by 
S01 and completed with data from NED\footnote{The NASA/IPAC Extragalactic Database 
(NED) is operated by the Jet Propulsion Laboratory, California Institute of 
Technology, under contract with the National Aeronautics and Space Administration} 
(see Appendix B for the list of references) at other wavelengths.
Indeed, the data from the NED show different precisions and different apertures,
leading to a large scattering.
The S01 data, however,  refer to the IR range and hardly include
radiation from the gas, so   NED data are included in our modelling in order
to determine the gas-to-dust ratio.

In most of the  objects of the sample 
different  fractions of flux  from the old star population background  are included in the 
near-IR (NIR) data,
as generally found in AGN and starburst galaxies (Viegas \& Contini 2000).
We will try to fit the data corresponding to the lowest flux by emission from the
clouds in the  objects (e.g. NGC 1614, NGC 2623, NGC 3690),
and those corresponding to the highest flux by the  black body emission.
In fact, the data from the NED  in the NIR cluster within the black body curve.

The models which explain the continuum SED  are constrained by the agreement 
with observations of the emission-line spectra (Armus et al. 1989; Veilleux et al. 1995).
The spectra are quite poor in number of emission lines, however,
being the most significant ones ([OIII] 5007, [OI] 6300, [NII] 6584, and [SII] 6716+6731)
 they are enough to constrain the models.

To obtain a basis for modelling, we have compared the observed emission-line ratios 
of each galaxy with the models presented in Contini \& 
Viegas (2001a, hereafter CV01), where 
model calculations of line spectra in the UV-optical-IR ranges are presented.
The results are obtained by the SUMA code (Viegas \& Contini 1994) which, besides 
the line and continuum spectra emitted by the gas, also calculates reradiation by 
 dust. Dust features are summarized in Sect. 2. 
The models in CV01 are all calculated adopting 
a dust-to-gas ratio  $d/g = 4\times 10^{-5}$ by mass. 
In many cases the $d/g$ value will be increased 
to fit the SED in the IR. The line ratios will then change accordingly, because a 
high $d/g$ in the postshock region, before complete destruction of the grains by 
sputtering, acts as an increased  density, speeding up the cooling processes. 
This will lead to a refinement of the models in order to obtain the best fit to 
the observational data. The models are described in Sect. 3.
Recall that modelling the line and continuum spectra simultaneously implies 
cross-checking them (Sect. 4) until a fine tuning  is obtained (Contini et al. 1998).
The models adopted in this work are described and referred to the models of CV01 in 
Appendix A (Table A1). Discussion of results and concluding remarks appear in Sect. 5.

\section{About dust}

Emission  from a starburst is generally
calculated by computing the radiative transfer in the 
galactic nuclei under the assumption of spherical symmetry (Siebenmorgen, Krugel, \& Zota 1999), or
treating the starburst (Efastathiou et al. 2000) as an ensemble
 of optically thick giant molecular clouds centrally illuminated by recently formed stars.
{\it The dust-to-gas ratio is not required for computing the radiation transfer models.}

Hildebrand (1983)   discussed direct methods to determine  cloud  and dust masses and derived a lower
limit ($\sim$ 60) for  g/d (where g=M$_{gas}$ and d=M$_{dust}$) calculating M$_{gas}$ from visual extinction   and  
M$_{dust}$ by thermal emission.  A rough g/d  $\sim$ 100  was estimated,
given  the errors for the implied coefficients (dielectric, etc). 

Generally, three population of grains are adopted  to fit the average IS extinction curve.
1) large grains of silicate and amorphous carbon, with a
 volume ratio of V$_{Si}$/V$_{aC}$ =1.4 (Siebenmorgen et al. 1999) and
a power-law size distribution, a$^{-q}$ with q=3.5 (100$<$a$<$2500 A)
 provide the  bulk of emission responsible for 9.7 and 18 \mm ~resonance. 
2) Small graphite particles corresponding to q=4 (10$<$a$<$100 A) emit primarily at    MIR wavelengths.
Being small,  they show dramatic fluctuations in temperature.
3) PAH carriers of IR features correspond to single molecules with 25 carbon atoms and clusters of 10-20
molecules.
The small graphite and PAH are underabundant with respect to ISM by factors of 10 or more.

Cold dust or cirrus emission  results from heating by the IS radiation field, warm dust is associated
with star formation regions, and hot dust appears  around an active nucleus (Helou 1986).
Rowan-Robinson \& Crawford (1989) found that the ratio of the starburst emission
to the cirrus emission  is 2 to 1.   PAHs  are responsible for  no more than 22 \% of the
total 3-1100 \mm ~emission (Dale et al. 2001), the exact percentage depending on the activity 
level of the interstellar medium.  

Diffuse ISM radiates strongly in 3.3, 6.2, 7.7, 8.6,  11.3,  12.7 \mm
~and in  atomic emission lines (Spoon et al. 2002),
the relative strengths of the different PAH emission bands  being a strong function of the PAH size:
small PAH radiates strongly at 6.2 and 7.7 \mm, while larger PAHs emit most of their powers
at increasingly long wavelengths.
Moreover,  laboratory simulations  predict a double peaked bump
in the 18-23 \mm ~wave band, due to the morphological transformation of ice
on grain mantles (Contini 1990).

A broad absorption band due to the Si-O stretching mode in amorphous
silicates, centered at 9.7 \mm ~is also commonly detected in galaxies (Spoon et al 2002).
The center of the silicate absorption coincides with the gap between
the 6.2 and 8.6 \mm ~and 11.3-12.8 \mm ~PAH complexes. Therefore, it is not clear
whether the 9.7 \mm ~flux minimum should be interpreted as the "trough" between
PAH emission features or as strong silicate absorption, or a combination of
the two.
A strong silicate feature is often accompanied by    MIR absorption features,
due to molecules frozen in ice mantles which leads to a
deep and broad 3.0 \mm ~water ice feature.
The interplay of 6.0 \mm ~water ice and 5.25,5.7, 6.25 PAH emission
is shown in many spectra (Spoon et al 2002).

Turbulent and hostile environments in galaxies
often destroy the molecular material that resides in or close to these regions.
PAHs are likely to be destroyed, or severely dumped in
localized regions of high heating intensity such as the regions near OB associations
(Dale et al. 2001), and may evaporate in a strong UV radiation field (Siebenmorgen et al. 1999).
Mid-IR flux may also be diminished toward regions of high extinction in AGN-dominated
galaxies, but it is unclear whether significant obscuration also occurs for
starburst galaxies.
Moreover, sputtering of grains throughout  shock fronts is  efficient
for shock velocities in the range  suitable to starburst galaxies and 
AGN (Viegas \& Contini 1994).
In conclusion, PAH emission comes from the extended region of the galaxy.

Coronal lines  eventually blended with the bands of PAH and
water ice (Contini \& Viegas 2001) are the following :
[MgVIII] 3.03 \mm, [SiIX] 3.9 \mm, [MgVII] 5.5 \mm, [MgV] 5.6 \mm, [NeVI] 7.6 \mm, [FeVII] 7.8 \mm, 
[ArIII] 8.99 \mm, [FeVII] 9.52 \mm, [SIV] 10.54 \mm,  [NeII] 12.8 \mm, etc. (see also Sturm et al 2002)
We have included lines which do not exactly correspond
to the bands,  because  large FWHM of  line profiles, corresponding to
high velocities,  lead to  blending  with the closer bands.
Mg and Si are easily locked up in dust grains. In this case
their lines in the    MIR would not interfere with the bands.
On the other hand, Ne is not adsorbed in grains due to its atomic
configuration, therefore, the  Ne lines coexist with dust features.

Model results (CV01) regarding black body radiation from stars 
show that [NeII] are  strong for shock velocities \Vs $>$ 100 \kms  and U$<$ 10,
while [ArIII]  may be detected for \Vs ~between  200 \kms and 500 \kms  
and decreases at higher  
velocities,  independently from the black body intensity flux.

Shock dominated models (Contini \& Viegas 2001b) show that 
[NeII] is strong for \Vs=100 \kms. 
High ionization lines  appear for \Vs $\geq$ 200 \kms, particularly [NeVI],
increasing at \Vs = 500 \kms,
[NeVI] and [NeII] being the strongest at \Vs=300 \kms. 
The high level lines [MgVIII], [SiIX], [MgVII] are  strong for \Vs=1000 \kms,
particularly, [SiIX] 3.9 \mm.

\section{Composite models}

\begin{figure*}
\includegraphics[width=78mm]{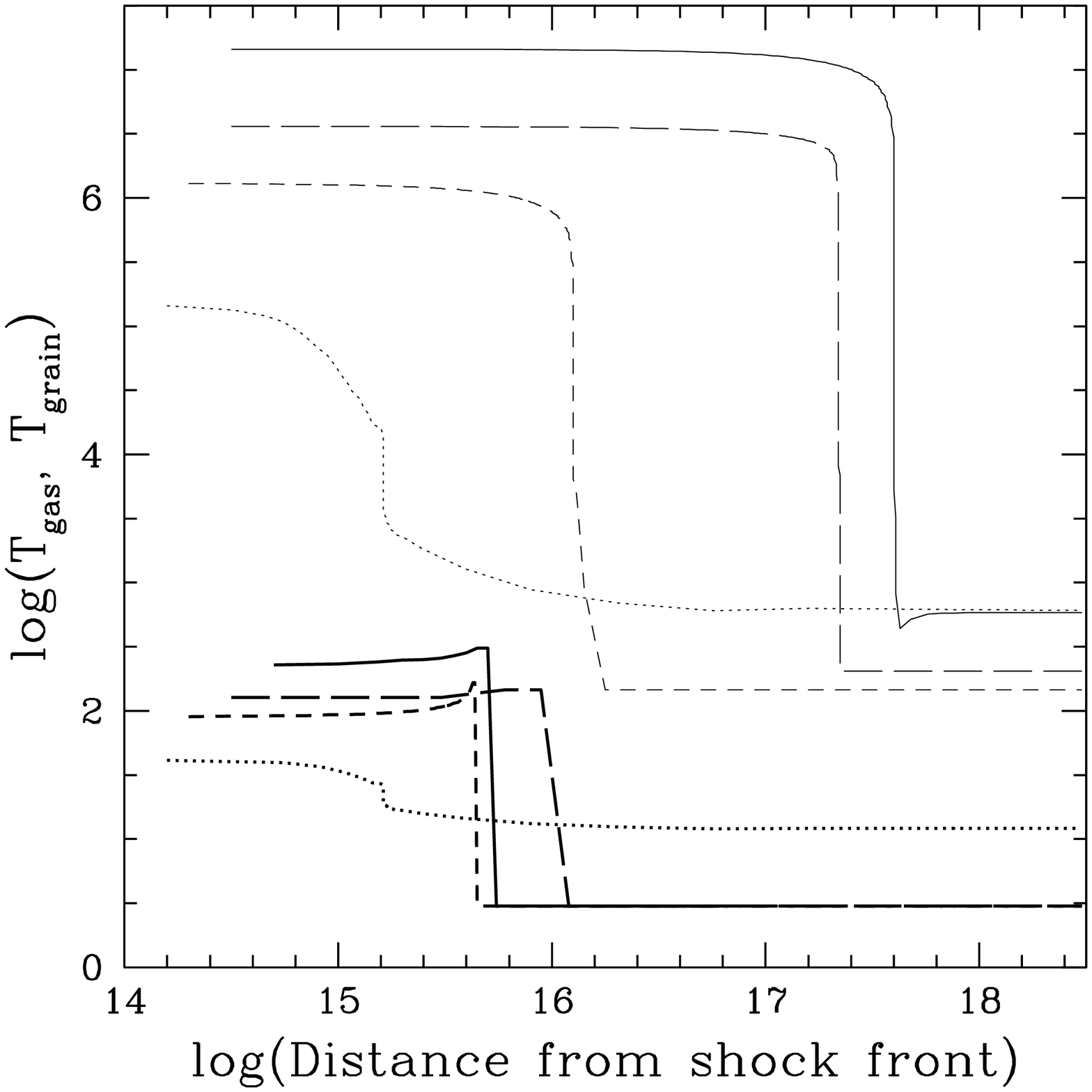}
\includegraphics[width=78mm]{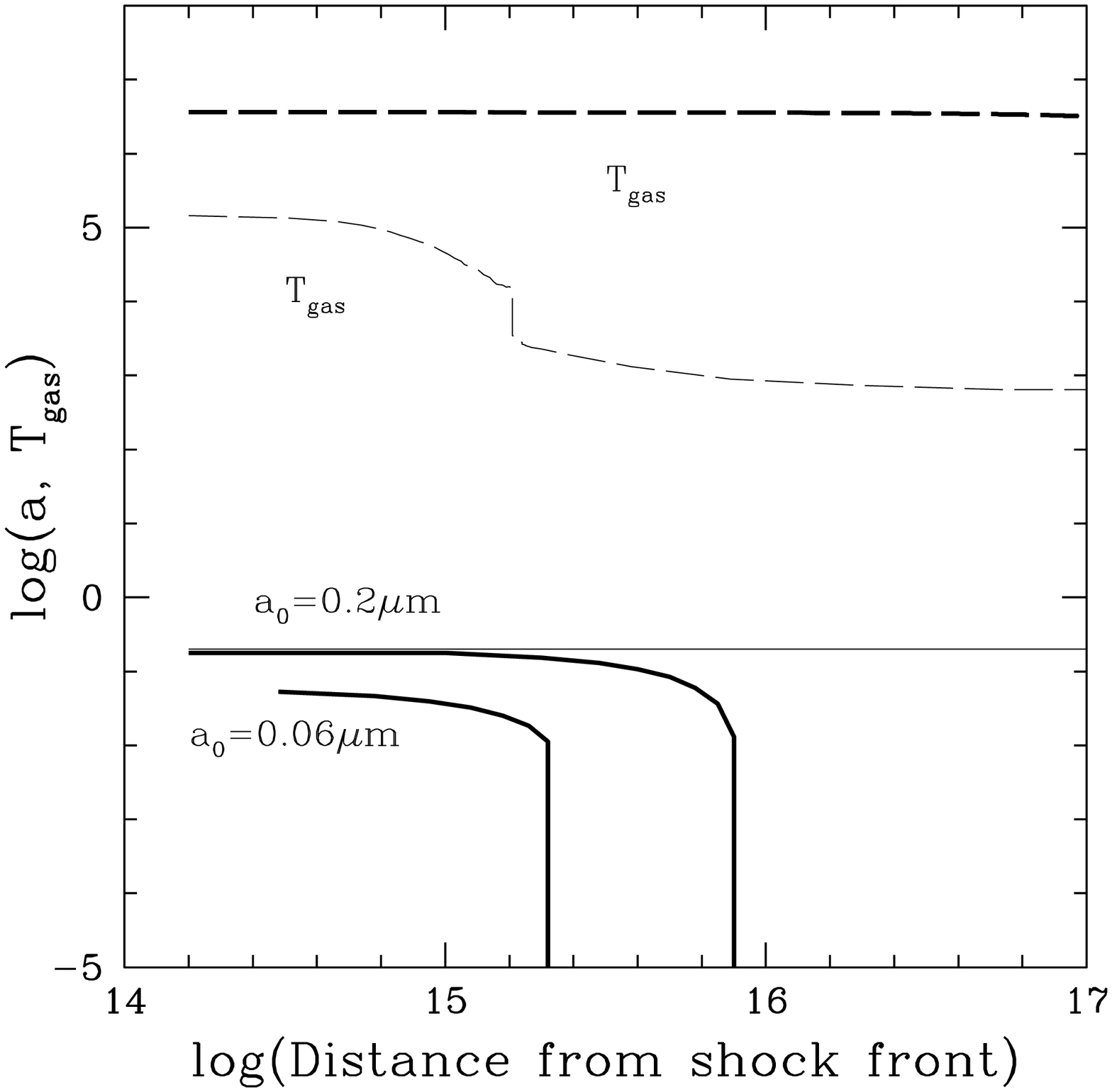}
\caption
    { {\it Left panel:} The profiles of the gas temperature (thin lines) and of
the dust grain temperature (thick lines) downstream for different models
(see text for symbols).
{\it Right  panel:} Grain radius, a (solid lines) for models with \Vs=500 \kms, 
\n0=300 \cm3 (thick lines) and \Vs=100 \kms, \n0=100 \cm3
(thin lines). The temperature of the gas (long-dashed lines) is also shown.}
\end{figure*}

The SUMA code is  described in CV01 and in Viegas \& Contini (1994).
The input parameters are the following:
the shock velocity, \Vs, the preshock density, \n0,
the preshock magnetic field, \B0,
the starburst age, $t$, the ionization parameter, $U$ (in number of photons / number of electrons),
the geometrical thickness of the clouds, D, and the dust-to-gas ratio,
$d/g$. Relative abundances to H of He, C, N, O, Ne,
Mg, Si, S, Ar, and Fe are cosmic (Allen 1973).

In our models we refer to clouds close to the starburst and therefore affected
by radiation and shocks with velocities $\geq$ 100 \kms.

Dust  is collisionally heated by the gas  across the shock front and downstream.
The temperature of dust follows the temperature of  gas and can reach very high
temperatures in the postshock region,
for \Vs\ $\ge$ 100 \kms (Figs. 1, left diagram).

The models in Fig. 1 (left)  are the following:
\Vs=1000 \kms, \n0=1000 \cm3 (solid lines);
\Vs=500 \kms, \n0=300 \cm3 (long-dashed lines);
\Vs=300 \kms, \n0=300 \cm3 (short-dashed lines); and
\Vs=100 \kms, \n0=100 \cm3 (dotted lines).
For all the models  $D$ = 1 pc,
\B0 = 10$^{-4}$ gauss and  $d/g = 4\times 10^{-4}$  are adopted.

The distribution on  grain size is automatically  derived by SUMA
which calculates sputtering in the different zone downstream of the shock.
The sputtering rate depends on the  gas temperature,  that is $\propto$ \Vs$^2$
in the immediate postshock region.
In Fig. 1 (right diagram) the distribution of grain radius downstream
in two cases of relative  strong (\Vs=500 \kms) and  weak (\Vs=100 \kms)
shocks  is shown for silicate grains with an initial radius a$_0$= 0.2 
and 0.06 \mm. 
In the high velocity case the sputtering rate is so high that grains with
a$_0$= 0.06 \mm ~are rapidly destroyed downstream.
So, only grains with large radius (a$_0$ $\geq$ 0.1 \mm) ~will survive
downstream, while small grains are  completely sputtered.
On the other hand, the grains survive downstream of low velocity shocks.
Graphite grains are more sputtered than silicate grains for T=10$^6$ K
(Draine \& Salpeter 1979)
Therefore,  in  regions where strong shocks  are created by the starburst,
we will consider relatively large grains, e.g.  silicate grains with an initial radius of 0.2 \mm.
Small grains (e.g. PAH) survive in the  extended galactic regions on scales of hundred pc and  lead to 
the  features  that  appear in the  spectral energy distribution (SED).

\begin{figure}
\includegraphics[width=78mm]{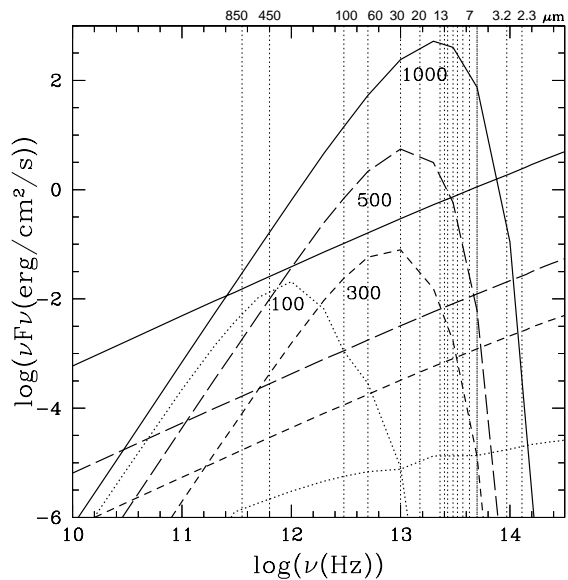}
\caption
 {The peaks of dust reradiation in the continuum
SED for different models. The corresponding bremsstrahlung radiation is also shown.
Small numbers refer to shock velocities (\Vs\ in \kms).}
\end{figure}

In Fig. 2   the  SED  of 
 bremsstrahlung radiation and emission by dust in the IR, corresponding to models in Fig. 1 (left diagram), is
presented. Some significant wavelengths (in \mm, top scale) ~are also plotted (thin dotted lines).

The peaks of dust reradiation in the IR correspond to maximum grain temperatures
of 314 K, 199 K, 143 K, and 41 K for \Vs\ = 1000, 500, 300, and 100 \kms,
respectively.

Notice that the bump corresponding to \Vs\ = 100 \kms\ is wide, because the
grains are less sputtered for low \Vs\ and a large region of grains emitting
at low temperature is present downstream (Fig. 1 left).
The   flux ratios between the dust bumps and the bremsstrahlung curve will
 change adopting different $d/g$ values.
In particular, the  flux from a starburst region at 2.3 \mm\  hardly corresponds to dust,
while for a high $d/g$  the flux at 3.2 \mm\  accounts for both gas and dust emission.

The  clouds in the
surrounding of a starburst  are heated and ionized by the light of the stars
with a distribution which depends on age (Cid Fernandes  et al. 1992),
and by secondary diffuse radiation from the  slabs of the gas heated by the shock.
The clouds move outwards from the starburst and shocks form on the outer edge of the clouds.
Radiation heating leads to  gas temperatures
below $\sim 3\times 10^4$ K, while gas heated by the shocks can reach
relatively high temperatures, depending on the shock velocity (Contini \& Viegas 2000).

The continuum emissions  by dust and gas  in the different frequency ranges  must be consistent.
Therefore, we will adopt the following procedure. First  we will  model the line emission spectra 
from the gas and then  will use the same models
to calculate the continuum SED in all the ranges. The emission of dust in the IR
depends on the intensity of the star cluster flux,  on mutual  cooling and heating
by collision with gas,  on the size of the grains which survive from sputtering,
and on the dust-to-gas ratio.

In previous papers (e.g. Viegas, Contini, \& Contini 1999, Contini, Viegas, \& Prieto 2002) 
it was shown that fragmentation of clouds in shocked regions
results from  R-T instability  in a turbulent regime.
 Therefore, the spectra emitted  from  the different regions of a galaxy
are modelled by multi-cloud models, i.e.  by models which account for the summed
spectra emitted from single clouds  adopting relative weights.
The relative weight, which are defined by the best fit to the data,
account for the relative number of clouds in the
conditions determined by the model, and for the  fraction of the emitting surfaces
Throughout  modelling 
we will  show the best  agreement to the data  of single cloud models 
in the different wavelength  domains in the SED, in order to explain the
role of each   type of cloud.  

Indeed, the models leak a small scale fit to dust IR features
corresponding to  cirrus and extended ISM regions which should be added.  
Nevertheless, modelling gas and dust in a consistent way, provides a rough information of
dust distribution in the different galaxies.
The dust-to-gas ratios will be derived by the fit of the dust infrared bump  
consistently with bremsstrahlung  at higher frequencies.

\section{Modelling the sample}

\begin{figure}
\includegraphics[width=68mm]{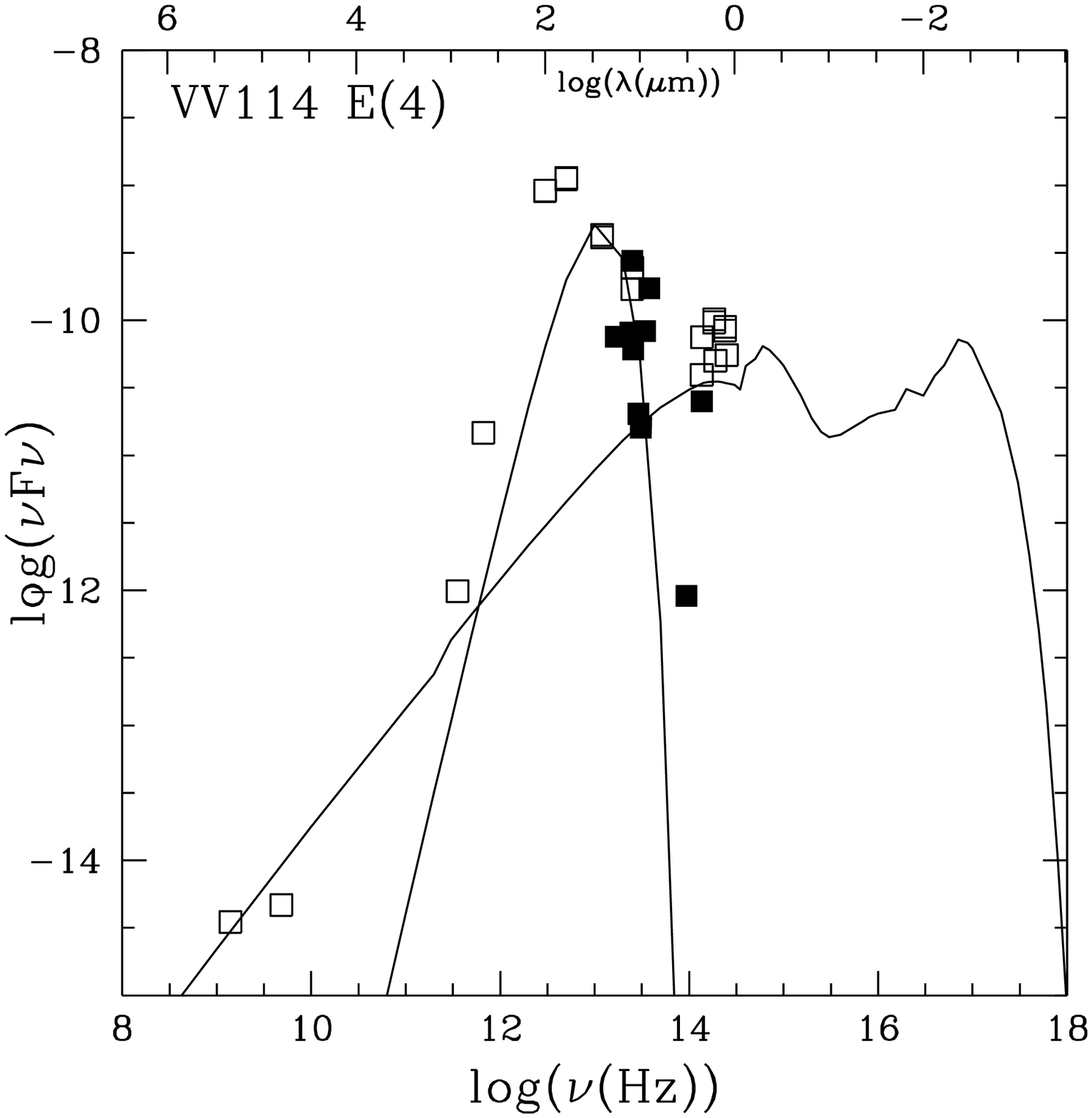}
\includegraphics[width=68mm]{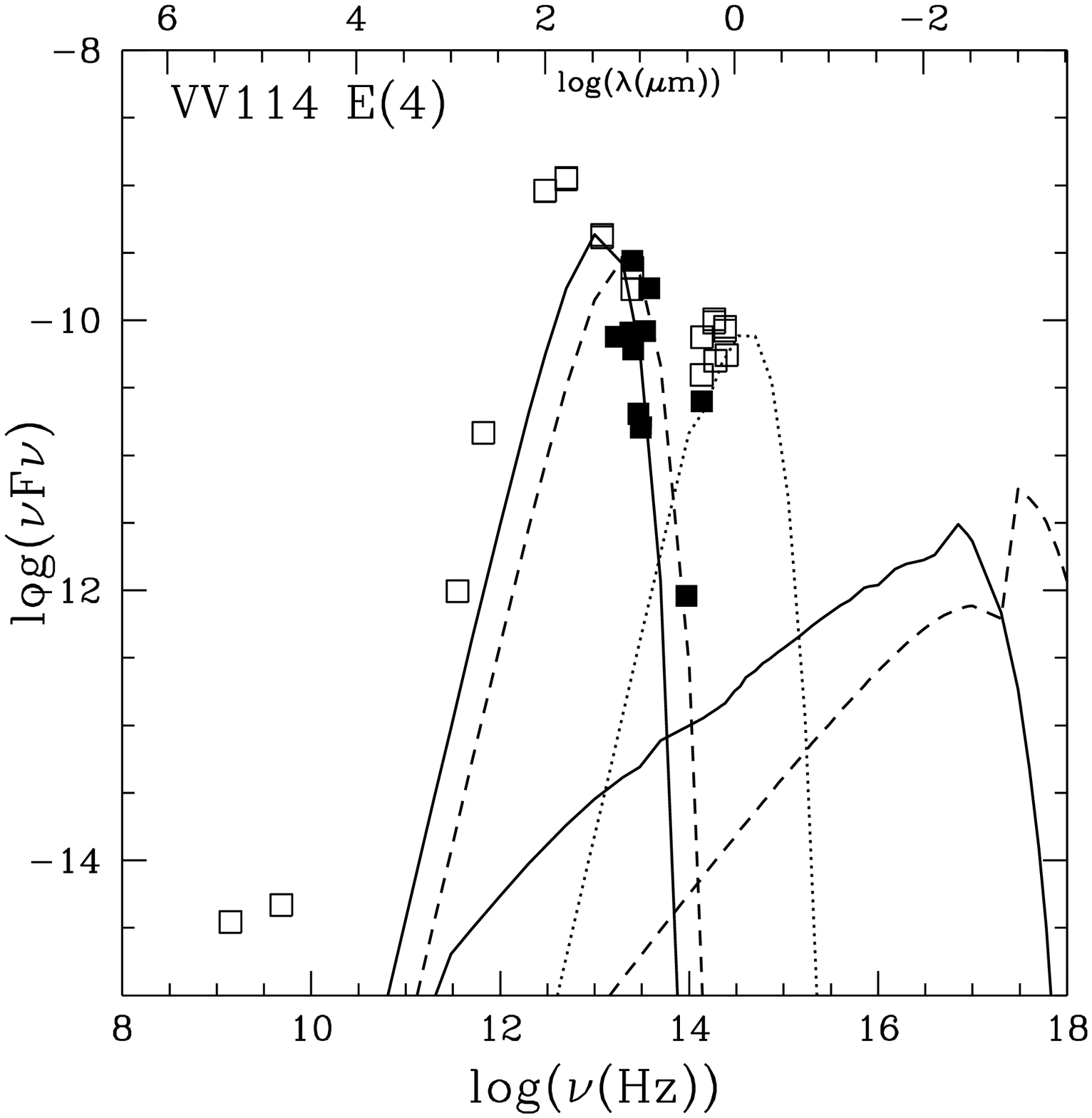}
\includegraphics[width=68mm]{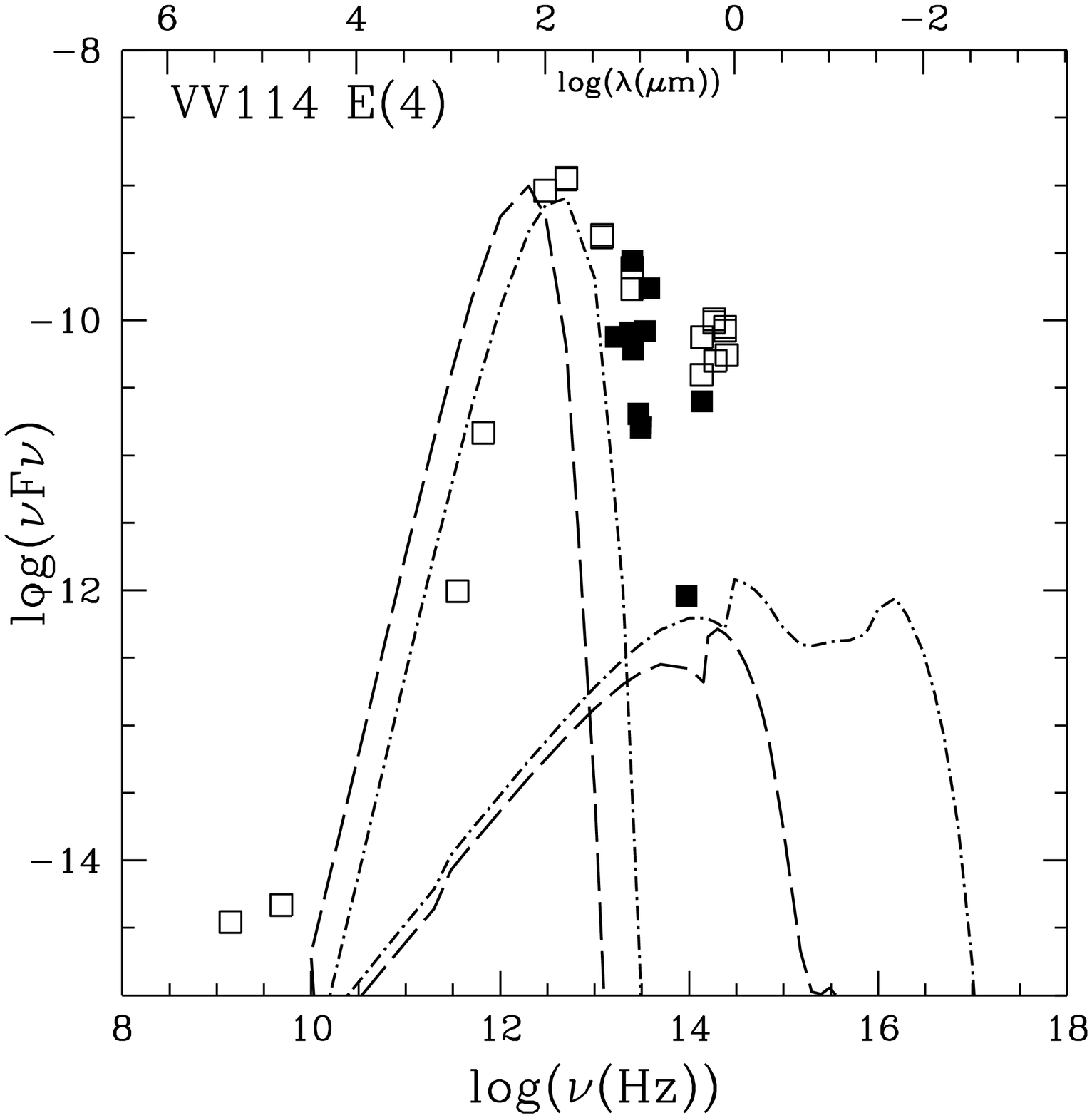}
\caption
{ VV 114 E(4) -- Single-component models. Filled squares refer to the MIR
photometry (Soifer et al. 2001), open squares to NED data (see text for details).
{\it top} The best fit of the MIR data by model M1. {\it middle} The fit by model M2.
{\it bottom} The fit of the FIR data by models M4 and M5. Different model values are
listed in Table 1  (see text).}
\end{figure}

\begin{figure}
\includegraphics[width=84mm]{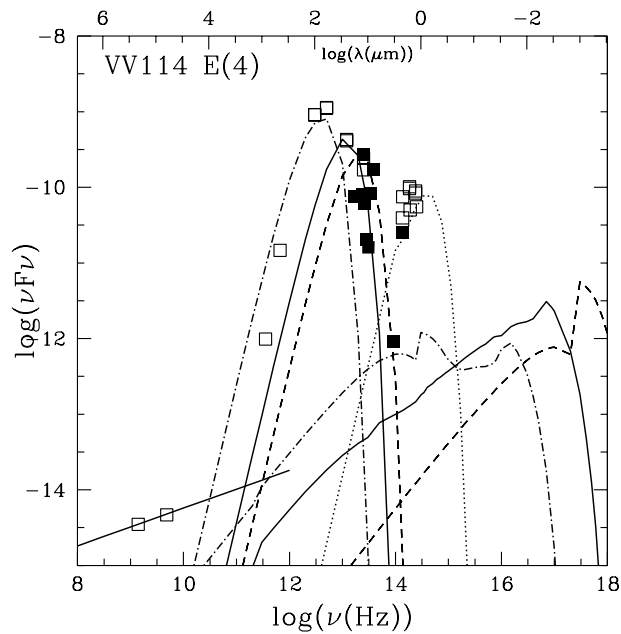}
\caption
{VV 114 E(4) -- All-component models.
The best fit of all the data. Same symbols as in Fig. 3}
\end{figure}

\begin{figure}
\includegraphics[width=84mm]{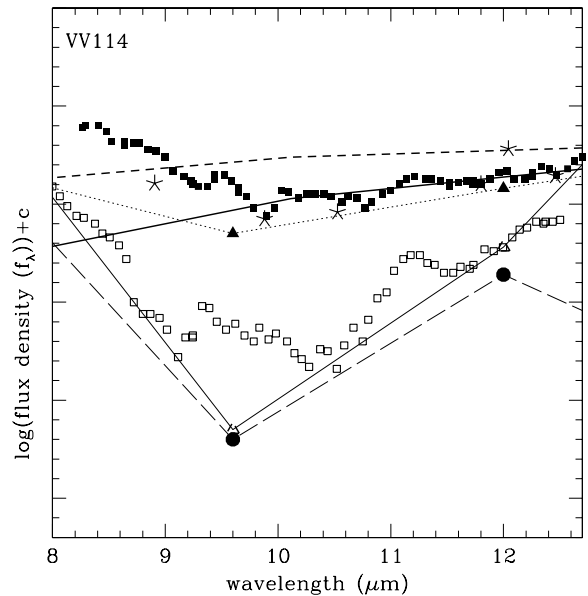}
\caption
{VV 114  -- 
open squares : VV 114E$_{NE}$ (Soifer et al 2002);
filled squares : VV 114E$_{SW}$ (Soifer et al 2002);
open triangles : broad band observations in the nucleus (Le Floc'h et al. 2002) ;
large filled circles : diffuse emission;
filled triangles : VV 114W, knot 2 (Le Floc'h et al. 2002);
asterisks : MIR data (Soifer et al 2001)
}
\end{figure}

In the following, we model the individual SED and optical emission-line ratios of 
S01's sample of galaxies.

\subsection{VV114}

\begin{table*}
\centering
\caption{VV 114 N. Line intensities relative to H$\beta$}
\begin{tabular}{llllllllll}
\hline
line &  obs &   M1       & M2      &M3    &M4    &M5     & SUMN\\ 
\hline
\ [OII] 3727 &  & 0.81 &9.(-4) &2.6 &2.7& 25.&0.52\\
\ [OIII] 5007+4959 &1.6  &1.44&1.63 &2.21&3. & 15.3&1.66  \\
\ [OI] 6300+6360&0.13 &0.043&0.0     &0.66&2.5&0.5&0.177\\
\ [NII] 6584+6548 &1.0  &0.87&2.3(-4)& 2.64&7.14&8.7&0.79  \\
\ [SII] 6716+6731 &0.71 &0.08&5.3(-6) & 3.47&7.8&7.6&0.58  \\
\ [SII] 6716/6731 &1/0.28:  &0.47 &0.47&1.06 &1.2&0.8&0.53\\
  \Hb\ (\erg)   & -  & 0.664&0.037&0.016&0.044&0.0025&0.25\\
   w         &-&       0.033 & 1.  &0.133&0.083 & 0.0167\\
\hline
\end{tabular}
\label{tab1}
\end{table*}

\begin{table*}
\centering
\caption{VV 114 S. Line intensities relative to H$\beta$}
\begin{tabular}{lllllllllll}
\hline
 line & obs&  M6   &M7  &M8     &M3   &SUMS  \\ 
\hline
\ [OII] 3727 & 0.14: & 0.8&0.013 &0.014&2.6&0.6    \\
\ [OIII] 5007+4959 &2.9  &3.48&3.97 &4.0&2.21 &3.5   \\
\ [OI] 6300+6360 &0.1  &0.064&0.0 &6.(-8)&0.66 &0.15 \\
\ [NII] 6584+6548 &0.8  &0.61&0.007 &0.0075&2.64&0.63    \\
\ [SII] 6716+6731 &0.67 &1.99&0.001 &0.0012&3.5 &0.9 \\
\ [SII] 6716/6731 &1/0.71  & 0.95&0.92&0.92& 1.06&0.95 \\
\  \Hb\ (\erg)   & -   &0.12&0.12&0.016&0.187& 0.12\\
\    w        &- &     0.067  &0.0167&0.0167&1.   &- \\
\hline
\end{tabular}
\end{table*}

VV114 is a pair of interacting galaxy, particularly interesting
for its  high infrared luminosity (10$^{11.3} L_{\odot}$) and high 
FIR-to-visible flux ratio.
VV 114 is composed of two stellar components separated by 15\arcsec, designated 
by VV 114 E and VV 114 W (Knop et al 1994). 
The extreme NIR colors of VV 114 E indicate the presence of a large concentration of dust, 
while VV 114 W is relatively unobscured at optical wavelengths (Knop et al. 1994).  
The near-infrared color-color diagram presented by Doyon et al. (1995) shows that 
"the intrinsic SED of VV114 is affected by three components, namely blue stars, 
nebular emission, hot dust, and extinction".

In Tables 1 and 2 the observed emission-line ratios in VV 114 N and 
VV 114 S, respectively, are 
compared with model calculations, and in Figs. 3 and Fig. 4 
the SED of the continuum for VV 114 E 
(see Soifer et al. 2001 for the nomenclature) is presented. Filled squares refer to 
the MIR photometry in a 4\arcsec\ diameter beam (Soifer et al. 2001), open squares to 
the NIR, IRAS and radio-cm data found in NED. 

We show in  Figs. 3 the modelling of data in the different wavelength ranges in three 
different diagrams (top, middle, and bottom), in order to explain our procedure,
while in Fig. 4 the best fit of the data by a multi-cloud model is shown.
The models are not summed up, in order to better understand  the role of each
single-cloud model.

Table 1 shows that model M1  explains roughly the emission-line ratios.
Indeed, the [OIII]/[OI] ratio is relatively high, but is lower than for other models.
This model could explain the SED (Fig. 3 top) adopting $d/g = 8\times 10^{-4}$, 
while the models in CV01 were all calculated with $d/g=4\times 10^{-5}$.
Notice, however, that the datum representing emission at 3.2 \mm\ strongly constrains the
modelling, and a more consistent fit to bremsstrahlung emission as well as to dust 
reradiation is presented in Fig. 3 (middle), where   model  M2 (solid lines) similar to M1 was run with 
$d/g = 0.032$ and $U=3.5$ to recover the fit of the line ratios.
Models  M1 and M2 correspond to a stellar population age of 3.3 Myr, a relatively high
shock velocity (\Vs\ = 500 \kms) and a preshock density \n0\ = 300 \cm3.
 Kim et al. (1995) give  FWHM $\sim$ 400 \kms   for the
[OIII] line profile  of VV 114.
Models with a higher age and/or with a reduced geometrical thickness give similar 
results  regarding the continuum,
because the effect of the shock prevails on radiation at high \Vs.

The NIR data from the NED show that there is a non negligible contribution from 
an old stellar population of about 5\,000 K (Fig. 3 middle, dotted line). 

Although model M2 explains satisfactorily S01 data both in the
bremsstrahlung and in the MIR ranges, the contribution of 
high velocity (\Vs\ = 1000 \kms)  high density (\n0\ = 1000 \cm3) dusty clouds 
with $d/g =0.02$ (short-dashed lines) is invoked to explain MIR 
data at higher frequencies. 
The maximum temperature of the grains corresponding to the high velocity
shock is 330 K. This model is radiation-bound and the temperature of the gas
in the downstream region behind the shock  never drops  below $1.2\times 10^7$ K. 
Therefore, these clouds do not contribute to the optical lines.

The IRAS FIR data indicate that lower shock velocities are also present in the emission 
regions. IRAS data at about 100 \mm\ can be  explained by models
with \Vs\ = 100 \kms\ and \n0\ = 100 \cm3\ (Fig. 3 bottom).
Model  M3 is selected because of  very high [OI]/\Hb\ and 
[SII]/\Hb\ emission-line ratios which could improve the fit of the summed 
spectrum (SUMN). 
Also in this case  model M4 similar to M3 but with a high $d/g = 0.016$ value has been adopted (long
dashed lines)),
in agreement with the high $d/g$ ratio reported by Frayer et al. (1999). 
The emission-line  ratios  which result from M3 and M4 are very different.
Moreover,  the emission at 60 \mm\ can be better explained  by emission 
from  hotter dust, i.e. a shock of \Vs\ = 200 \kms\ must be used 
(short dash-dot lines in Fig. 3 bottom).
The corresponding  line ratios appear in column 7 (M5) of Table 1 .
The low \Vs-\n0\ models  correspond to a younger population ($t$ = 0.0 Myr), while 
model M1 corresponding to an older population shows high \Vs-\n0\, in agreement 
with Viegas, Contini \& Contini (1999), namely, that an old starburst corresponds to 
high \Vs.

The two sub-mm data points at 450 \mm\ and 850 \mm\ (from Frayer et al. 1999) 
show that the FIR bump intensity corresponding to the low velocity models must 
be reduced. Considering that there is a single datum (at 3.2 \mm) which constrains 
the models,  and consistently with the line spectra, 
we reduce both gas and dust emission of model M4. The 
fit of the dataset available for VV 114 E is presented in Fig. 4. 
The two radio data follow the trend of synchrotron emission,
confirming the presence of strong supernova and stellar winds in the galaxy. 

Model M5 overpredicts the datum at 850 \mm.
Dust becomes optically thick for $\tau$ = $\pi a^2$ (d/g) n $\cal D$ Q$_{abs}$ $\geq$ 1.
Adopting a $\sim$ 0.02 \mm (radius of the sputtered grains), d/g= 6 10$^{-13}$ (by number), n= 3 10$^3$ \cm3
(density downstream for model M5), and Q$_{abs}$ =2 (Contini 1992), the absorbing length, $\cal D$, 
results  $\sim$ 2 10$^{19}$ cm, similar to the geometrical thickness of the clouds.
Notice that the grain radius, the temperature, the density, etc change downstream
with distance from the shock front, therefore,  absorption changes accordingly.

In Fig. 5 we compare the      MIR spectra of VV 114 separate nuclei observed by Soifer et al (2002, fig. 1))
with    MIR ISOCAM broad-band photometry by Le Floc'h et al.  (2002).
The depression at 10 \mm ~in the  spectrum of  VV 114E$_{NE}$  
(open squares) is  deeper than for VV 114E$_{SW}$ (filled squares) (Soifer et al 2002).
VV 114$_{NE}$ also shows the peak at 11.3 \mm ~and the shoulder at 8.6 \mm,
characteristic of PAH. Le Floc'h et al. broad band observations in the nucleus (open triangles,
connected by a solid lines) and diffuse emission (large filled circles connected by a long-dashed line) 
are directly compared
with the data for VV 114E$_{NE}$ and show a rough high scale similarity, however, they
leak the substructures at about 9.5  and 11 \mm.
Le Floc'h et al observations of VV 114W (knot 2) which correspond to  filled triangles,
connected with a dotted line, fit  Soifer et al (2002) VV 114E$_{SW}$ data at wavelengths
$>$ 9 \mm ~but leak the shoulder by PAH at lower wavelength.

The    MIR data  from the Keck telescope by Soifer et al. (2001) VV114E(4)  (asteriks) are given for comparison
in Fig. 5   as well as the models which better explain them  (Fig. 3, middle diagram). 
The data follow the trend of Soifer et al (2002) for VV 114E$_{SW}$  at large wavelength.
Recall that  absorption reduces the emission corresponding to the high velocity model (\Vs=1000 \kms)
(short dashed line)  more than  the emission corresponding to \Vs=500 \kms (solid thick line),
because densities downstream  increase with \Vs ~and \n0.

PAH emission is a consistent  fraction f$_{PAH}$    of dust emission in the IR because their features are observed.
Considering that the  flux  of silicate dust, integrated in the IR domain, is evaluated   by  models  to $\sim$ 10$^3$ \erg,
and adopting that PAH emission is  20\%  of the  total IR luminosity, 
10$^{11.3}$ L$_{\odot}$ (Sect. 2), the radius of  PAH  extended region is $\sim$ 37/(f$_{PAH}$)$^{1/2}$ pc.
PAH emission is observed up to some hundred pc scales, therefore f$_{PAH}$ $\sim$ 0.01 and hardly affects model
physical processes in our model calculations.

About the MIR domain, PAH flux from Siebenmorgen, Kr\"{u}gel, and Laureijs (2001, fig. 6) integrated between 5 and 12 \mm
~is $\sim$ 10$^{-9}$ \erg, while the flux in the same wavelength range calculated with our high
velocity model which peaks in the MIR, is $\sim$ 2 10$^{-9}$ \erg. the discrepancy is negligible,
considering the approximation of modelling and indicates that PAH emission and silicate dust emission
are similar.

Optical spectroscopic data for VV 114 S (from Veilleux et al. 1995) are compared 
with models in Table 2. 
Notice that the [OIII]/\Hb\ emission-line ratio is higher than 
in VV 114 N. To reproduce such a ratio a stellar population with an age of 2.5 Myr is 
adopted (models M6, M7, and M8). The model corresponding to a young population (M3) 
is also added to better fit the data.
Tables 1 and 2 refer to VV 114 N and S respectively, 
while the continuum data refer to VV 114 E. So we cannot refer to  well 
located regions and the continuum relative to star with $t=2.5$ Myr is also shown 
in Fig. 3 (long dash-dot).

The weighted sum of the single contributions appears in the last column of 
Tables 1 and 2.
We have added also the models calculated with a low $d/g$, as they appear in CV01.
In fact, dust has a clumpy distribution and cannot affect in the same way the emission 
from the overall galaxy.
The weight $w$ adopted for each single model are given in the last row of 
Tables 1 and 2.
They  correspond  roughly to those obtained to fit the
SED of the continuum. Actually, the large range of emission-line ratios, 
corresponding to the different models, indicate that many different physical 
conditions coexist in VV 114.

To determine the amount of dust  in the different regions of VV 114.
 specific emission in the different regions are compared 
with model results in Fig. 4. The data in the NIR domain are few, moreover, they are explained
by black body emission from the old stellar population background rather than 
by bremsstrahlung. So,  we will use the
relative weights, which will indicate in this case the intensity of dust emission in the NIR.
Therefore, we normalize the  relative weight w of  model M2,  which best fit the data
relative to dust emission,  to that of model BB representing the old stellar 
population, corresponding to black body emission of $T=5\,000$ K, in the different regions.
Table 3 shows w in columns 2 and 3 for M2 and  BB, respectively,
 and the ratio of the weights in column 4.
In the second row  we refer to the VV 114 E region represented in Fig. 3.
Rows 3, 4, and 5  refer to regions whose SED is represented in Fig. 4 diagrams.
The  w referring to the IRAS data  appear in the last row.

We consider now the ratios  w(M2)/w(BB) (Table 3, last column), in order to  
compare the relative amount of dust in the different regions of VV 114. 
Adopting a homogeneous distribution of old stars, the results  
suggest that VV114 E SW and NE regions show about 
the same  amount of dust, while W4 region is less dusty, in agreement with 
the results of Knop et al. (1994), namely that VV 114 W is relatively 
unobscured at optical wavelength. IRAS data (Fig. 4 right diagram) show a 
higher flux in all domain, besides  less dust than for VV114 E.

\begin{figure*}
\includegraphics[width=42mm]{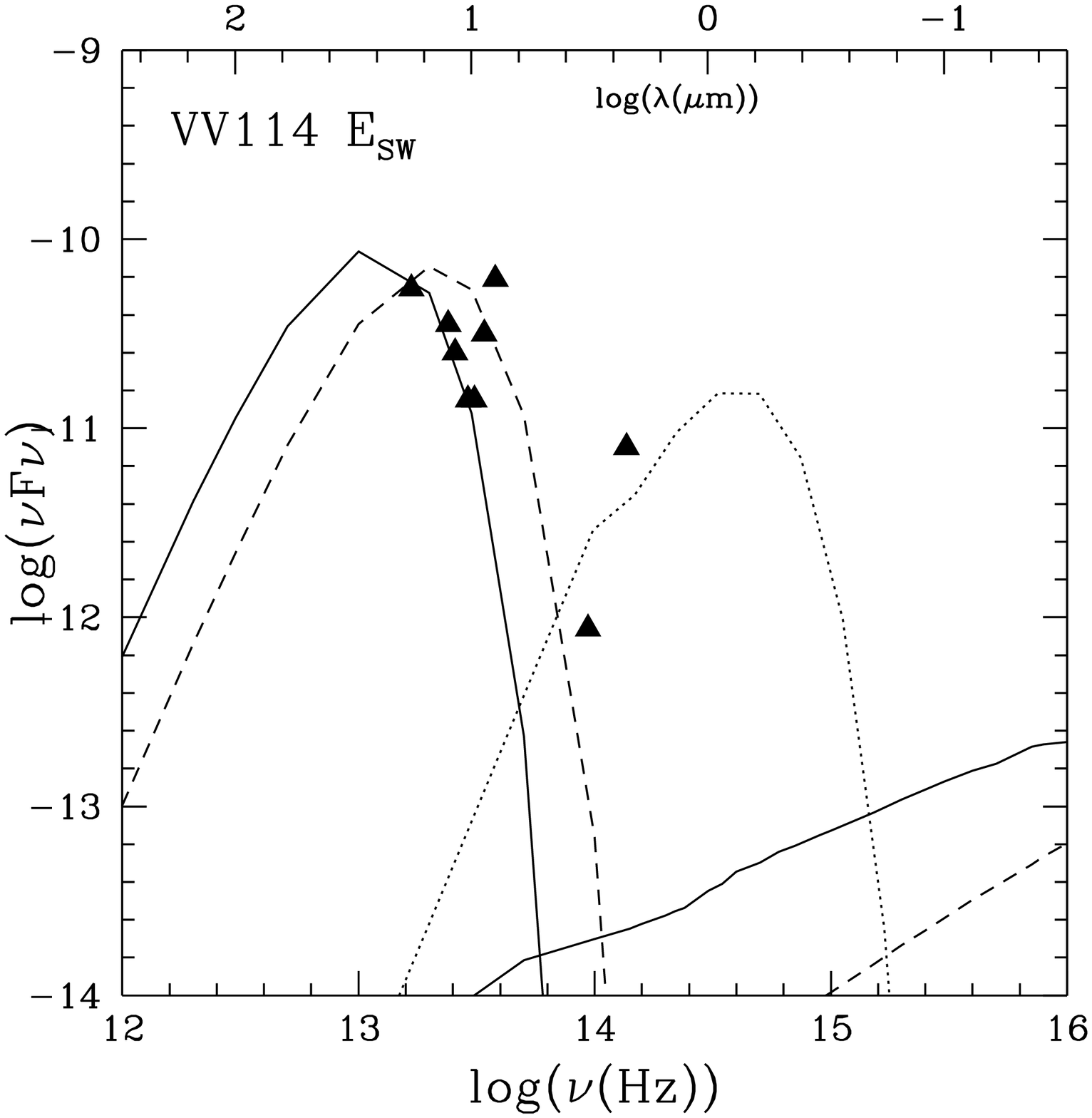}
\includegraphics[width=42mm]{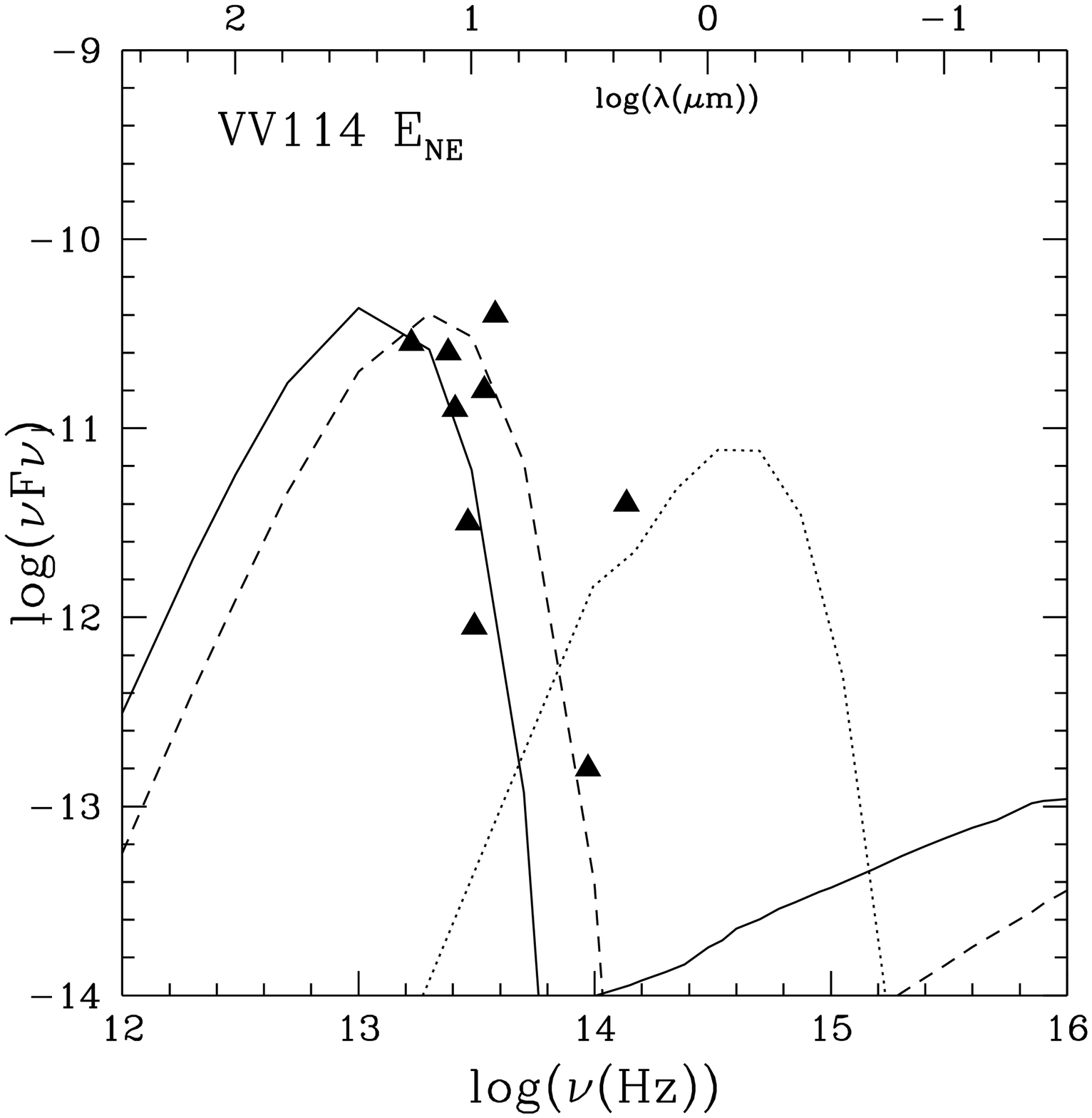}
\includegraphics[width=42mm]{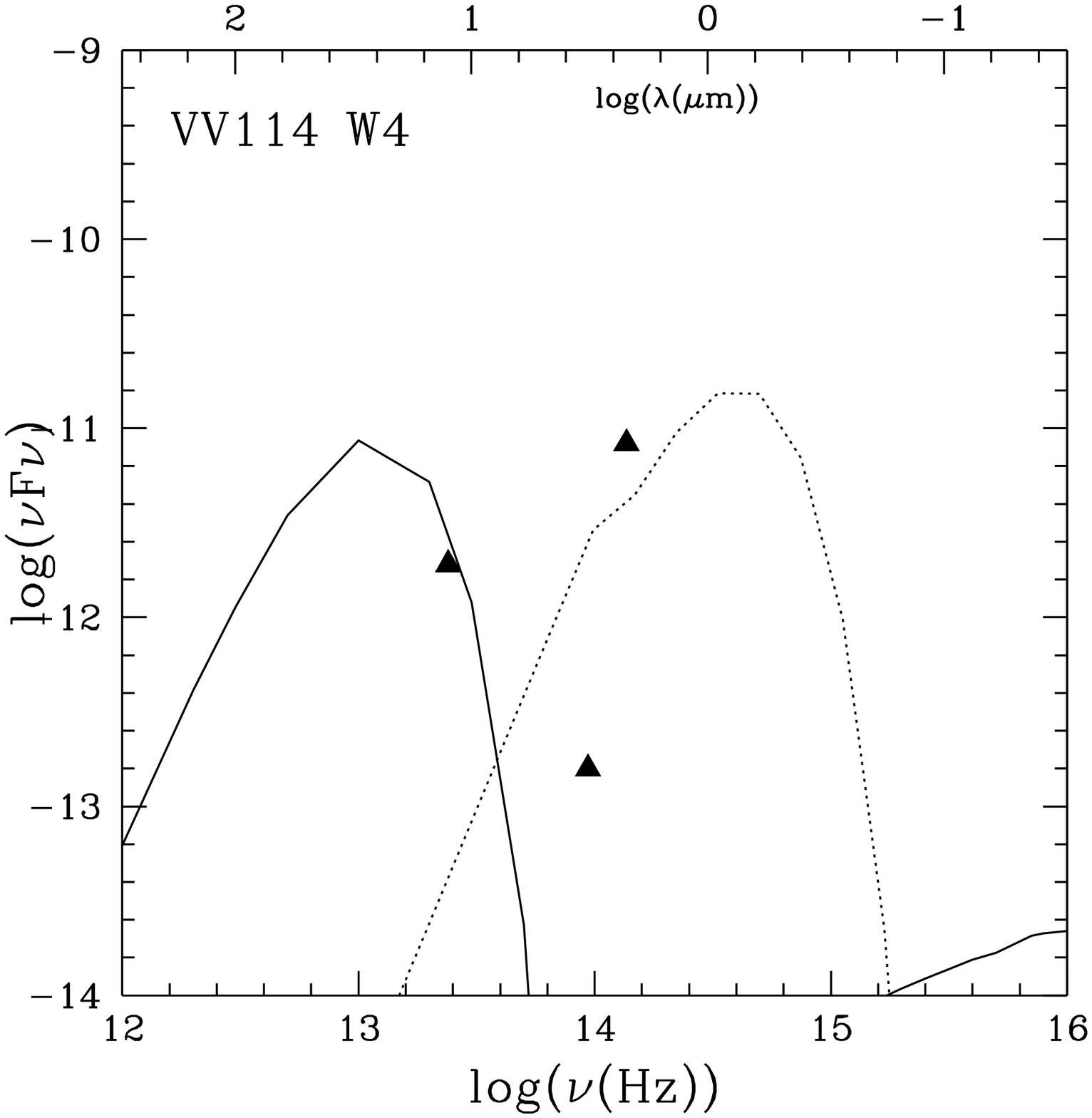}
\includegraphics[width=42mm]{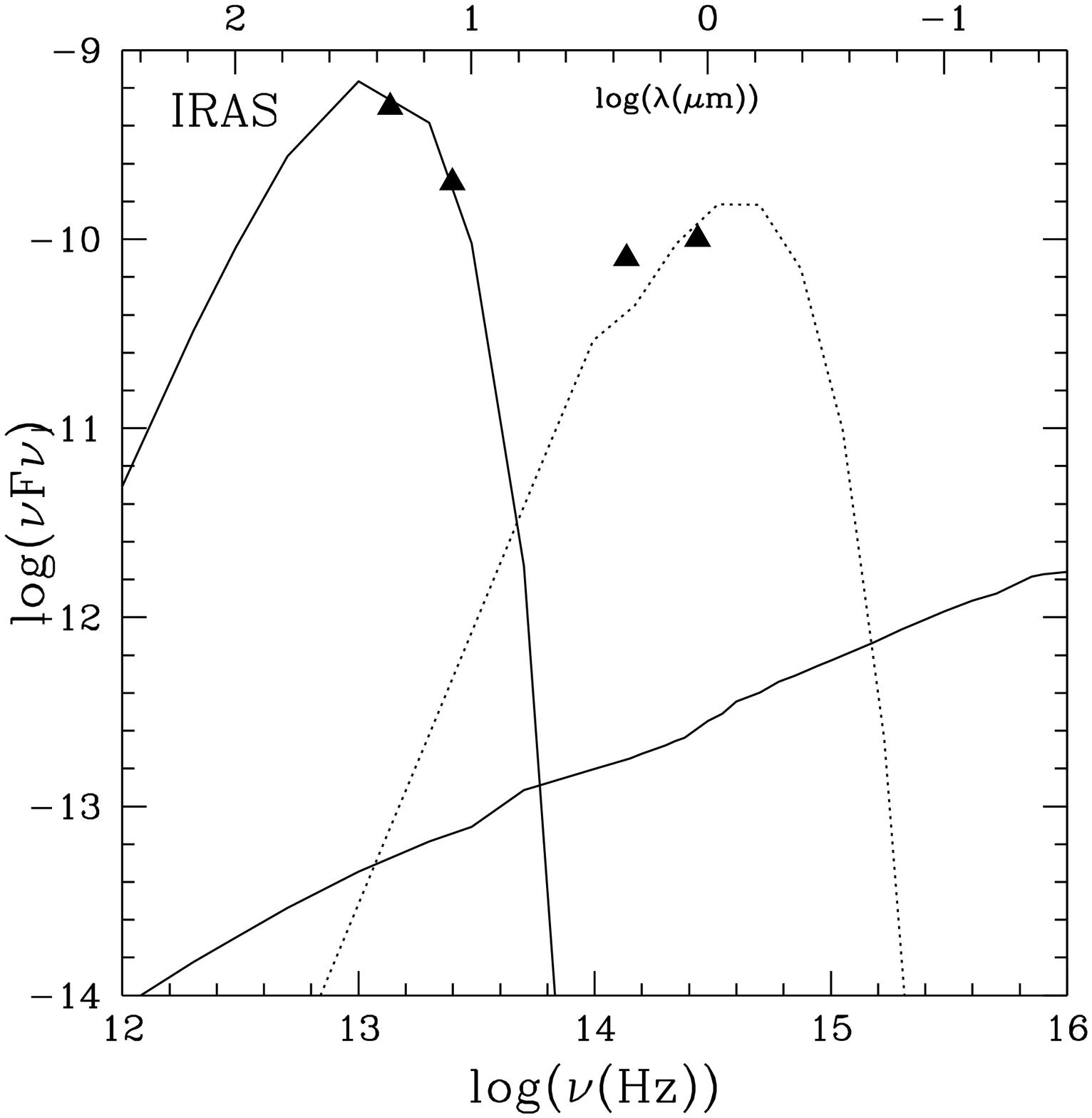}
\caption
{The fit in different regions of VV 114: E$_{\rm SW}$ -- E$_{\rm NE}$  -- W4 and  IRAS data.}
\end{figure*}

\begin{table}
\caption{Relative weights}
\begin{tabular}{lccclllll} 
\hline
\ region   & log(w(M2)) & log(w(BB$^1$))& w(M2)/w(BB$^1$)  \\ 
\hline
\ VV 114\,E  & -11.05  & -20.5 &  2.82(9) \\  
\ E$_{\rm SW}$  & -11.75  & -21.2 & 2.82(9) \\
\ E$_{\rm NE}$ & -12.05   & -21.5 & 2.82(9)  \\
\ W4       & -12.75  & -21.2  & 2.82(8) \\
\ IR       & -10.85  & -20.2  & 2.24(9) \\ 
\hline
\end{tabular}

$^1$  $T$ = 5\,000 K
\label{tab3}
\end{table}

Images of VV 114 taken at different wavelengths  (see Fig. 1 of Soifer et al. 2001)
show  different characteristics.
For example, at 2.2 \mm\ the NIR emission is black body from old stars 
(\Ts\ = 5\,000 K, see Fig. 3) with perhaps some contribution from 
bremsstrahlung  from gas heated and ionized by the starburst and by 
diffuse emission downstream. Therefore we see a quite complex picture at 2.2 \mm\ 
and the shock fronts disrupted by R-T instability can also be recognized. 
At 3.3 \mm\ the image reveals dust reradiation downstream  the high velocity shock. 
At 12.5 \mm\ the image shows the maximum of dust emission from clouds near a rather 
aged starburst. At this wavelength dust emits  from the postshock region of a high velocity 
shock. The emission has therefore a clumpy distribution, considering that the high 
velocity shock fronts are disrupted. Moreover, the clumps are relatively small, because 
the grains are rapidly sputtered in the post shock region.
Unfortunately there is no image at 
longer wavelengths which could show the distribution of low velocity gas and dust.
 Frayer et al. (1999) show images of VV 114 at 450 and 850 \mm. 
The images are different indicating that the two data in Fig. 3 belong to
different models. The datum at 450 \mm\ represents dust
corresponding to a model with \Vs\ = 200 \kms\ (model M5) and the other one
corresponds to a lower shock velocity (100 \kms). In fact the image
is less tormented and more homogeneous.

The continuum SED in the radio  range in Fig. 4 
shows that radio emission at 8.4 GHz is synchrotron radiation created  near the shock front. 
So, the image at 8.4 GHz shows some clumpiness.

\subsection{NGC 1614}

%\begin{figure}
%\includegraphics[width=78mm]{figure11.eps}
%\includegraphics[width=78mm]{figure12.eps}
%\caption
%{NGC 1614 -- Multi-component models. Filled squares refer to the MIR
%photometry (Soifer et al. 2001), open squares to NED data (see text for details).
%{\it top} The fit of the NIR data. {\it bottom} The fit of the FIR data. Different model
%values are listed in Table 4.}
%\end{figure}

\begin{figure}
\includegraphics[width=84mm]{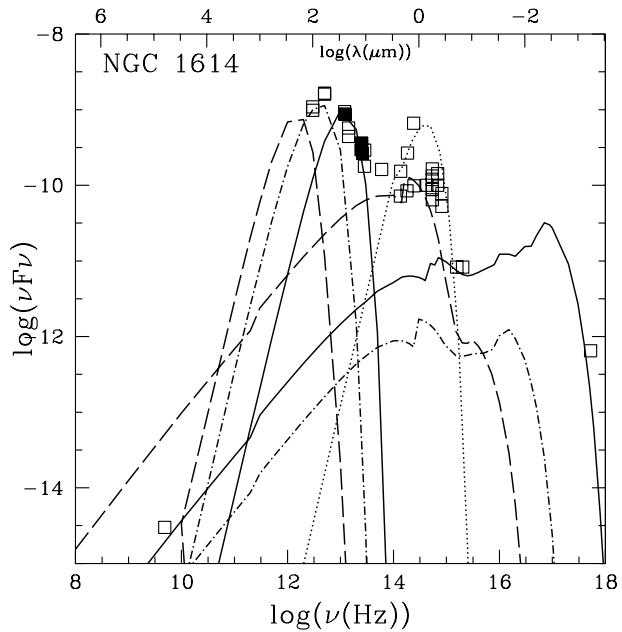}
\caption
{NGC 1614 --  The best fit of all the data by a
multi-cloud model. Same symbols as in Fig. 2}
\end{figure}

\begin{table*}
\caption{NGC 1614 N. Line intensities relative to H$\beta$}
\begin{tabular}{llllllll} 
\hline
 line &  obs$^1$ & M9     &  M10       &   M3    &M5    &SUM\\ 
\hline
\ [OII] 3727 & 0.8&0.14&5.(-6)&  2.6&25.&0.9 \\
\ [NeIII] 3869 & 0.51&10.15&0.1&0.33&3.1&0.4\\
\ [SII] 4071& 0.11: &1.3&0.0 &0.16&1.0&0.08\\
\ [OIII] 5007&0.85&0.84&0.11 &2.21 &15.&0.87 \\
\ [NI] 5200 & 0.033&1.(-4)&0.0&0.03 &0.019&0.011\\
\ HeI  5876 & 0.07&0.002&0.008 & 0.29& 0.13&0.1\\
\ [OI] 6300&0.093 &0.002&0.0&0.66&0.5&0.23 \\
\ [NII] 6584&2.0  &0.086&0.0   & 2.64&8.8&1.0   \\
\ [SII] 6716+6731 &0.56+&0.10&0.0& 3.47 &7.7&1.2 \\
\ [SII] 6716/6731 &1/1.43&0.47 & &  1.06&0.79&- \\
\ [OII] 7325 & 0.053&0.014&3.(-7)&0.06&1.32&0.02\\
\  \Hb\ (\erg)   & 3.3 (-13)  &0.44& 0.14 &0.016&0.003&-\\
\  w          & -& 0.002& 0.2 & 1. &0.02&-\\
\hline
\end{tabular}

$^1$ from Vaceli et al. (1997), observed at earth  
\label{tab4}
\end{table*}

NGC 1614 is a strongly interacting galaxy showing crossed tails on photographic 
plates (Arp 1966). High-resolution NIR images and spectra (Alonso-Herrero et al. 2001) 
reveal a strong nuclear starburst ($\sim 100$ pc diameter) surrounded by a ring 
($\sim 400$ pc diameter) of brigh HII regions. The detection of numerous Wolf-Rayet 
stars in the integrated nuclear spectrum of this galaxy (e.g. Vacca \& Conti 1992) 
indicates that the starburst is very young (between 2 and 6 Myr).
The optical spectrum  observed by Vaceli et al. (1997)  
(see Table 4)  is used to cross-check the  continuum. 
The few line ratios  by Veilleux et al. (1995) are in agreement with those 
of Vaceli et al. (1997). Model 
M9 (CV01) is selected because i) the line ratio [OIII]/\Hb\ is in agreement with the data, ii) 
the [OIII]/[OI] ratio, even if high, is lower than for other models, and iii) 
particularly, because  the MIR bump for VV 114 is nicely fitted
by models with \Vs\ = 500 \kms.  
Black body radiation from stars with \Ts\ = 5\,000 K (dotted line) is also revealed between 
1 and 3 \mm. 

The dust peak in the FIR indicates that  clouds with
lower shock velocities  contributes to  the  
spectra, i. e. model M3 (CV01, $d/g=4\times 10^{-5}$,
dashed lines) and clouds corresponding
to  \Vs\ = 200 \kms\ and a higher $d/g = 0.024$ 
(M5, dash-dot lines). A $d/g = 4\times 10^{-5}$ is a higher limit for model M3; a lower $d/g$ will not 
change the emission-line ratios. The high $d/g$ used for clouds at \Vs\ = 200 \kms\ 
is also constrained by the data at $1.57-2\times 10^{15}$ Hz.

Fig. 7  shows that bremsstrahlung radiation from a cool gas overpredicts
the datum in the radio range.  Indeed, absorption of free-free  radiation 
increases with wavelength and can reduce drastically the bremsstrahlung from cool gas.
At 5 GHz gas with a density of 10$^3$ \cm3 and T between 10$^3$ - 10$^4$ K becomes
optically thick for distances of 4 - 90 pc (Osterbrock 1989).

 The high velocity model 
represents a stellar population with an age of 3.3 Myr, \Vs\ = 500 \kms, \n0\ = 300 \cm3, 
$U$ = 10, and $d/g = 4\times 10^{-3}$ (M10, represented by solid lines), 
 similar to model 12(10a) in CV01, but with a high $d/g$ (4 10$^{-3}$).
This model  explains both the data in the soft X-ray from {\sl Einstein} 
satellite and the radio measurement at 4.85 GHz (Griffith et al. 1995). 
Model M9 with a low $d/g$ contributes to the emission-line ratio and is included in Table 4. 
Both high and low d/g clouds are accounted for.

Kim et al. (1995) give  [OIII] line FWHM of 310 \kms, which is lower than the
high velocity found. This could be a rough signal of galaxy interaction, namely,
in some region of the system the shock velocity is higher than the real gas velocity,
due to head-on collision of clouds. Low shock velocities,
on the other hand,   correspond to regions farther from the starburst bulk,
or  shocks resulting from head-on-back collisions.

The emission-line ratios of all the models used to explain the continuum and 
the optical line spectrum are presented in Table 4 and  the weighted sum appears in
the last column. The corresponding relative weights are given in the last row.
The [OI]/\Hb\ emission-line ratio is overpredicted by a factor of $\sim$ 2 and the 
[NII]/\Hb\ emission-line ratio is underpredicted by the same factor. 
[NI]/\Hb\ is also underpredicted indicating a  N/H
higher than cosmic ($9.1\times 10^{-5}$). Indeed, NGC 1614 is one of the 
starburst nucleus galaxies showing a particularly high nitrogen-to-oxygen 
abundance ratio (Coziol et al. 1999), 
which could be due to a sequence of starbursts over the past Gyrs. 

Soifer et al. (2001) claim that the 2.2 \mm\ continuum image does not reveal 
a ring-like structure, but a strong central peak of emission. This results from the fact
that the 2.2 \mm\ light traces the old stellar populations which peak at the nucleus, 
while the MIR, radio  and P$\alpha$ images trace the ring of current star formation.
Indeed Fig. 7 shows that the datum at 2.2 \mm\ is explained by black 
body radiation from stars with \Ts\ = 5\,000 K.

\subsection{NGC 2623}

\begin{figure}
\includegraphics[width=78mm]{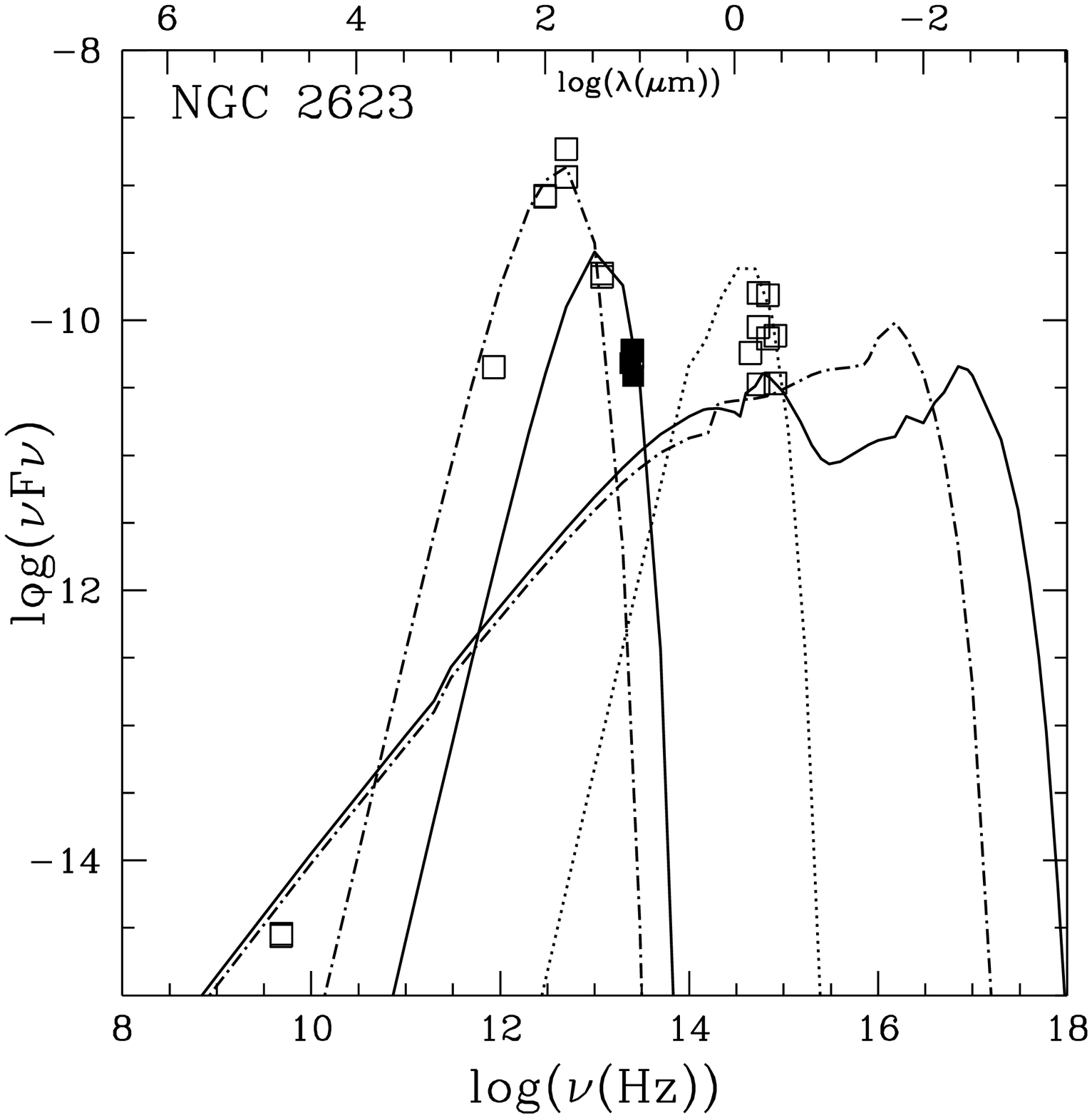}
\caption
{NGC 2623. The best fit of the SED. Symbols as in Fig. 3.
Different model values are listed in Table5.}
\end{figure}

\begin{table}
\caption{NGC 2623. Line intensities relative to H$\beta$}
\begin{tabular}{llllll}
\hline
 line &  obs$^1$ & M11    & M3      & M12    &SUM  \\ 
\hline
\ [OI] 6300& 0.9   &0.0012&0.66& 4.0& 1.0\\
\ [NII] 6584&3.8   &0.096&2.64& 11.7 & 3.1\\
\ [SII] 6716+6731 &1.6  &0.043&3.47& 7. &1.9\\
\ [SII] 6716/6731 &  -     &0.47&1.06& 1.0 &-  \\
\  \Hb\ (\erg)   &-  &0.23  & 0.016& 6.(-4)&- \\
\  w          &- &        0.007&0.005&1.  &- \\
\hline
\end{tabular}

$^1$ from Armus et al (1989)
\end{table}

This is a strongly interacting galaxy (Soifer et al. 2001), exhibiting strong Balmer 
absorption lines in its integrated optical spectrum 
(Armus, Heckman, \& Miley 1989). Because of the
steep emission-line Balmer decrement, the \Hb\ and H$\gamma$ lines 
appear in absorption, while H$\alpha$ appears in emission. The
strong absorption features indicate a large intermediate-age ($10^8 - 10^9$ yr) 
stellar population. 

The optical spectrum is presented in Table 5. The observed emission 
lines are  few,  hardly constraining the models. Therefore, we have used  models
similar to those   adopted for VV 114 and NGC 1614.
The continuum SED is presented in Fig. 8. 
The  MIR bump and a few optical data are well explained by model M11 
(solid lines), similar to 11(10A) from CV01, with $d/g= 4\times 10^{-4}$, while reradiation 
by dust at wavelengths longer than 60 \mm\ is explained by model M12 (dash-dot lines), namely a 
model corresponding to \Vs\ = 200 \kms, and a very young stellar population. 
Also for this model $d/g=4\times 10^{-4}$.
% Model M3 is also included to fit the FIR data points.
Only one datum is available in the radio range. So, we cannot decide whether 
radio emission is synchrotron or bremsstrahlung from a relatively
cool gas. Stars with \Ts\ = 5\,000 K also contribute to the SED at about 
$\nu = 10^{15}$ Hz. Concluding, NGC 2623 shows a lower amount of dust relative 
to VV 114.

\subsection{NGC 3690 + IC 694}

\begin{figure}
\includegraphics[width=78mm]{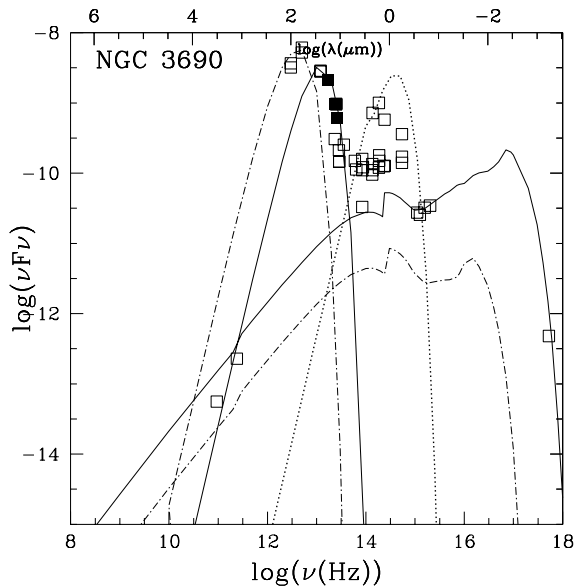}
\caption
 {NGC 3690\,+\,IC 694. The best fit of the SED. 
Symbols as in Fig. 3.
Different model values are listed in Table 6.
}
\end{figure}

\begin{figure*}
\includegraphics[width=43mm]{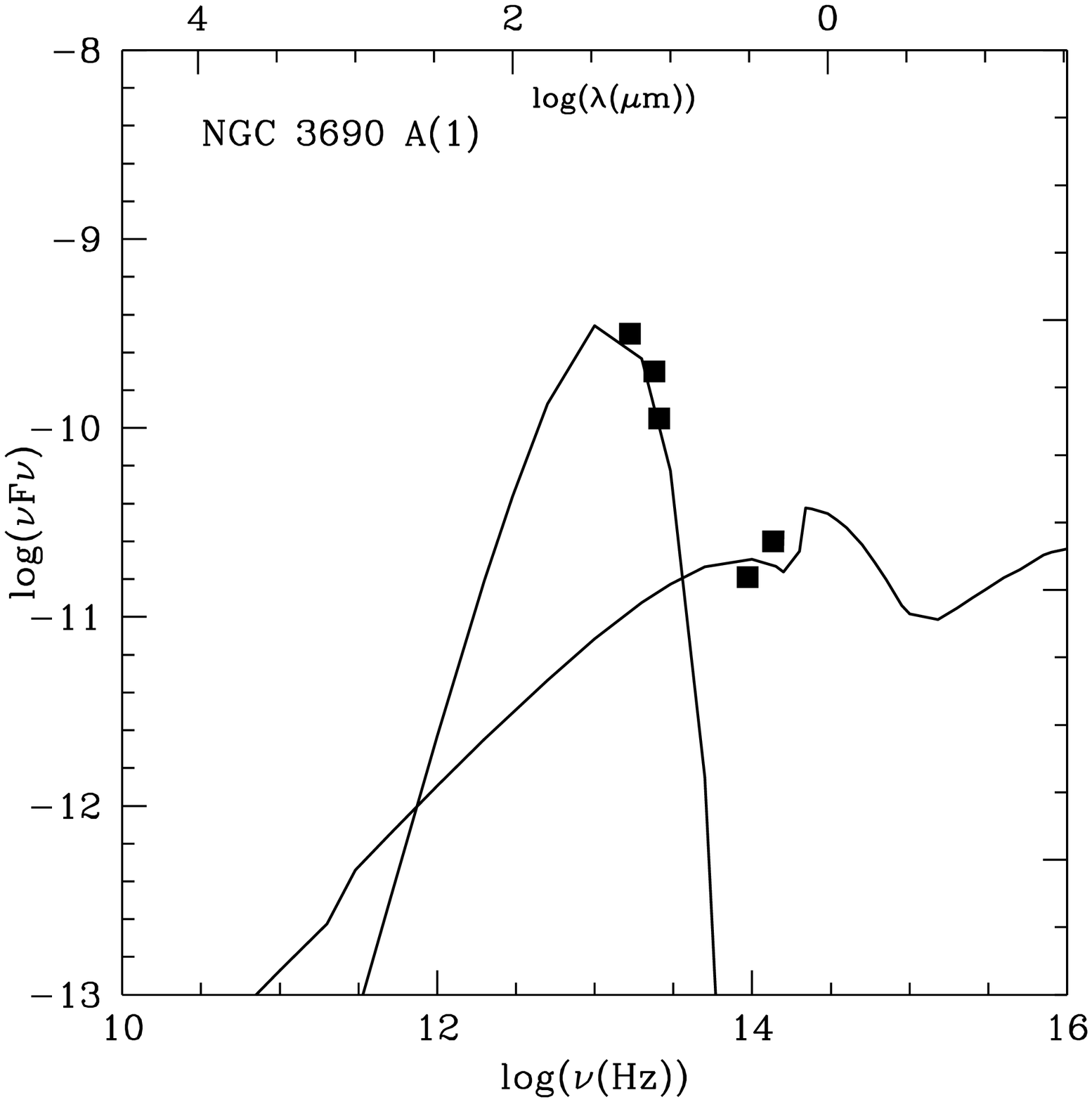}
\includegraphics[width=43mm]{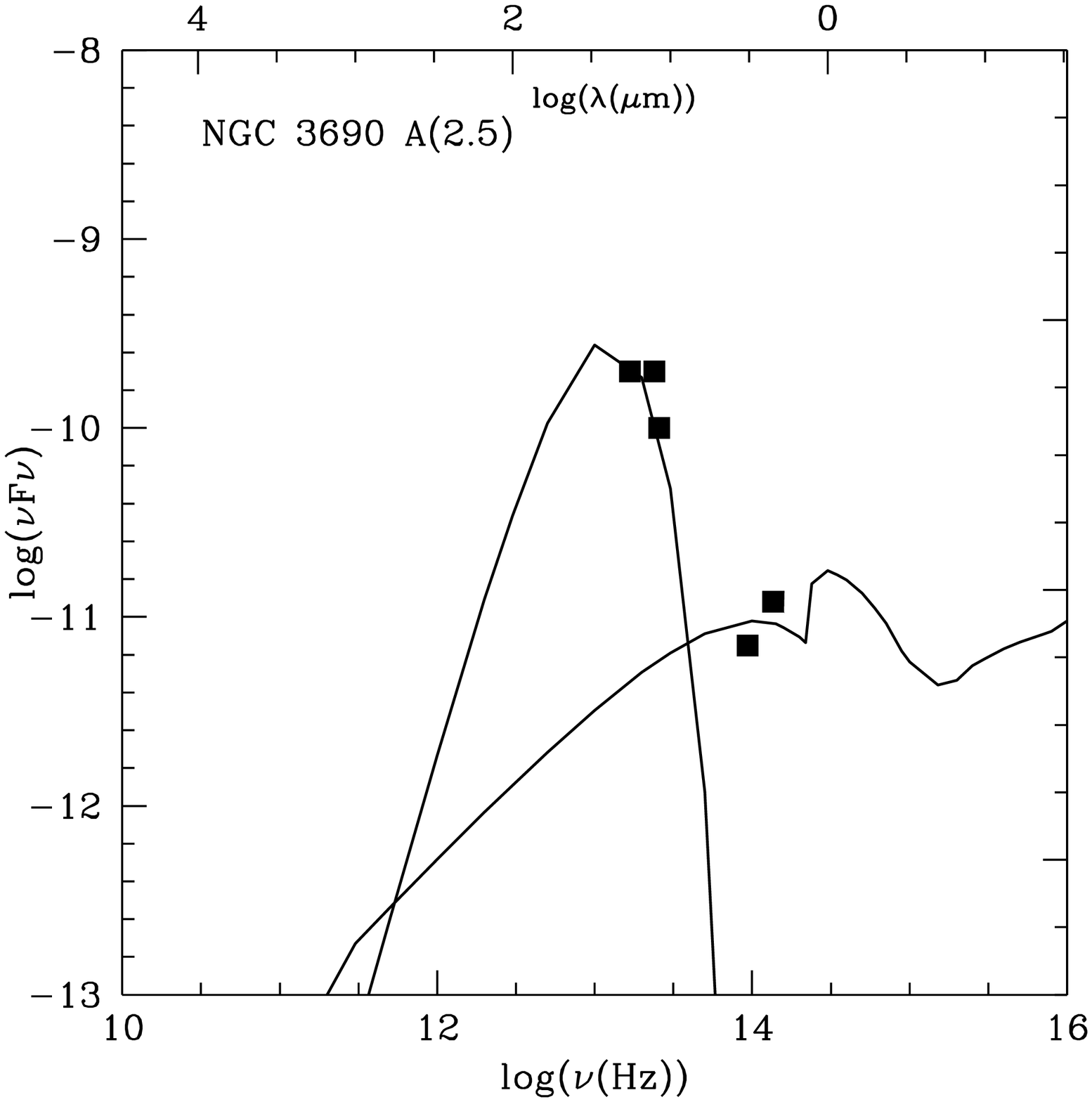}
\includegraphics[width=43mm]{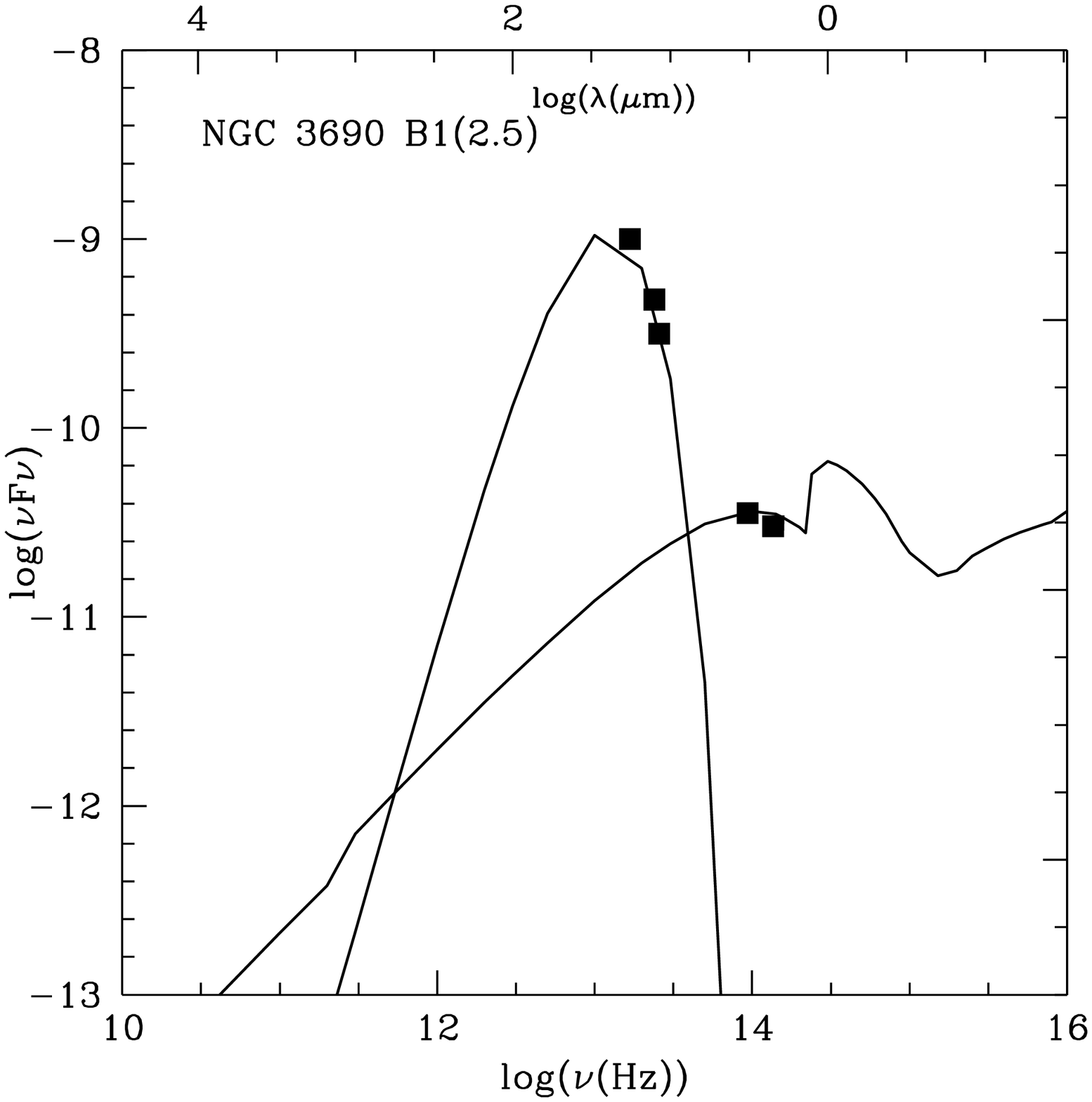}
\includegraphics[width=43mm]{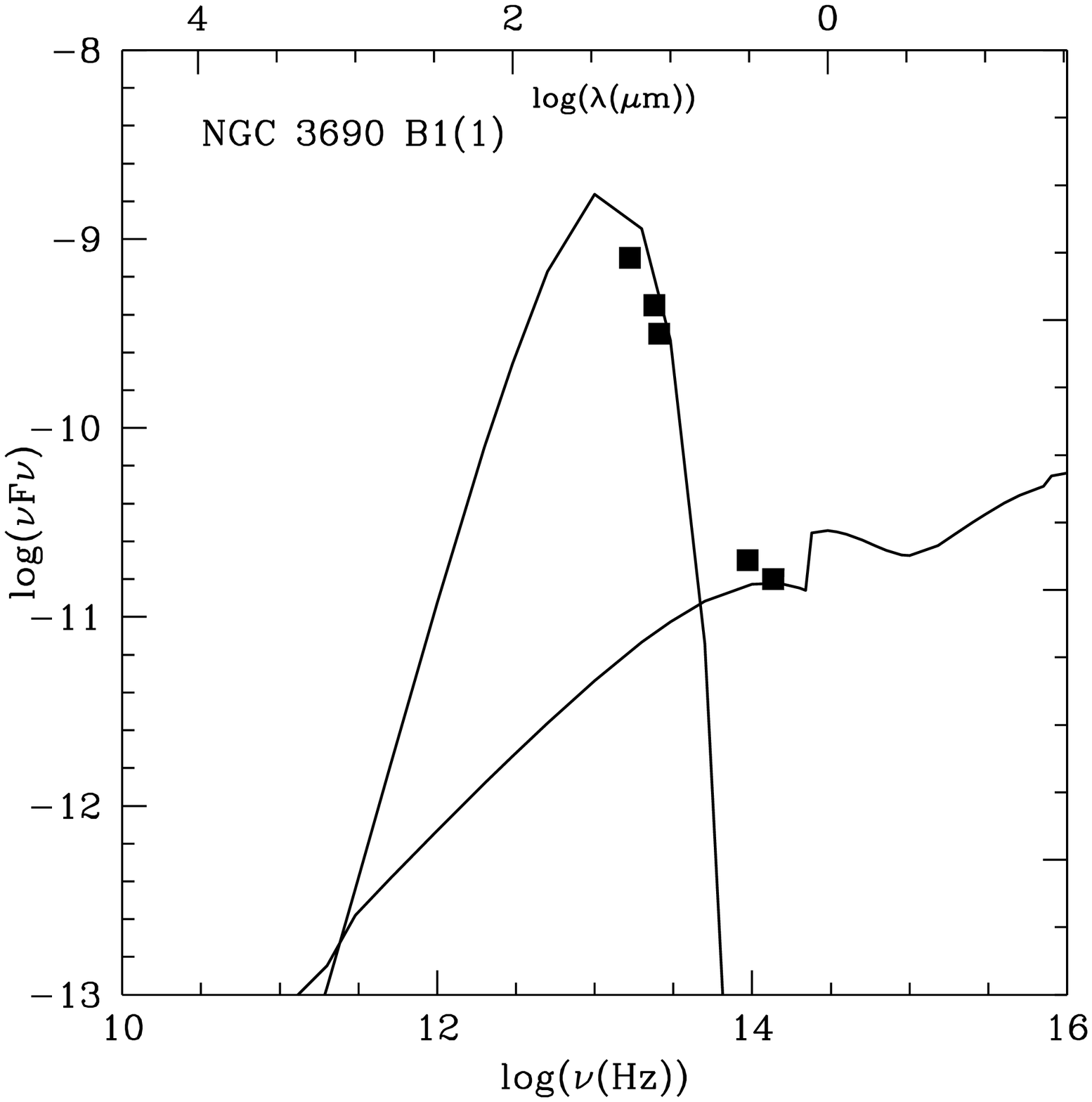}
\includegraphics[width=43mm]{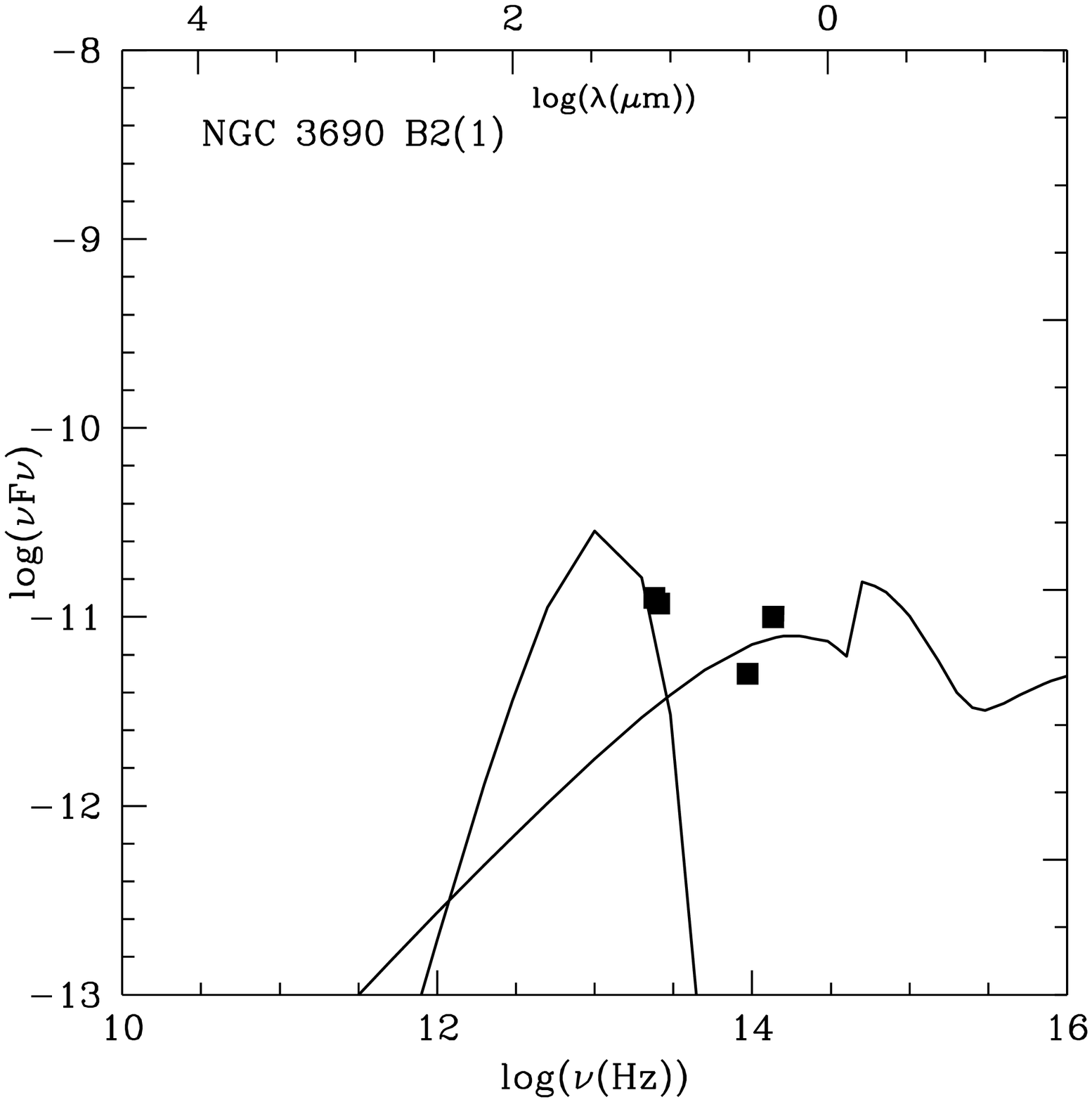}
\includegraphics[width=43mm]{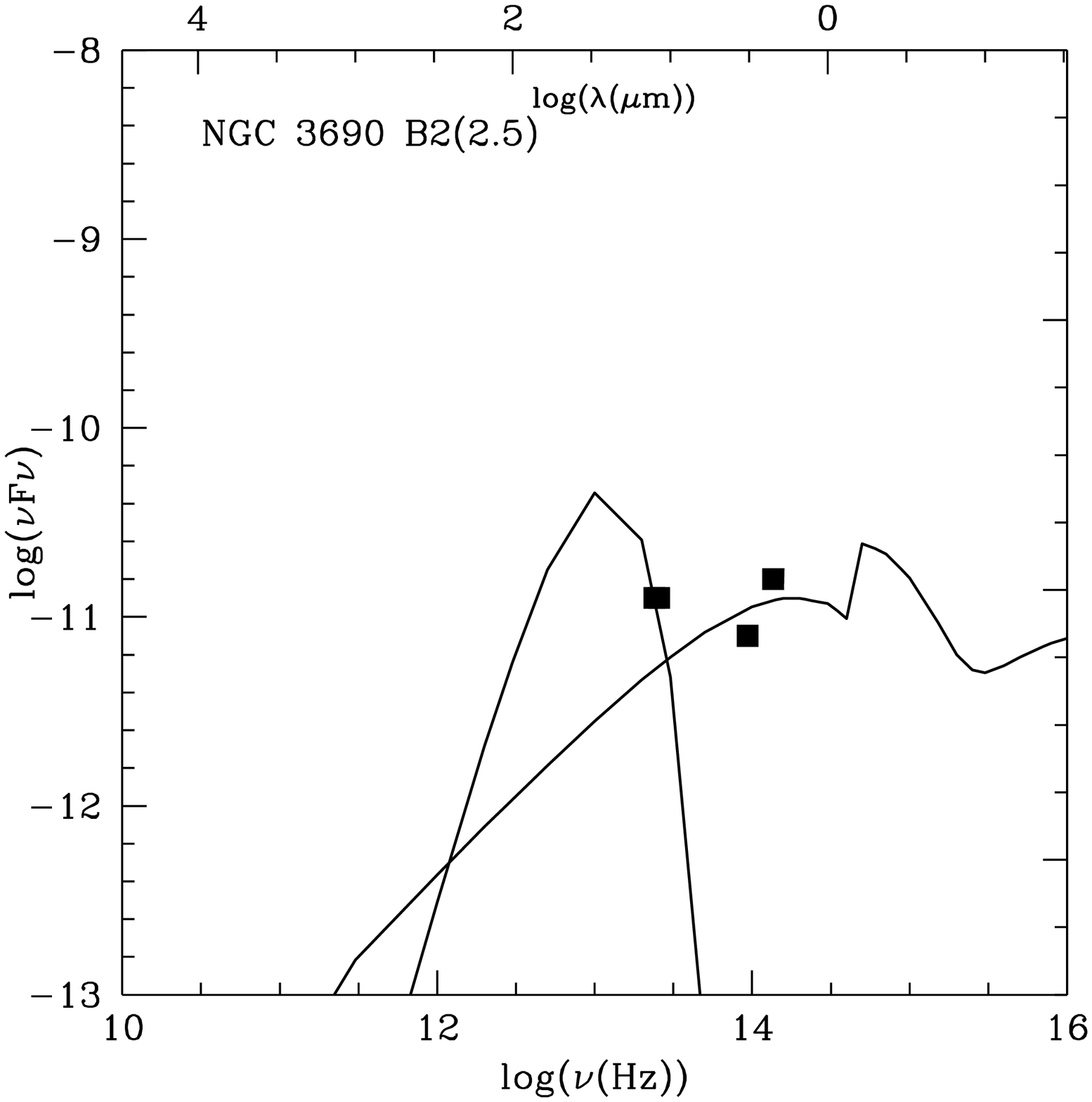}
\includegraphics[width=43mm]{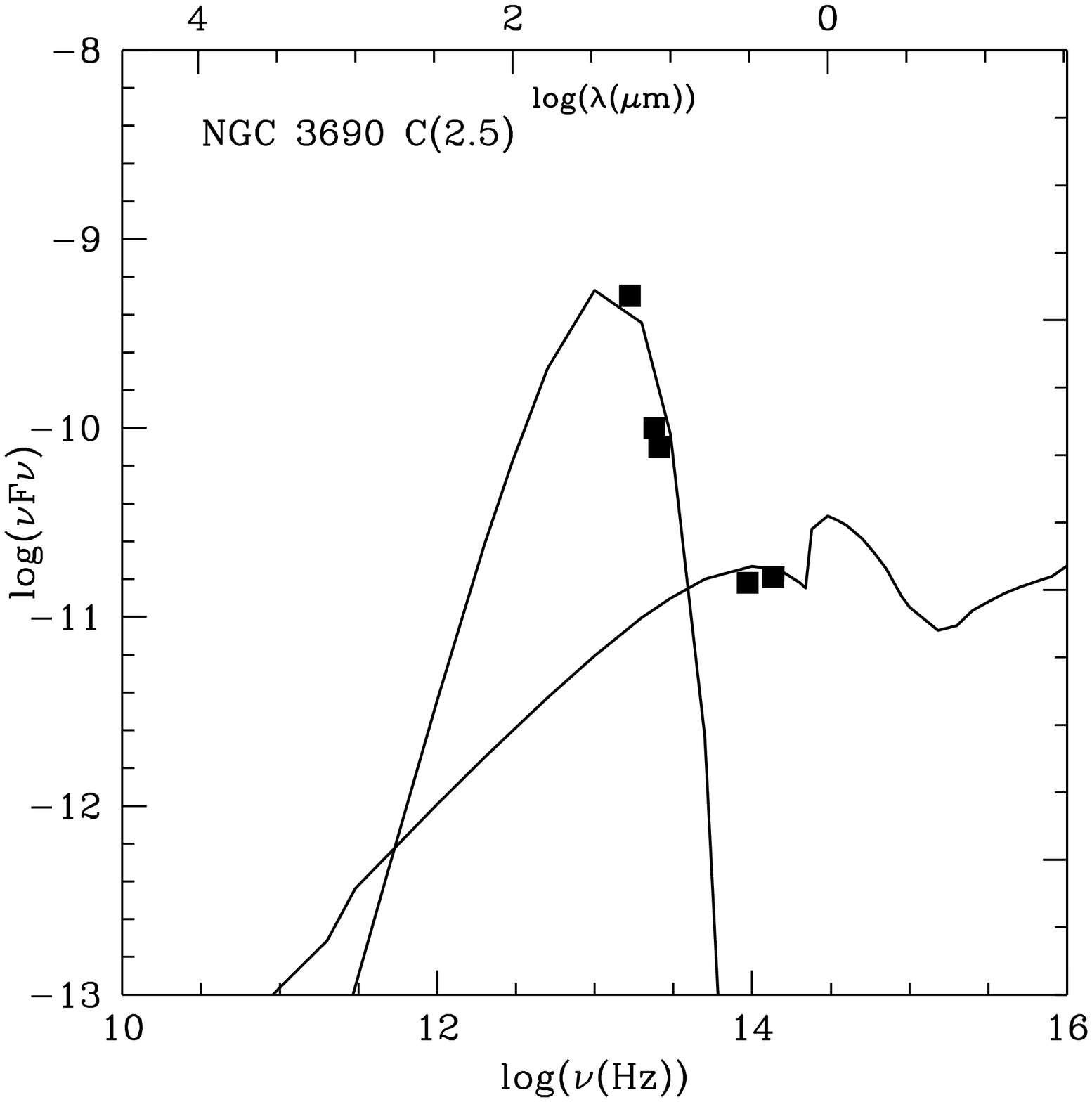}
\includegraphics[width=43mm]{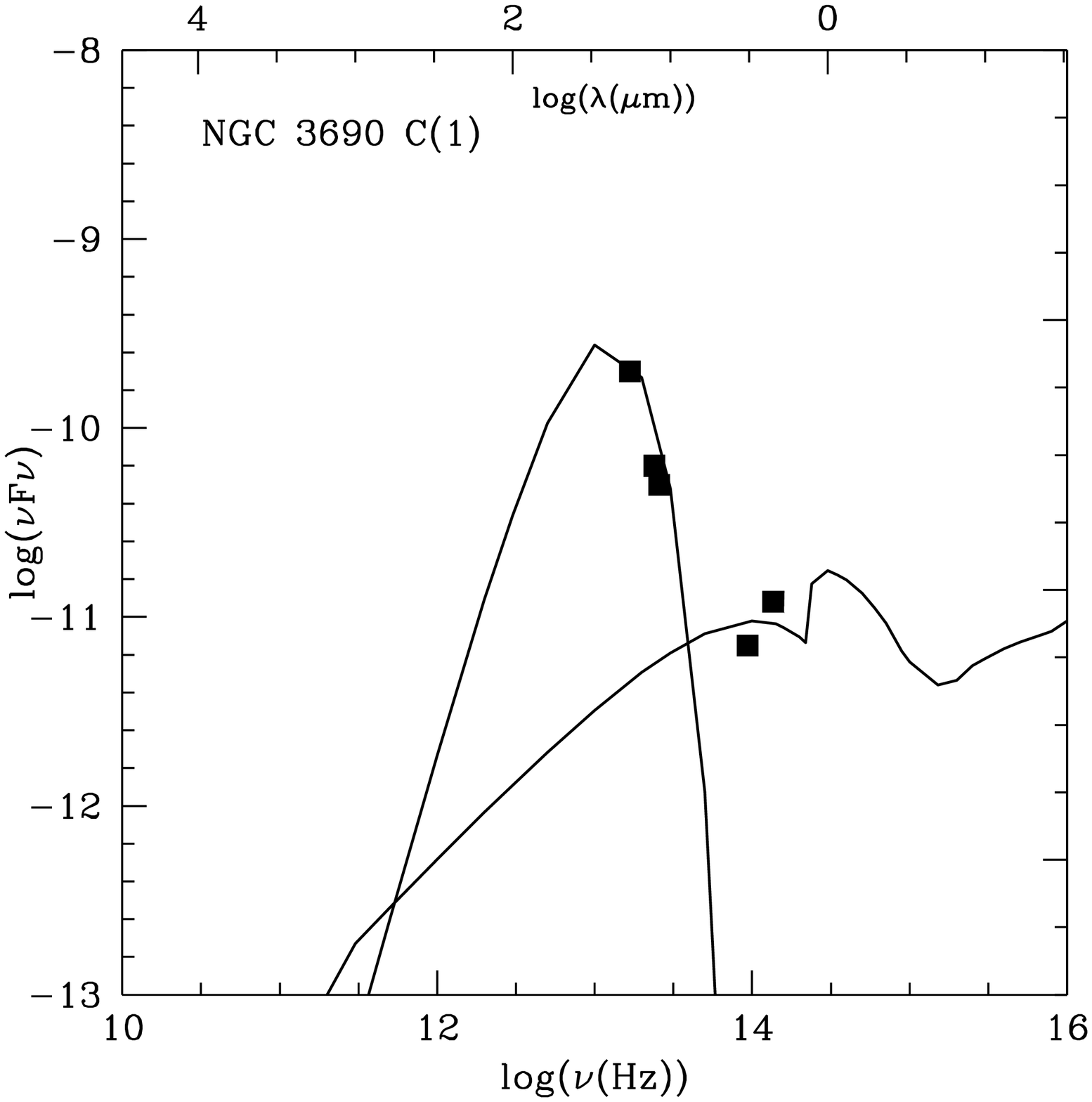}
\includegraphics[width=43mm]{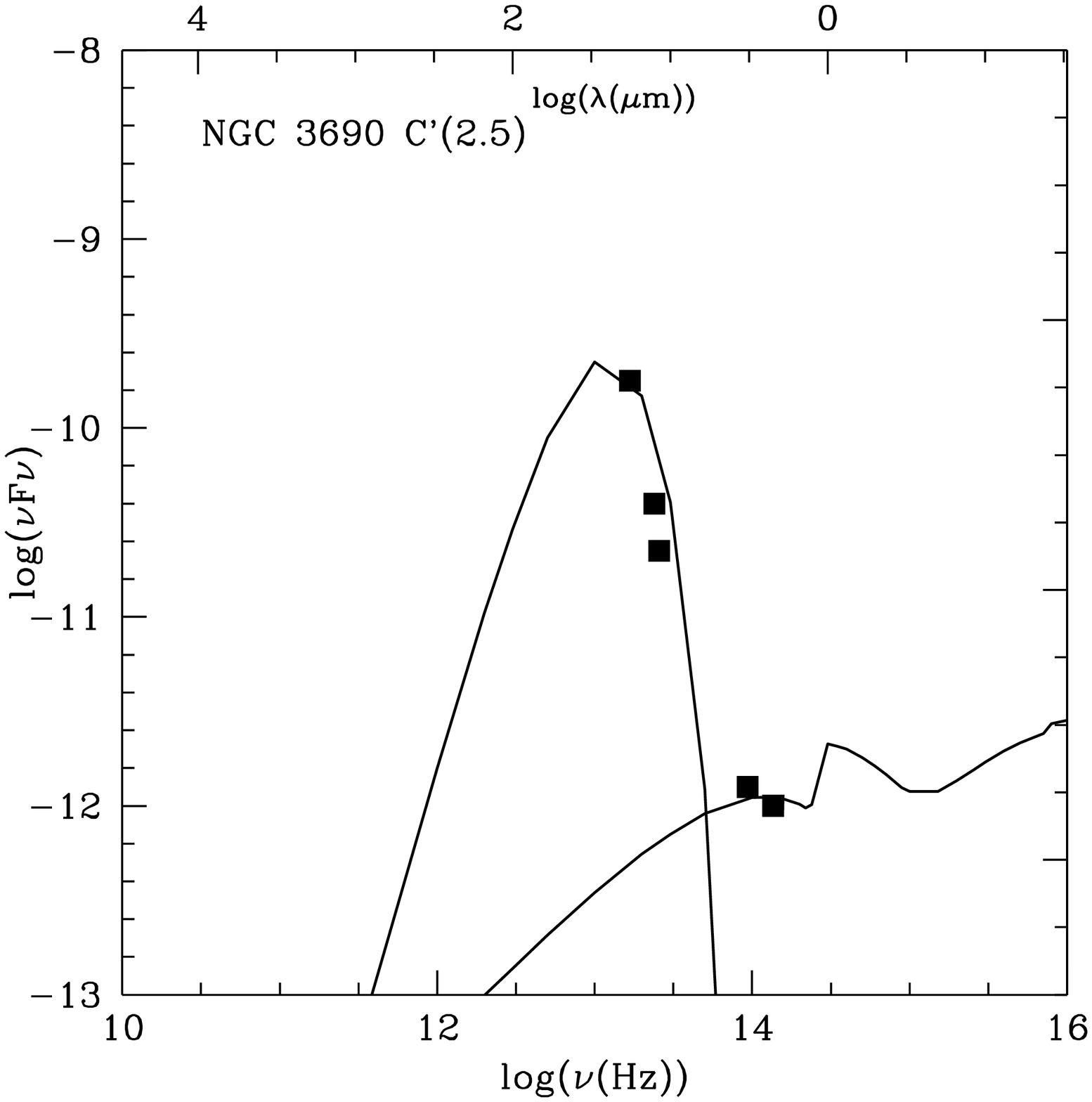}
\includegraphics[width=43mm]{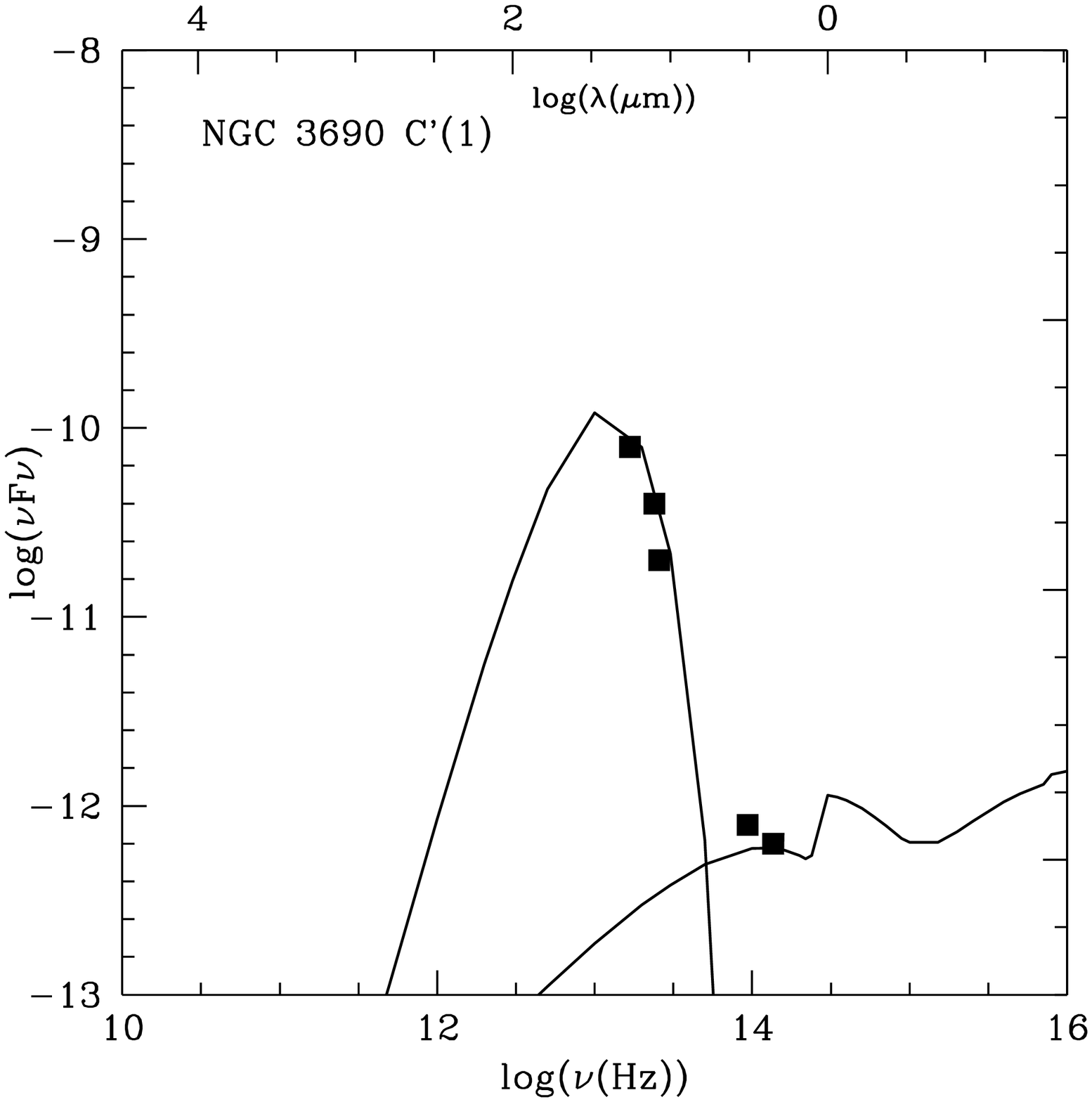}
\caption
{NGC 3690. The best fit of the MIR data in the different regions (see
text for details).}
\end{figure*}

\begin{table*}
\caption{NGC 3690. Line intensities relative to H$\beta$}
\begin{tabular}{llllllllllllll} 
\hline
 line &  obs(C)$^1$ &obs(A)$^1$ &obs(B)$^1$ &M13    &M10     &M5    &M14   &M3  & SUM1 & SUM2 \\ 
\hline
\ [OIII] 5007 & 1.25 & 1.11& -&    2.86 & 0.11 & 15. &0.84 &2.2& 1.37& 1.4 \\
\ [OI] 6300& 0.067 &0.31  &0.21&  3.(-5)&0.0  &  0.5 & 0.002& 0.66& 0.057 & 0.2    \\
\ [NII] 6584&1.34  &1.83 &1.83&     0.007&0.0  & 8.8 &  0.086 & 2.64& 0.51 & 1.0    \\
\ [SII] 6716+6731 &0.57 &1.28 &0.95& 0.8& 0.0 & 7.7   & 1.02  &   3.5 & 1.4 & 1.8 \\
\  \Hb\ (\erg)   & - & -  &- & 0.11  & 0.14  & 0.0025 & 0.44  & 0.187&- &- \\
\ w1          & - &- &- & 3.3(-3)&3.3(-4)&1.&0.167& 0.027&- &- \\
\ w2          & - &-  &-  &3.6(-3)&3.3(-4)&1.  & 0.27&0.27&- &-  \\       
\hline
\end{tabular}

$^1$ from Armus et al.  (1989)
\label{tab6}

\centering
\caption{NGC 3690: $d/g$ ($4\times 10^{-4}$) in the different regions}
\begin{tabular}{lllllllllll} 
\hline
\ A(1) &A(2.5)&B1(2.5)&B1(1)&B2(1)&B2(2.5)&C(2.5)&C(1)&C'(2.5)&C'(1) \\
\ 3.&6. &  6.  & 6.5  & 0.5  & 0.5&  6.  & 6.  & 20.& 20. \\
\hline
\end{tabular}
\label{tab7}
\end{table*}

NGC 3690\,+\,IC 694 show a system of five nuclei. The MIR photometry has been 
performed by S01 in single regions and with different apertures. It is thus 
possible to determine the $d/g$ ratios for each nucleus.
The optical spectra (Armus et al. 1989) are available in the A, B, and C 
regions (see S01 for nomenclature).  The number of lines is very small allowing only a 
rough modelling. 

In Table 6 we compare model calculations with the observed emission-line 
ratios from Armus et al. (1989). The models are taken from CV01 and some of them 
are run with different $d/g$ values. Model M5 which refers to \Vs\ = 200 \kms\ is 
constrained by the fit of the SED. The summed spectra SUM1 and SUM2 appear in the 
last two columns of Table 6. The fit is not very good, particularly the 
[NII]/\Hb\ emission-line ratio is underpredicted and [SII]/\Hb\ line ratio is 
overpredicted. 
A higher than cosmic N/H abundance ratio has been found in  many galaxies
containing a starburst.
Moreover, in dust rich galaxies
sulphur  could be locked up into  CS diatomic molecules.

The SED of the continuum is shown in Fig. 9 for the summed spectrum. 
Notice that most of the data in the radio-to-X-ray range are well explained by a model
with \Vs\ = 500 \kms\ and $t=3.3$ Myr (solid lines). However, lower velocities are 
also important to explain the FIR bump (dash-dot lines). For the summed SED shown 
by S01 in their fig. 6,  $d/g$ =  $4\times 10^{-3}$.

We adopted  the same ratio of relative weights  for models M5 and M14
as for the line ratios (Table 6), although 
 dust emission from model M5 overpredicts the data in the FIR and indicates
strong absorption by dust (see sect. 4.1).

In Fig. 10, the data from S01 (fig. 4f) for single regions are compared 
with models. The $d/g$ found for each region are presented in Table 7. 
B2 shows the lowest $d/g$ and C' the highest. In fact, [SII]/\Hb\ line ratios are 
low in region C, indicating that sulphur is locked in dust.

The image at 2.2 \mm ~shows a more complete and tormented structure, corresponding to 
the complex of  contributions by the starburst radiation,  by the old stellar 
population and  by the strong shock. 
On the contrary, the image at 3.2 \mm\ is smoother 
because  shows only the  black body emission of the old stellar population. The 
images at 12.5 \mm\ and that at 6.4 GHz are similar because they show the flux at 
different frequencies on the distribution of dust emission from the same cloud.

\subsection{IC 883}

\begin{figure}
\includegraphics[width=78mm]{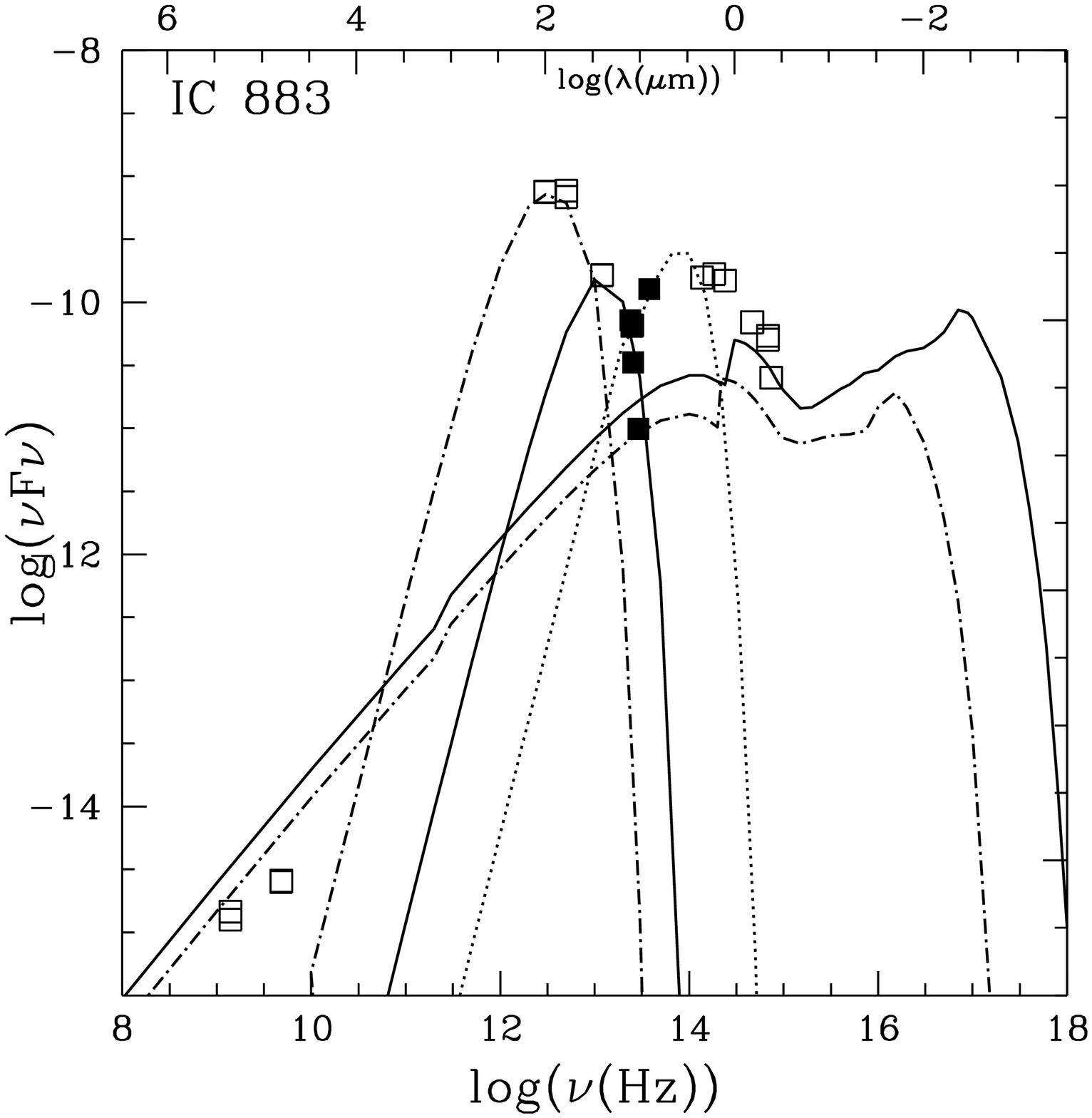}
\includegraphics[width=78mm]{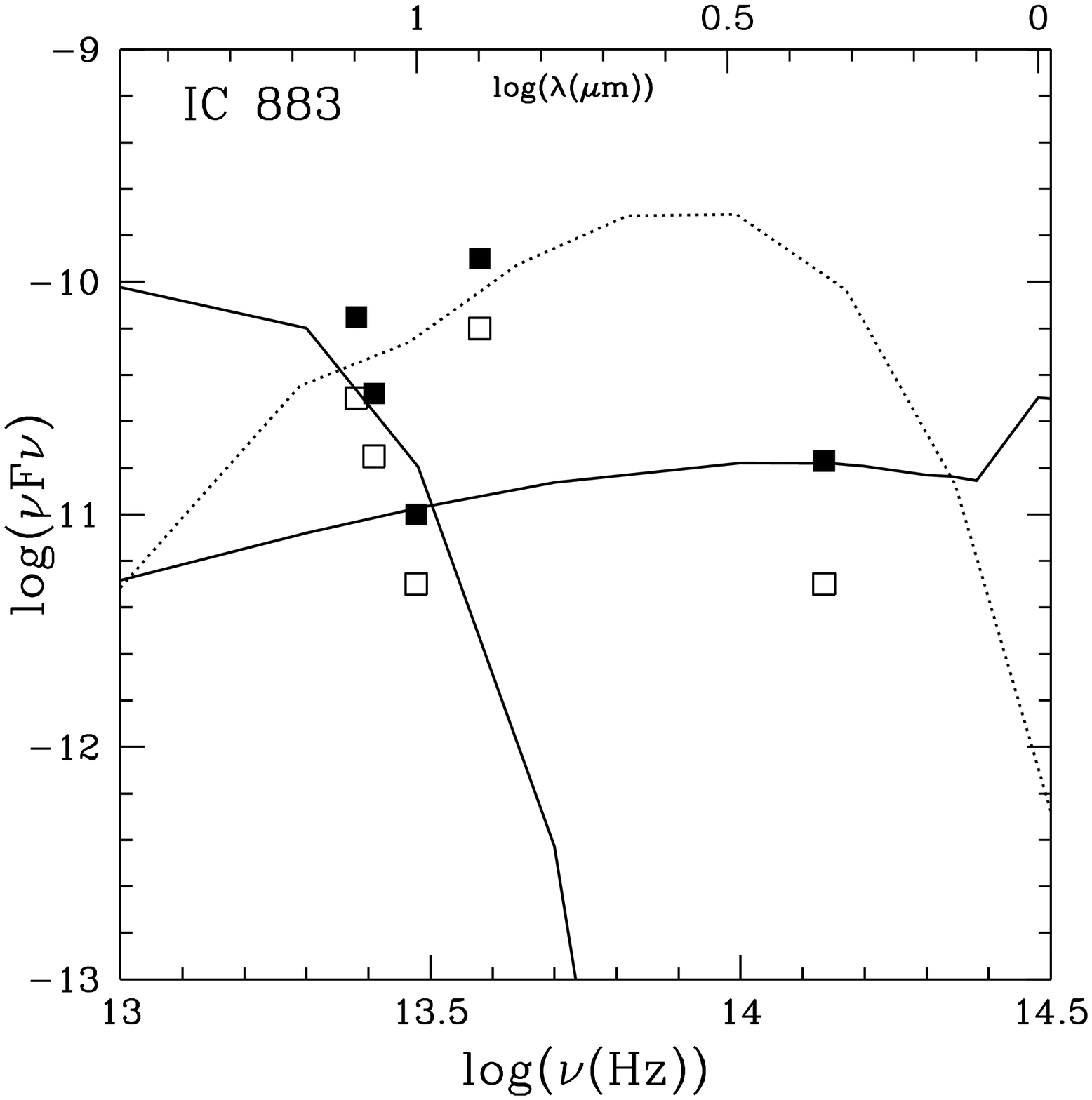}
\caption
{IC 883. The best fit of the SED. 
Symbols as in Fig. 3.
Different model values are listed in Table 8.}

\end{figure}

\begin{table}
\centering
\caption{IC 883. Line intensities relative to H$\beta$}
\begin{tabular}{llllll} 
\hline
 line &  obs & M9     &M15      &M16     & SUM\\
\hline
\ [OIII] 5007&0.86& 0.8& 0.82&2.7 &0.81         \\
\ [OI] 6300&0.44  & 0.002&4.55&9.(-4)  &0.6      \\
\ [NII] 6584&2.7  & 0.086&12.7&0.039 &1.9\\
\ [SII] 6716+6731 &1.23 & 0.10 &6.6 &0.6  &1.0      \\
\ [SII] 6716/6731 &1/1.23:&0.47&1.&0.46  & -      \\
\  \Hb\ (\erg)   & -  & 0.44&0.004& 0.24& \\
\   w                 &- &0.05& 1.  & 1.6(-4) &-  \\
\hline
\end{tabular}
\label{tab8}
\end{table}

The optical spectrum of IC 883 is classified as that of a LINER by Veilleux et al. 
(1995). However, we could fit the few line ratios presented in Table 8 with 
an averaged sum of starburst models, as for the other galaxies of the sample.
Nevertheless, checking the results of composite models presented  by 
Contini \& Viegas (2001b) which refers to AGN, we found that the observed line 
ratios could be also explained by a model with \Vs\ = 100 \kms, \n0\ = 100 \cm3, 
and \Fh\ = 10$^9$ cm$^{-2}$ s$^{-1}$ eV$^{-1}$ at 1 Ryd. However, such a model 
cannot explain the emission by dust in the MIR, because \Vs\ is too low.

The SED of the continuum is presented in Fig. 11 top. 
We show  both the contributions of a model corresponding to a stellar age of 3.3 Myr
with \Vs\ = 500 \kms\ and $d/g=4\times 10^{-5}$ (M9 represented by solid lines) 
and a model also corresponding to $t=3.3$ 
Myr with \Vs\ = 200 \kms\ and $d/g=8\times  10^{-4}$ (M15 represented by dash-dot lines). 
As the galaxy is more compact than the former ones it was reasonable to consider 
that the  clouds with different \Vs\ correspond to the same starburst age.
Interestingly, some data in the NIR are explained by black body radiation
from stars with \Ts\ = 1\,000 K (dotted line). This is justified by the assumption that
there is strong silicate absorption at 10 \mm\ (S01).

In the bottom diagram of Fig. 11, the photometry in different apertures 
(filled squares: 4\arcsec; open squares: 1\arcsec) are explained by the same 
starburst model with \Vs\ = 500 \kms. Notice that with a lower aperture the 
flux is reduced for all the data by about the same factor.

The images at 2.2 \mm\ and 12.5 \mm\ (S01) are similar because corresponding to the
same distribution of dust emission.

\subsection{NGC 6090}

\begin{figure}
\includegraphics[width=84mm]{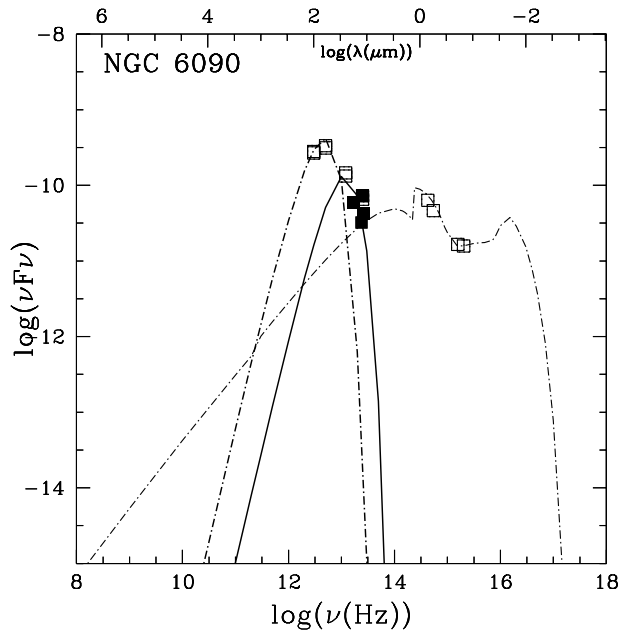}
\caption
{NGC 6090. The best fit of the SED. 
Symbols as in Fig. 3.
Different model values are listed in Tables 9 and 10.}
\end{figure}

 The results of Acosta-Pulido et al (1996) show that 84\% of the NGC 6090
emission comes from the wavelength regime
shortward of 120 \mm ~and argues for very efficient heating of the dust. It implies
temperatures higher than about 25 K which cannot be accounted for by the interstellar
radiation field alone.

The models are compared with observed emission-line ratios 
(Veilleux et al. 1995) in Tables 9 and 10 for the SW and NE regions 
respectively.
With cross-checking considerations, the SED of the continuum which appears in 
Fig. 12 shows that model M17 (solid lines) with \Vs=500 \kms ~and d/g=1.6$\times 10^{-4}$ explains some 
significant data in the MIR from NED, in agreement with Keck data (S01) 
Reradiation by dust in clouds with 
\Vs\ = 200 \kms\ and $d/g = 2.8\times 10^{-4}$ (M18) explains the IR data (dash-dot lines)
and bremsstrahlung. 

The fit of the summed optical line ratios shows that  other models 
contribute.
Clouds  photoionized by a $t=3.3$ Myr starburst and with \Vs\ = 500, 200, and 100 \kms\ 
explain the summed line ratios in the NE region, while the low velocity cloud contribution
in region SW corresponds to a younger age ($t=2.5$ Myr) and appears in the sum with a 
high relative weight (model M19). Notice, however that \Hb\ absolute value is very 
low and  is compensated by the high relative weight.
Also for this galaxy, [NII]/\Hb\ line ratio is underestimated  and [OI]/\Hb\ is 
overestimated in region NE.

In Fig. 13 the diagrams correspond to the different continuum SED in the regions 
indicated by S01 (fig. 6c). A $d/g < 1.6\times 10^{-3}$ is revealed in E(6), even if 
the flux is high. The fluxes decrease in E$_{\rm N}$ and E$_{\rm S}$ and mostly in 
W(4) and W(1.5).

The data point at 2.2 \mm\ corresponds always to bremsstrahlung emission, while the 
other data are associated  with reradiation by dust in a relative high velocity regime.
Consequently, the image at 2.2 \mm\ (S01) is complex and more complete. 
It seems that emission in the radio is bremsstrahlung by cool gas.

\begin{figure*}
\includegraphics[width=48mm]{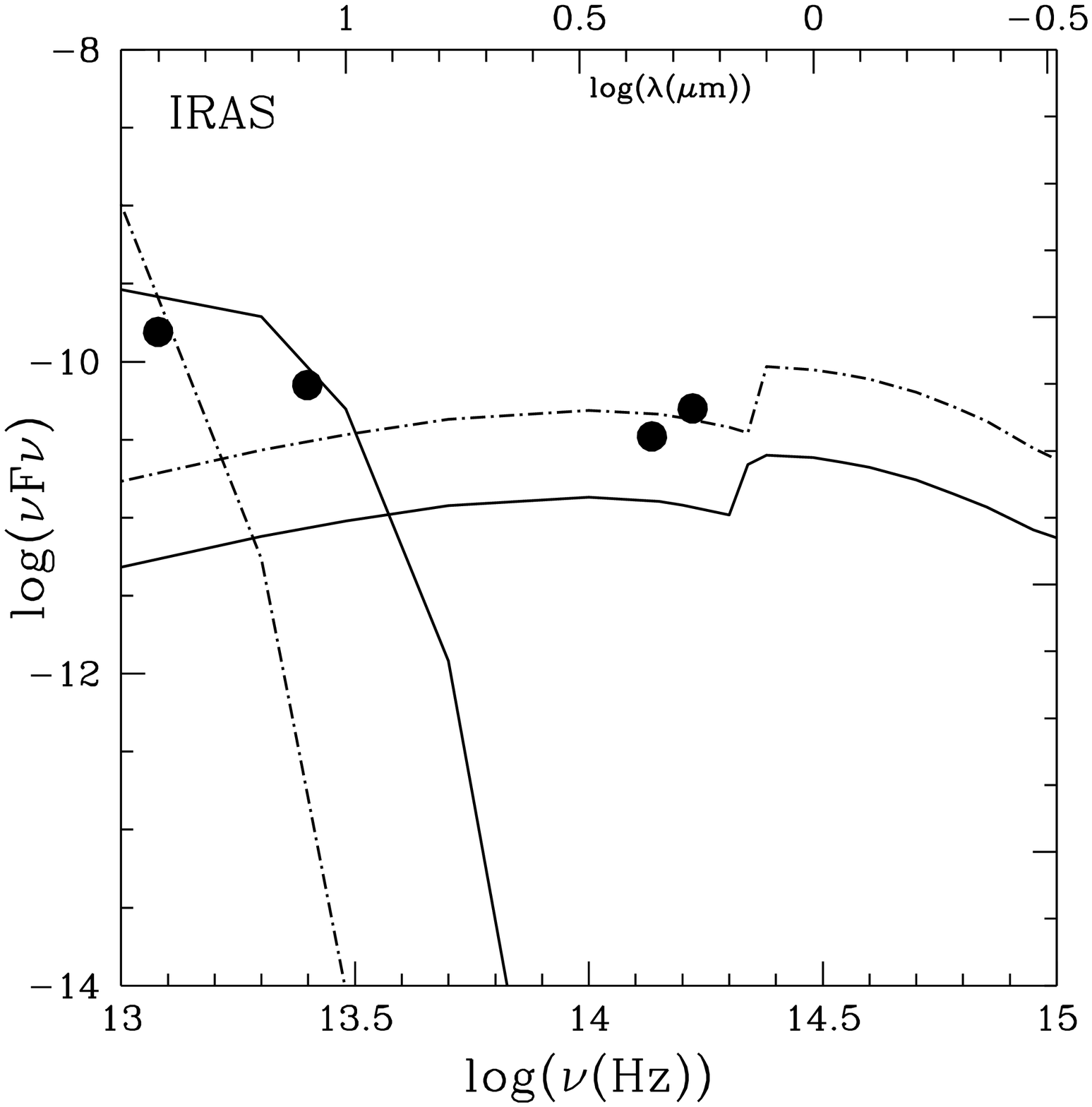}
\includegraphics[width=48mm]{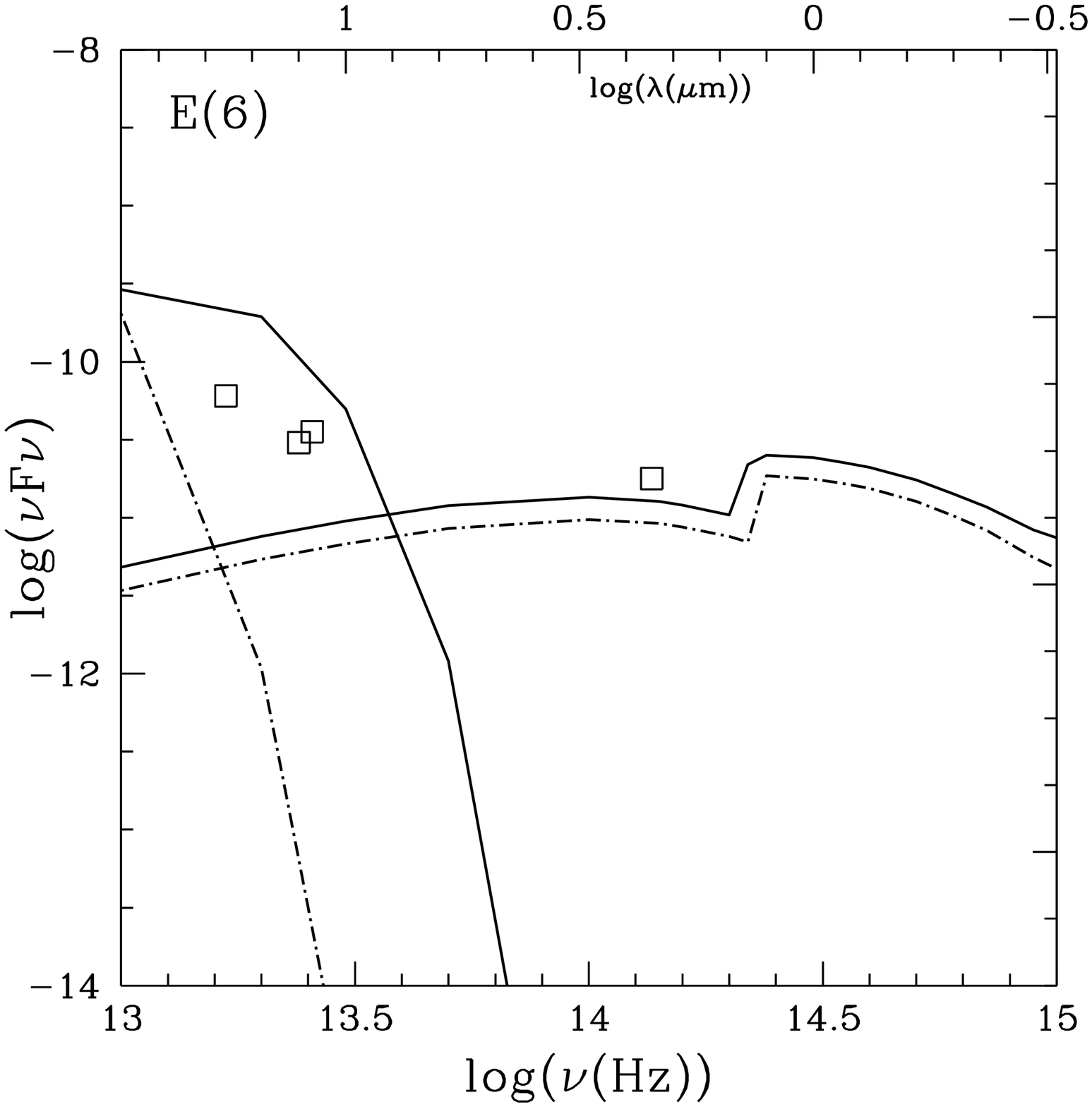}
\includegraphics[width=48mm]{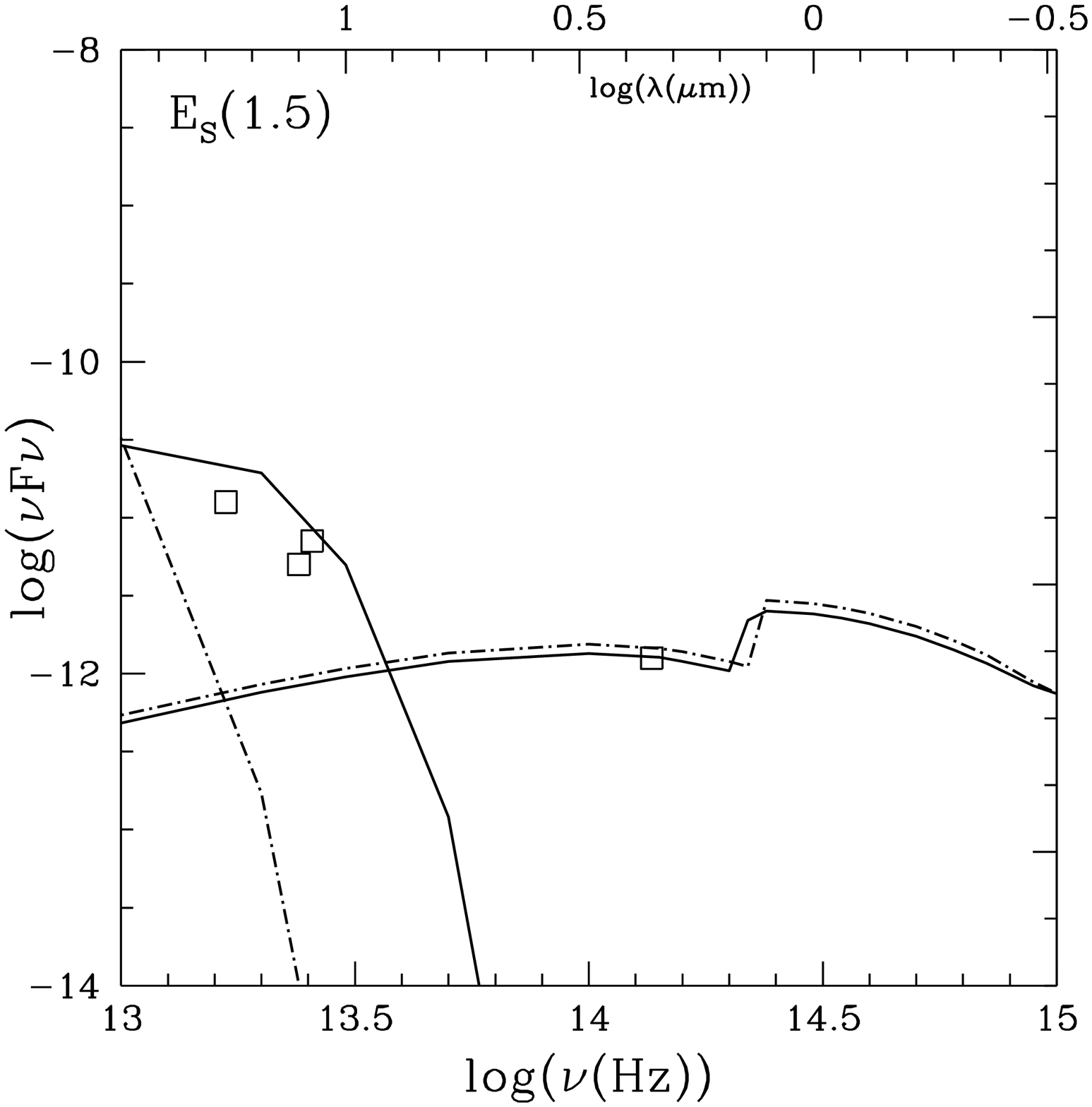}
\includegraphics[width=48mm]{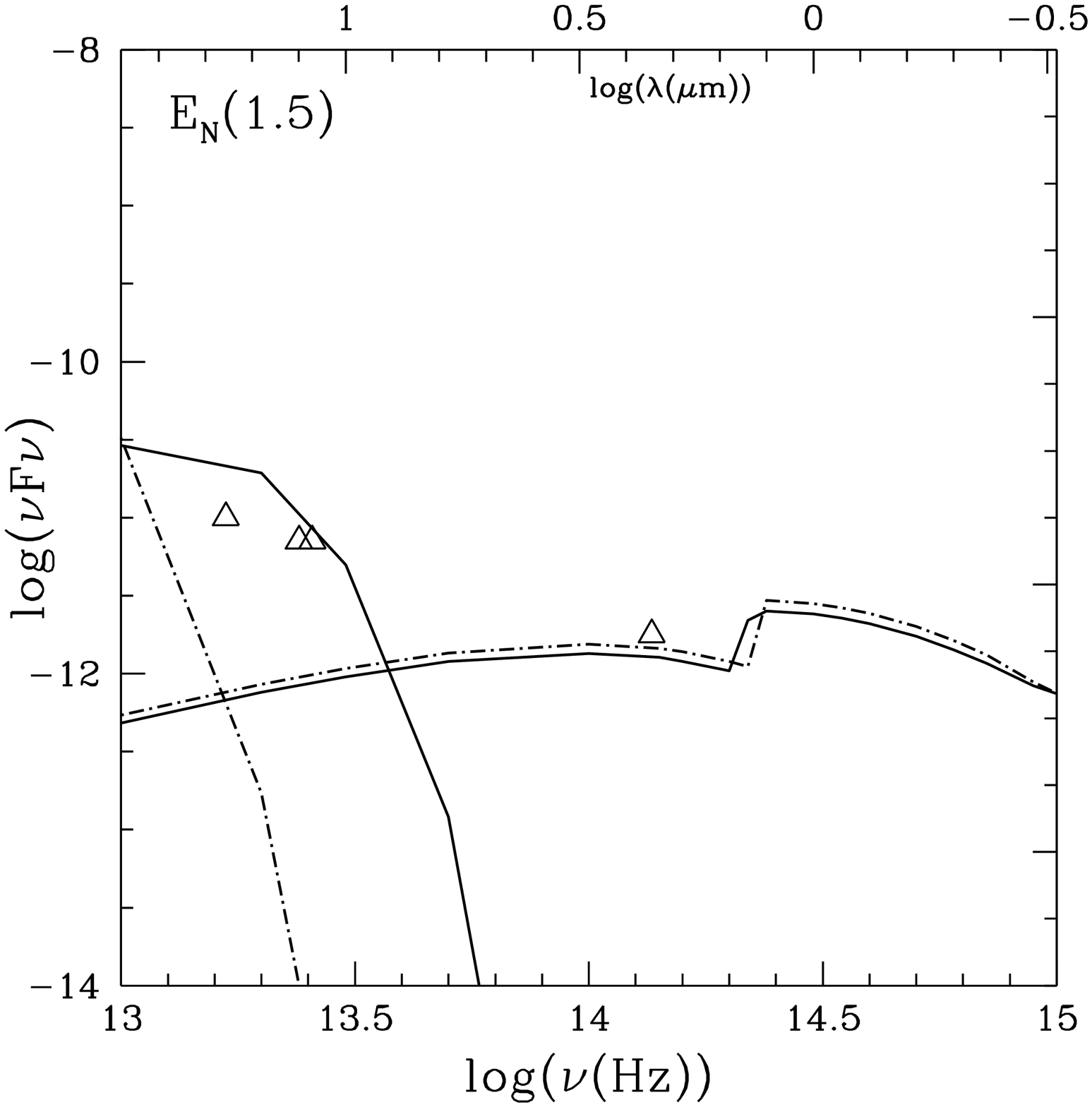}
\includegraphics[width=48mm]{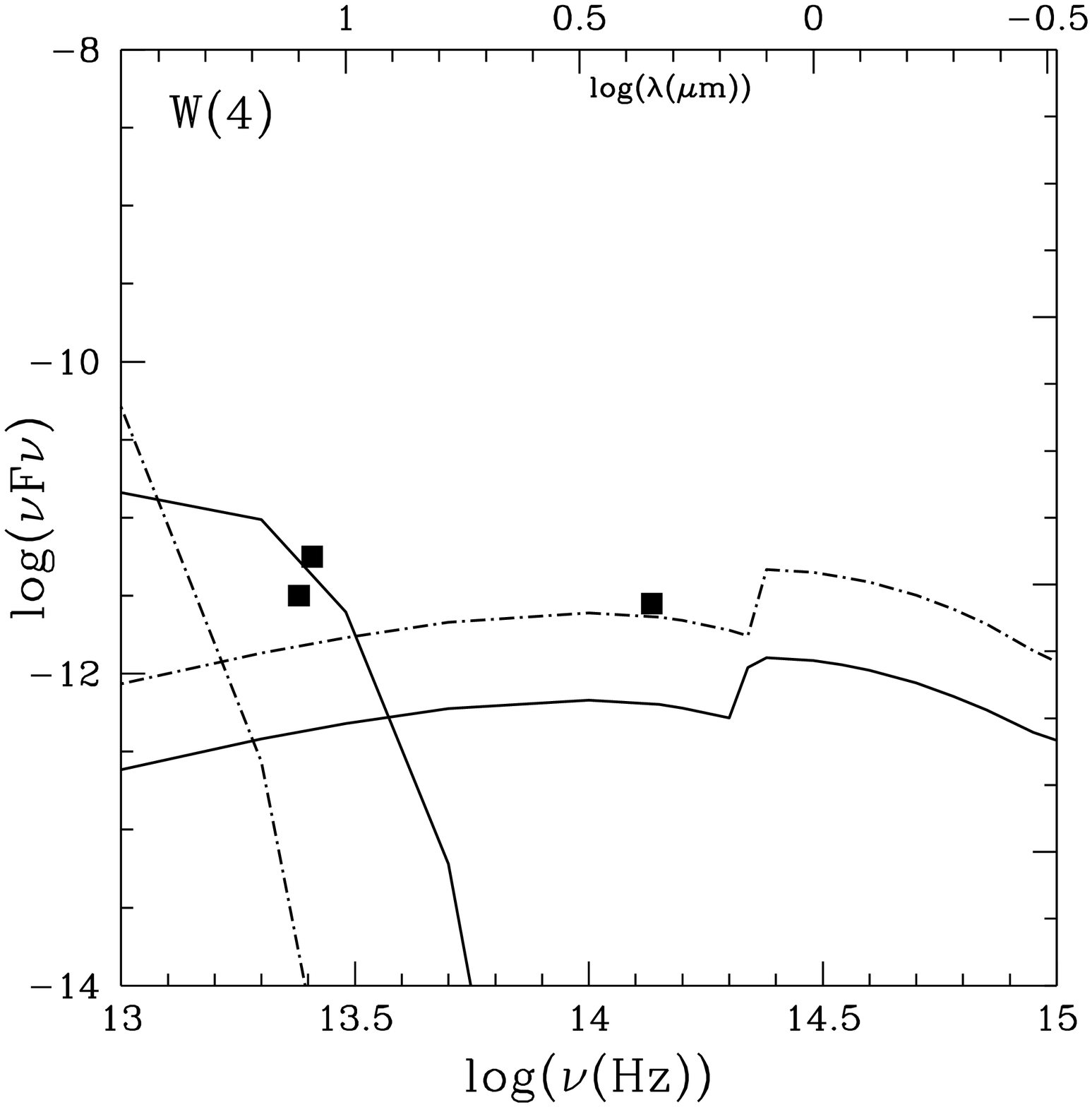}
\includegraphics[width=48mm]{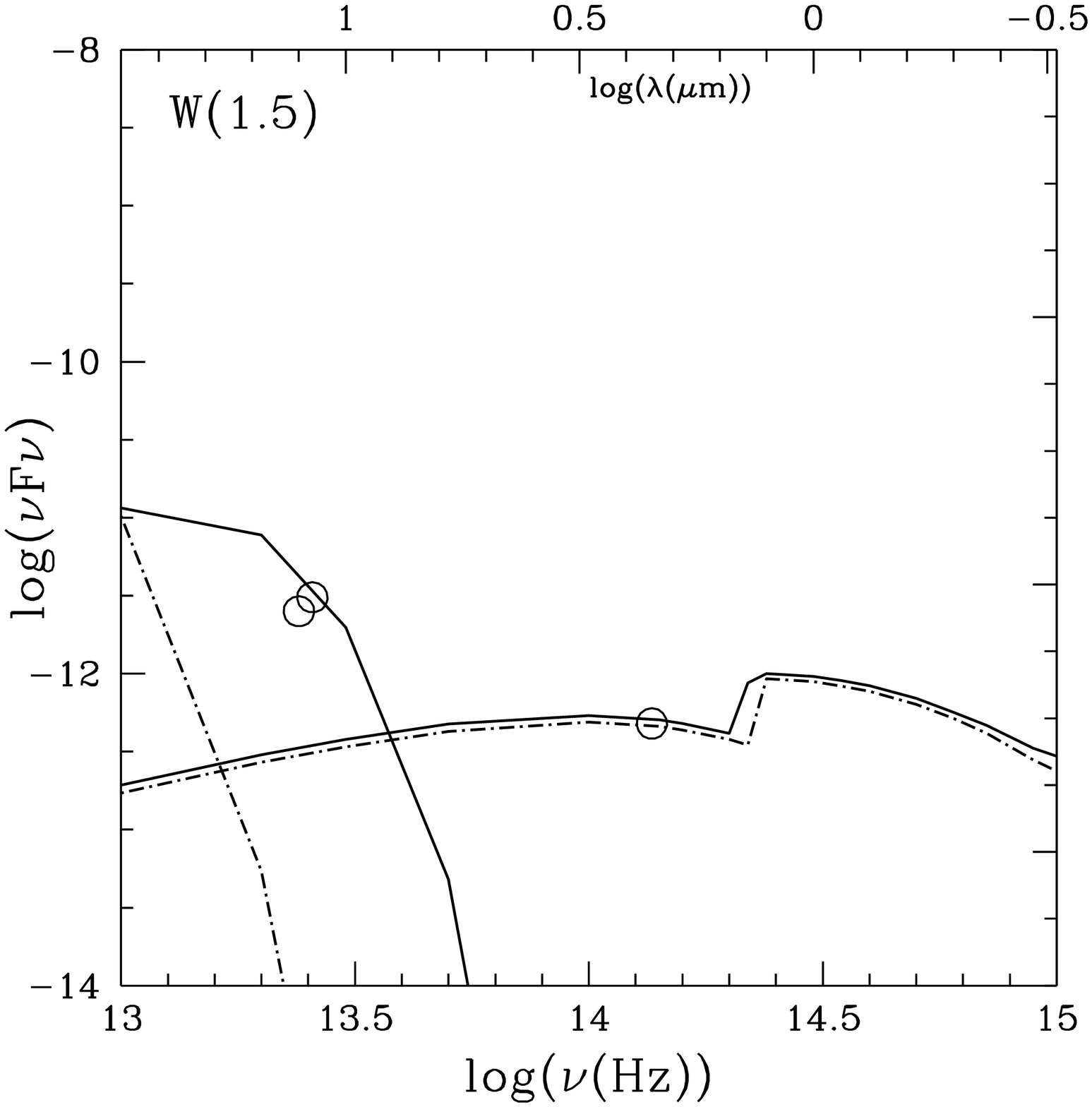}
\caption
{NGC 6090. The best fit of the MIR data in the different regions (see
text for details).}
\end{figure*}

\begin{table*}
\centering
\caption{NGC 6090. Line intensities relative to H$\beta$}
\begin{tabular}{lllllllll}
\hline
 line &  obs (SW)& M17     & M18      & M14   & M19  & SUM(SW)   \\
\hline
\ [OIII] 5007 &1.5 &2.76 & 0.76  &   0.84 & 2.24& 1.7 \\
\ [OI] 6300& 0.066 &6.5(-5) & 2.62 & 0.002 & 0.0054 & 0.05      \\
\ [NII] 6584& 1.3  & 0.008  & 10.8  & 0.0086 & 1.3 & 0.9    \\
\ [SII] 6716+6731 & 0.64  & 0.8 & 6.0 &  1.02 & 0.35& 0.7\\
\  \Hb\ (\erg)   & -& 0.317 & 0.007 & 0.44 & 0.0066 & \\
\   w         & - & 0.004 & 0.032 & 0.012 & 1.   & \\
\hline
\end{tabular}
\label{tab9}
\end{table*}

\begin{table*}
\centering
\caption{NGC 6090. Line intensities relative to H$\beta$}
\begin{tabular}{lllllllll}
\hline
 line &  obs (NE)& M17     & M18      & M14   &   M20    &  M21   &SUM(NE)   \\
\hline
\ [OIII] 5007 &0.56 &2.76 & 0.76  &   0.84 & 0.05 &  3.27 &  0.56  \\
\ [OI] 6300& 0.063 &6.5(-5) & 2.62 & 0.002 & 0.0    & 14.0 &  0.23      \\
\ [NII] 6584& 1.2  & 0.008  & 10.8  & 0.0086 & 0.0 & 24.5&   0.8    \\
\ [SII] 6716+6731 & 0.5   & 0.8 & 6.0 &  1.02 & 0.0 & 9.0 &   0.8\\
\  \Hb\ (\erg)   & -& 0.317 & 0.007 & 0.44 & 0.83   &0.44&  -   \\
\   w         & - & 0.025 & 1.   & 0.125 & 0.087 & 0.05  & - \\
\hline
\end{tabular}
\label{tab10}
\end{table*}

\subsection{Mrk 331}

\begin{figure}
\includegraphics[width=84mm]{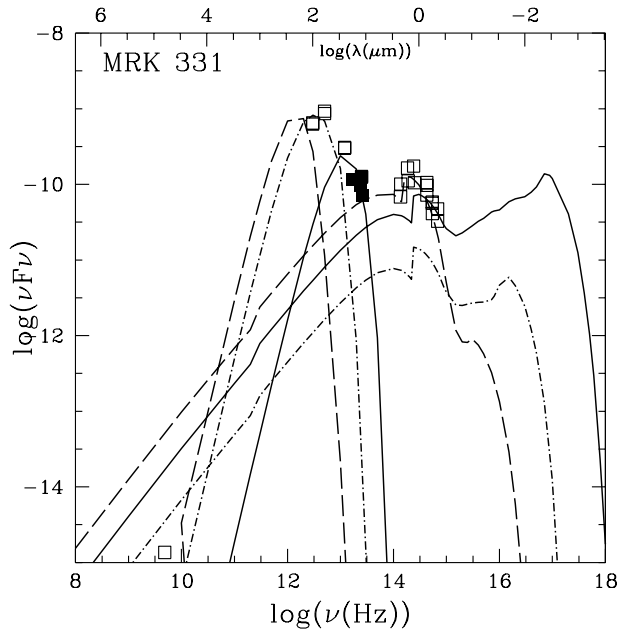}
\caption
{Mrk 331. The best fit of the SED. 
Symbols as in Fig. 3.
Different model values are listed in Table 11.}
\end{figure}

\begin{figure*}
\includegraphics[width=43mm]{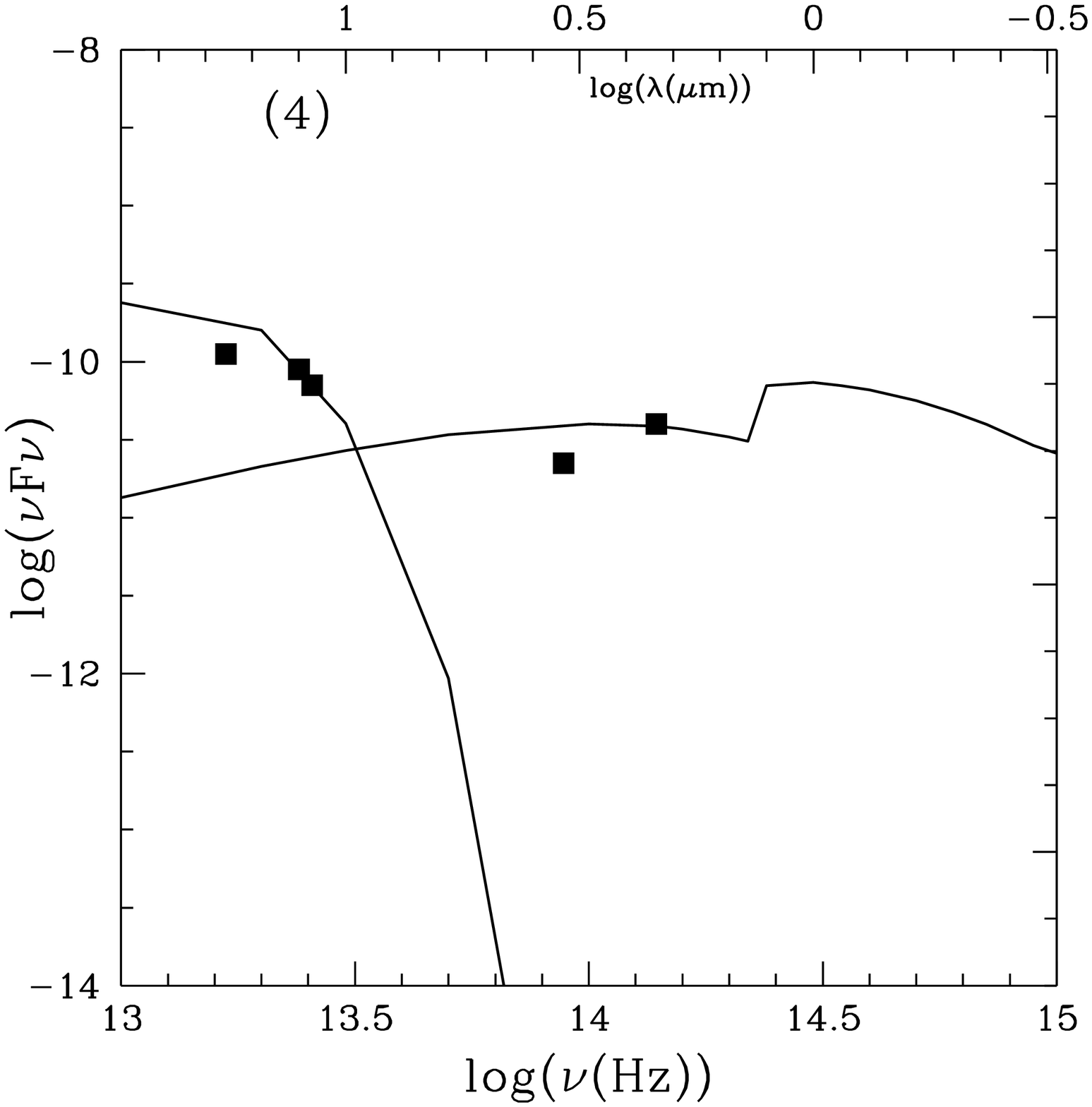}
\includegraphics[width=43mm]{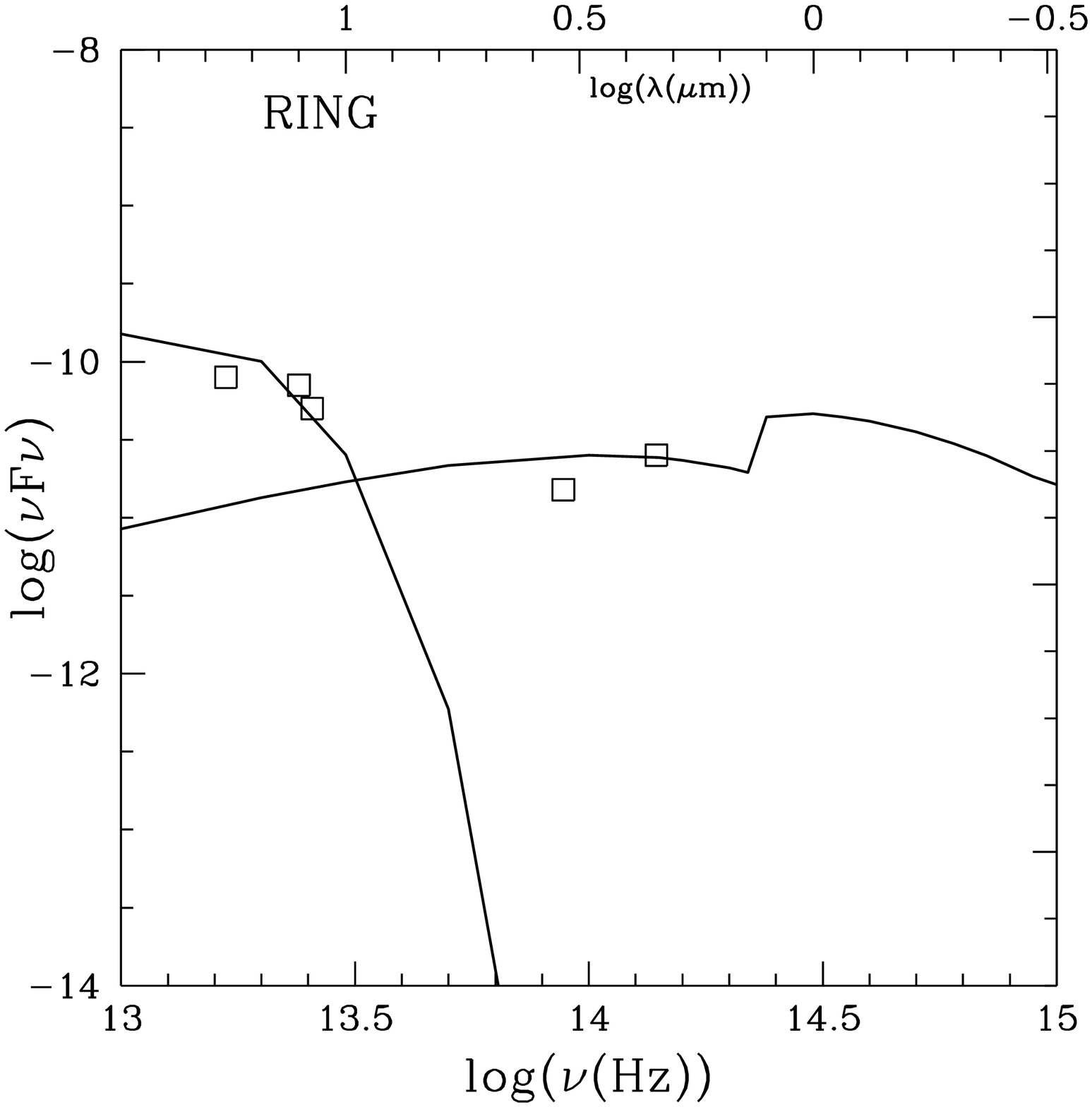}
\includegraphics[width=43mm]{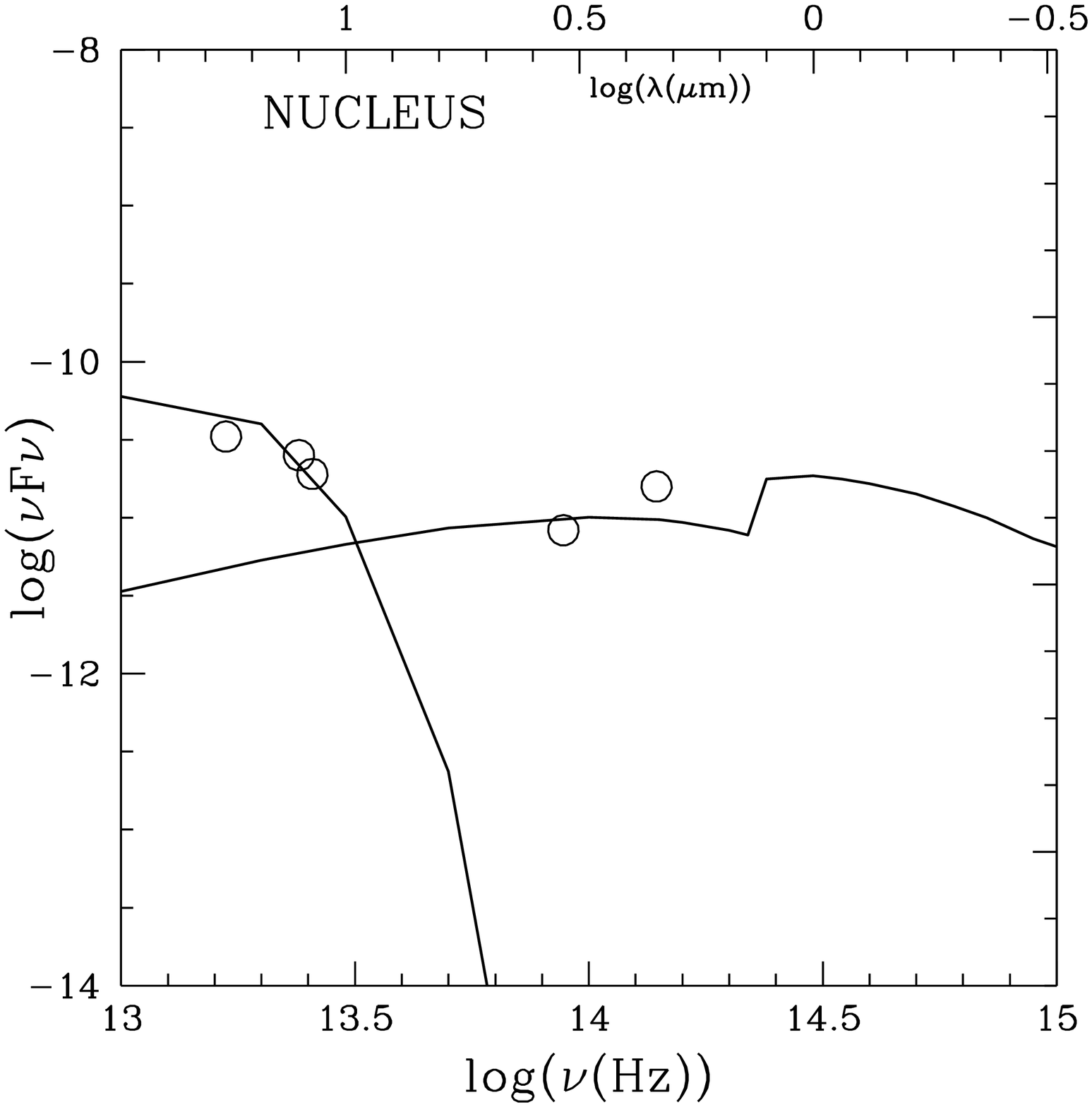}
\includegraphics[width=43mm]{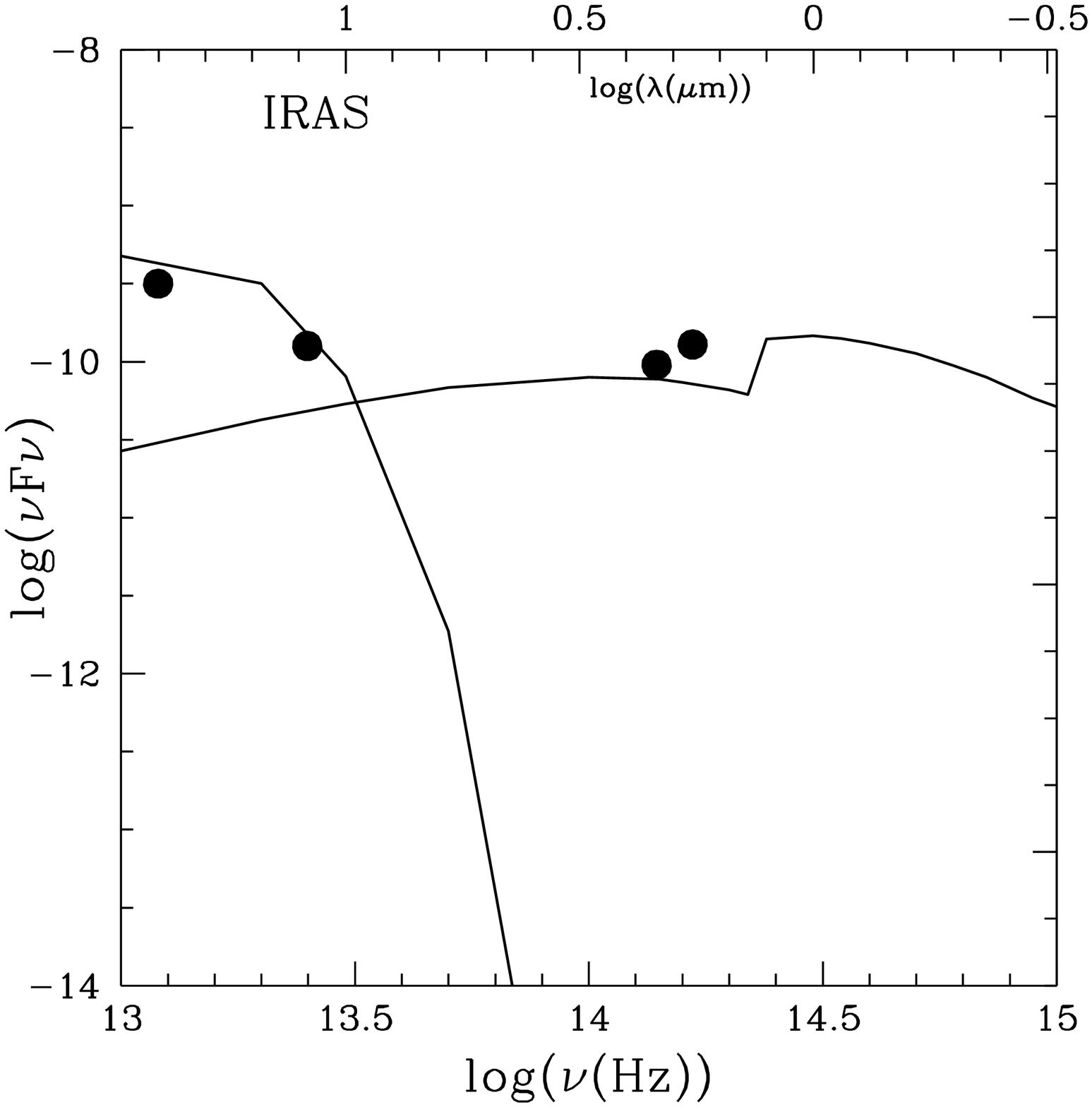}
\caption
{Mrk 331. The best fit of the MIR data, except bottom right panel (IRAS data)
in the different regions (see text for details).}
\end{figure*}

\begin{table*}
\caption{Mrk 331. Line intensities relative to H$\beta$}
\small{
\begin{tabular}{lllllllll}
 \hline
 line &  obs & M22     & M18      & M3   & M23  & SUM\\
\hline
\ [OIII] 5007 & 0.43 & 2.21  & 0.76&2.21& 0.12& 0.5 \\
\ [OI] 6300 & 0.1 & 9.(-4)& 2.62& 0.66&0.005 & 0.2 \\
\ [NII] 6584 &2.0 & 0.037& 10.8&2.64& 0.34 & 1.0 \\
\ [SII] 6716+6731 &0.77& 0.84& 6.0&3.5& 0.13 & 0.7 \\
\ [SII] 6716/6731 &1/0.98 & 0.47&  1.& 1.06 & 0.5&- \\
\  \Hb\ (\erg)   & -  & 0.26& 0.007&0.016& 68.4 &- \\
\   w                 &- &0.056 & 1.  & 0.11&1.1(-3)&- \\
\hline
\end{tabular}}

\label{tab11}
\end{table*}

The continuum SED of Mrk 331 which is shown in Fig. 14 shows that the data 
are well explained by model M22  (solid lines) with $d/g=4\times 10^{-4}$, \Vs\ = 500 \kms\ and $t=3.3$ Myr. 
However, a model corresponding to $t=0.0$ and \Vs\ = 100 \kms\ better fits the data 
in the optical range (M3). The corresponding $d/g$ is 10 times lower.
As we found for the other galaxies, a model with \Vs\ = 200 \kms\ is needed to
fit the data in the IR (M18, represented by dash-dot lines). 
We use a starburst corresponding to $t=3.3$ Myr, because 
the one corresponding to $t=0.0$ Myr shows too high [OIII]/[OI] line ratios.
Fig. 14 shows that bremsstrahlung emission is strongly absorbed at 5 GHz,
as  was found for NGC 1614 (sect. 4.2).

In Table 11 the optical line ratios observed by Veilleux et al. (1995) 
are compared with model calculations. To obtain an acceptable fit, the summed 
spectrum  accounts  for  model  M23 which corresponds to \Vs\ = 300 \kms
and to  black body radiation from stars with \Ts\ = 10\,000 K.
The SED of Mrk 331 does not show the emission from  old stars.
A possible explanation comes from the presence of the ring. In Fig. 15 the flux 
from different regions are given. It can be noticed that the flux from the ring is higher
than the flux from the nucleus. Then, radiation from the old stellar population is
reprocessed by the gas and dust in the ring. Fig. 15 shows that 
$d/g = 4\times 10^{-4}$ nicely fits all of the diagrams out of region (4) which shows a 
reduced $d/g$. All emission at wavelengths longer than 10 \mm\ comes from dust, 
at smaller values from gas.

\section{Discussion and concluding remarks}

\begin{table}
\centering
\caption{$d/g$ ratios in the galaxies of the sample}
\begin{tabular}{lllllllll} 
\hline
\ galaxy & $d/g$  \\
\hline
\ VV 114\,N & 0.016 -- 0.032 \\                    
\ VV 114\,S  & 0.008 \\
\ NGC 1614 & 0.001 -- 0.024\\
\ NGC 2326 & 4. (-4)\\
\ NGC 3690 & 0.004 -- 0.024 \\
\ IC 883 & 4.(-4) -- 8.(-4) \\
\ NGC 6090 (SW) & 1.6(-4) -- 2.8(-4)\\
\ NGC 6090 (NE) & 1.6(-4)-- 2.8(-4)\\
\ Mrk 331 & 4.(-4) -- 0.0028 \\
\hline
\end{tabular}
\label{tab12}
\end{table}

In previous sections we have modelled both the continuum SED and optical emission-line 
ratios of a representative sample of IR luminous starburst galaxies observed in the MIR 
by S01 using the Keck Telescope.

Some  significant results, obtained by modelling the optical line spectra, are 
summarized in Fig. 16, where the distribution of the dust-to 
gas ratios, in units of 4$\times 10^{-4}$, 
the age of the starburst  in Myr, and the  contribution 
in \% to the [OIII] 5007+4959 line (P$_{[OIII]}$) are  given
  in clouds corresponding to different velocities. 
Notice that VV 114 S  (Table 2) does not appear in Fig. 16 because
all the clouds therein have  the same  velocity  (\Vs=100 \kms), with ages between
0.0 and 2.5 Myr. 

The diagrams show that generally lower velocities
correspond to  early ages (see Viegas et al. 1999), out of IC 883 and NGC 6090,
which are dominated by relative old starbursts.

The d/g ratio peaks  for clouds with \Vs = 200 \kms. 
This is in agreement with the results
of Fig. 1 (left panel) which shows that for shock velocities
$<$ 300 \kms dust grains are less sputtered. 
In  high velocity  environments d/g is  partly reduced, 
although grains with a radius $>$ 0.2 \mm ~can survive
in a small region downstream, close to the front of a  \Vs = 500 \kms shock.
Moreover, IC 883, NGC 6090, and Mrk 331 show that d/g ratios are lower
in older starbursts  also when \Vs = 200 \kms.  Generally, 
low  velocity clouds  (\Vs=100 \kms) in young (and old, e.g. NGC 6090) 
starbursts are less dusty. Recall that d/g in the corresponding models are
often upper limits (e.g. NGC 1614).
In fact, these clouds are generally far  from the starburst bulk, already
merging with the ISM.
Accordingly, IC 883 that is a compact object, does not show a low velocity component
neither in the line spectrum nor in the continuum.

The relative contribution to the [OIII] lines from clouds corresponding
to \Vs = 500 \kms prevails in VV 114 N, NGC 3690, and NGC 6090 NE.
Interestingly, in Mrk 331 the contribution increases  smoothly with \Vs.
We would expect a relatively large FWHM of the [OII] line profile.

Moreover, the comparison of model calculations with the emission-line ratios shows 
that  sulphur is depleted. Generally, N/H abundance ratio 
higher than cosmic is indicated.

The main results  regarding the IR  domain  of the continuum SED, are the following:

Emission from clouds in the neighbourhood of starbursts with $t=3.3$ Myr and 
with shock velocities of $\sim$ 500 \kms\ explains both the bremsstrahlung and 
reradiation from dust in the MIR.
 Clouds with lower velocity (100 \kms) and corresponding to a lower age 
($t=0.0$ Myr) which contribute to  the line  spectra give a low contribution
to the SED in the FIR, while clouds with \Vs\ = 200 \kms\ are crucial to explain 
the FIR bump due to dust emission.

By fitting the continuum SED we found that dust-to-gas ratios change in the 
different regions of single galaxies. 
We found that  d/g ranges  between $1.6\times 10^{-4}$ and $3.2\times 10^{-2}$
(see Sect. 2). More specific results are given in Table 12.

An old stellar population with \Ts\ = 5\,000 K is revealed by black 
body emission in the optical-NIR range of a few objects.

Finally, the data which correspond to the SED are not resolved enough to
permit an analysis of absorption by ice and emission by PAH. However, 
small dust grains corresponding to PAH and graphite, and ice mantles are easily destroyed
in a shock dominated regime close to the starburst, therefore, they are emitting from  a farther extended region. 
The modelling shows that absorption by silicate   at $\sim$ 9.7 \mm ~is strong for VV 114 E(4), MRK 331, 
 and NGC 3690. Moreover, absorption by water ice at $\sim$ 3 \mm ~is suggested for
VV 114.

In summary, the frequencies corresponding to the maxima of dust reradiation depend on 
the shock velocity. However, the radiation from the starburst affects gas bremsstrahlung 
emission in the optical range. Gas and dust are mutually heating and cooling,
therefore, from the ratio between the intensity of dust emission and the bremsstrahlung 
it is possible to deduce the dust-to-gas ratio in each observed region of single galaxies.

\begin{figure*}
\includegraphics[width=55mm]{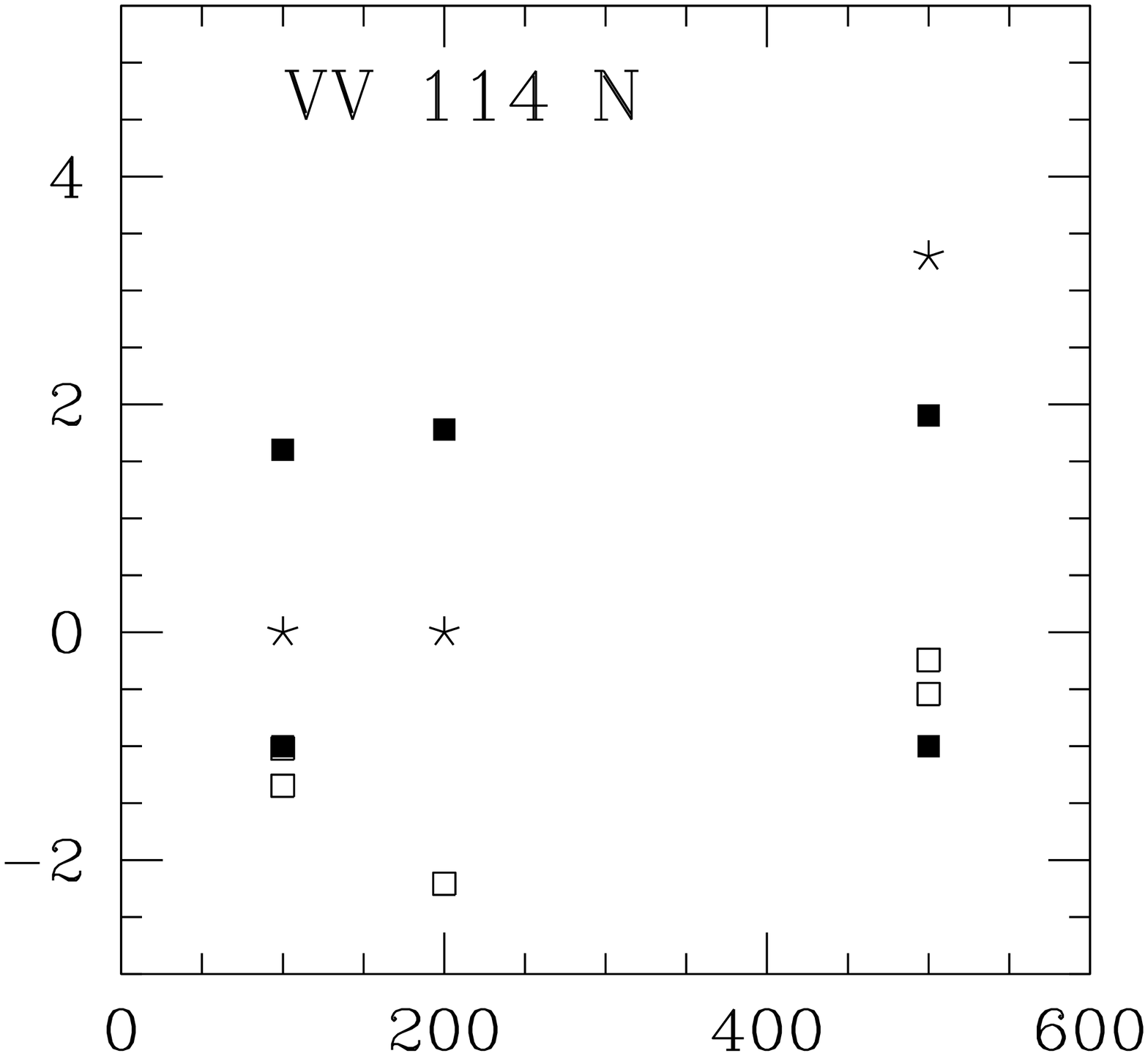}
\includegraphics[width=55mm]{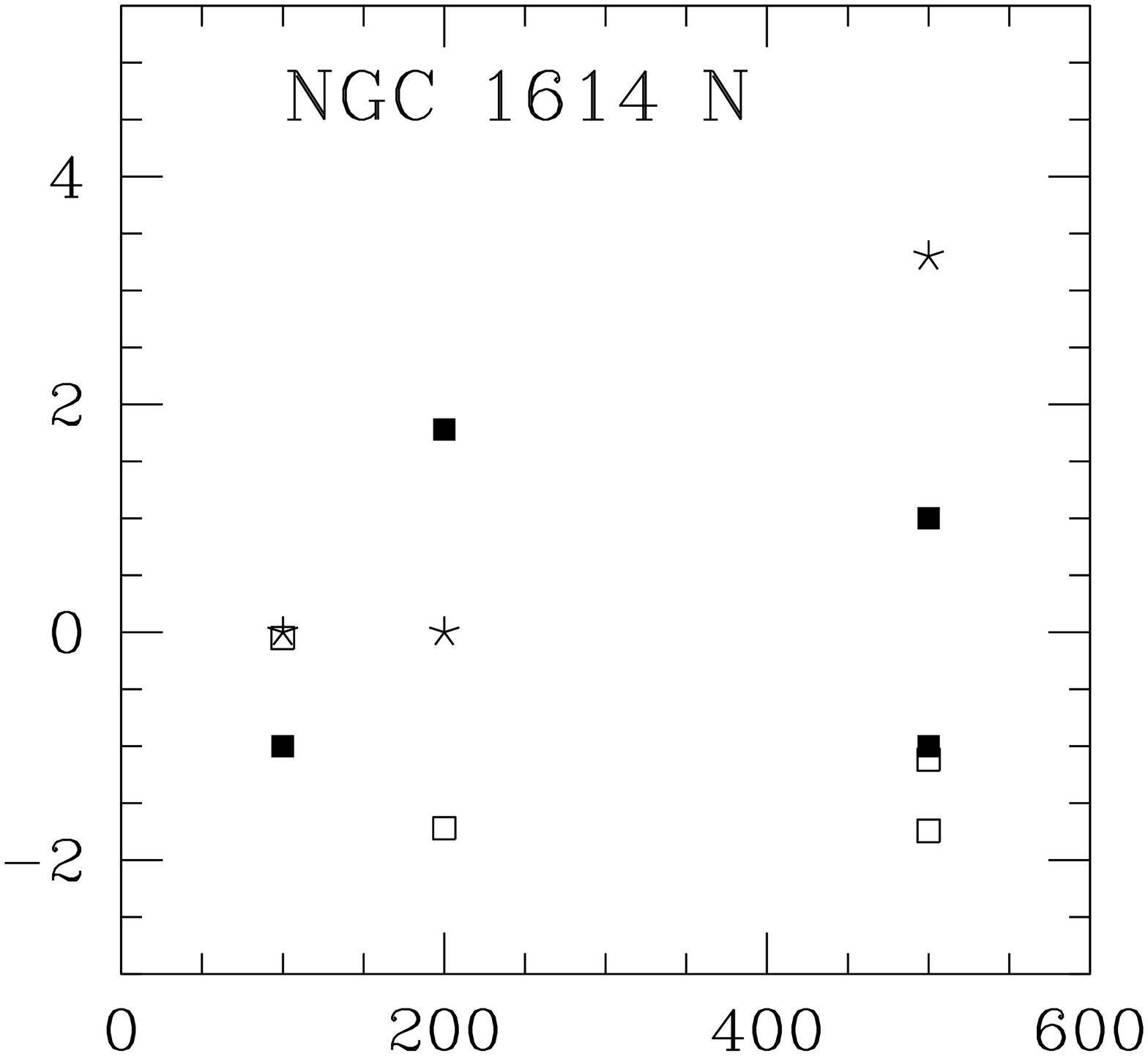}
\includegraphics[width=55mm]{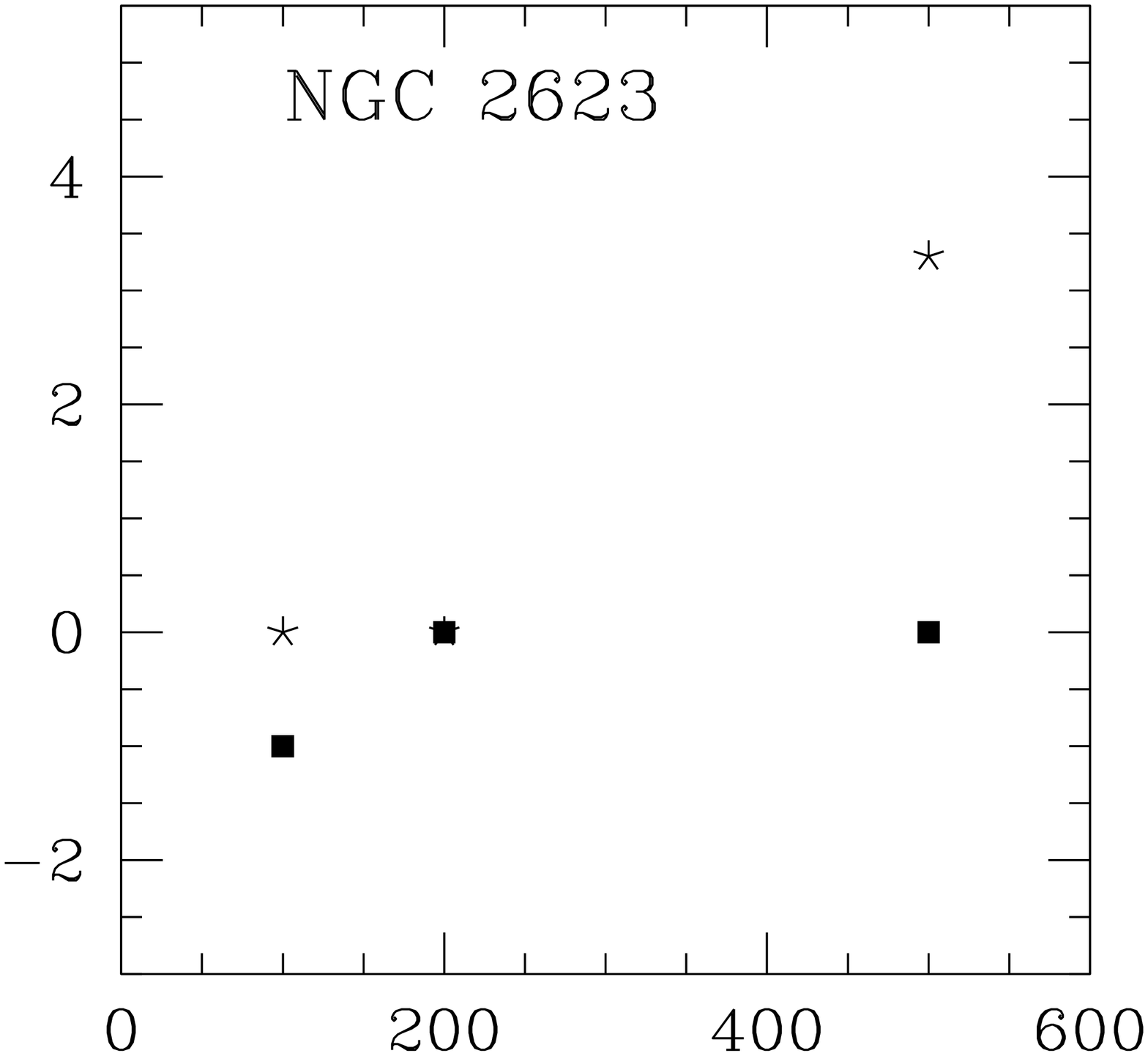}
\includegraphics[width=55mm]{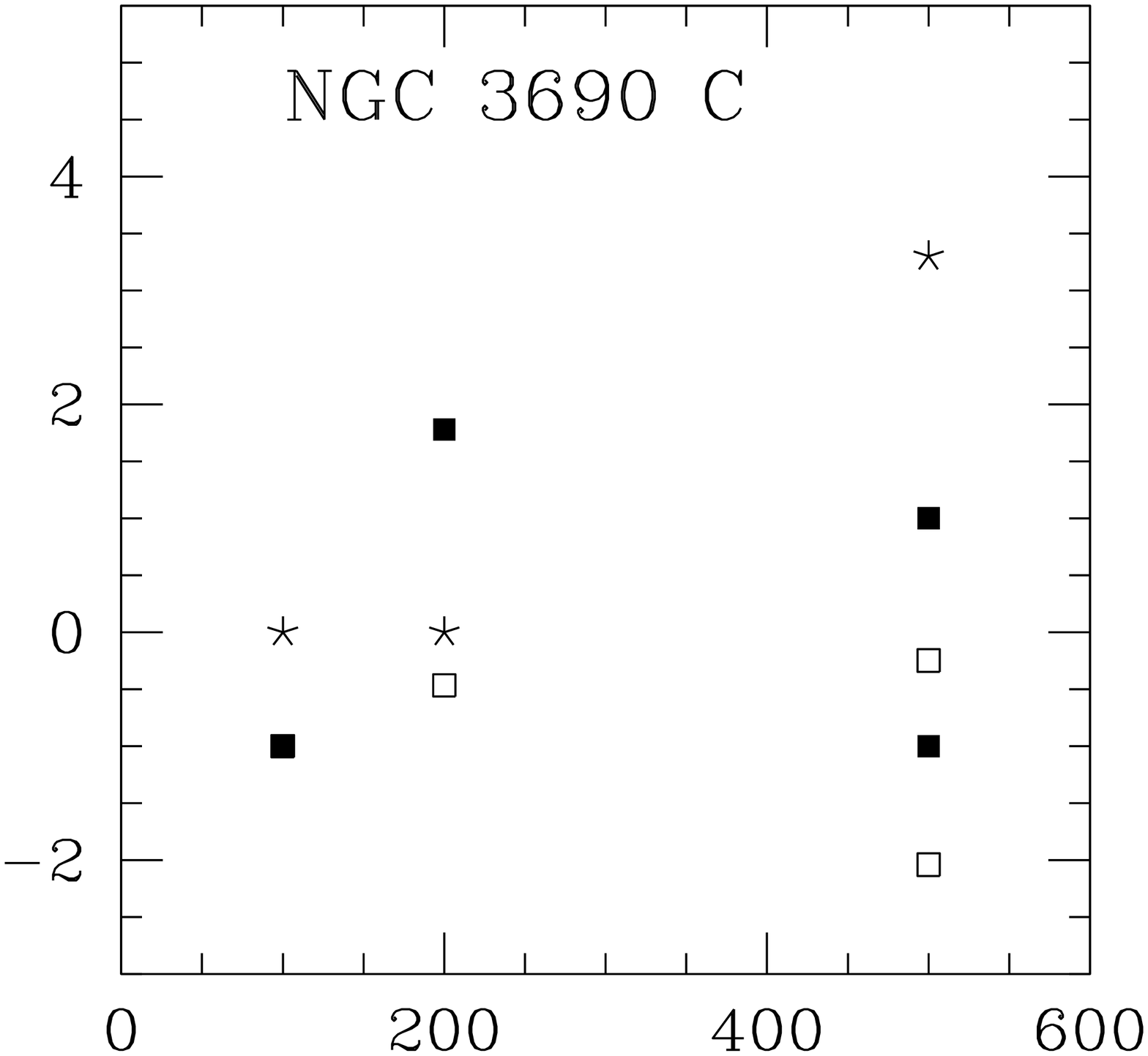}
\includegraphics[width=55mm]{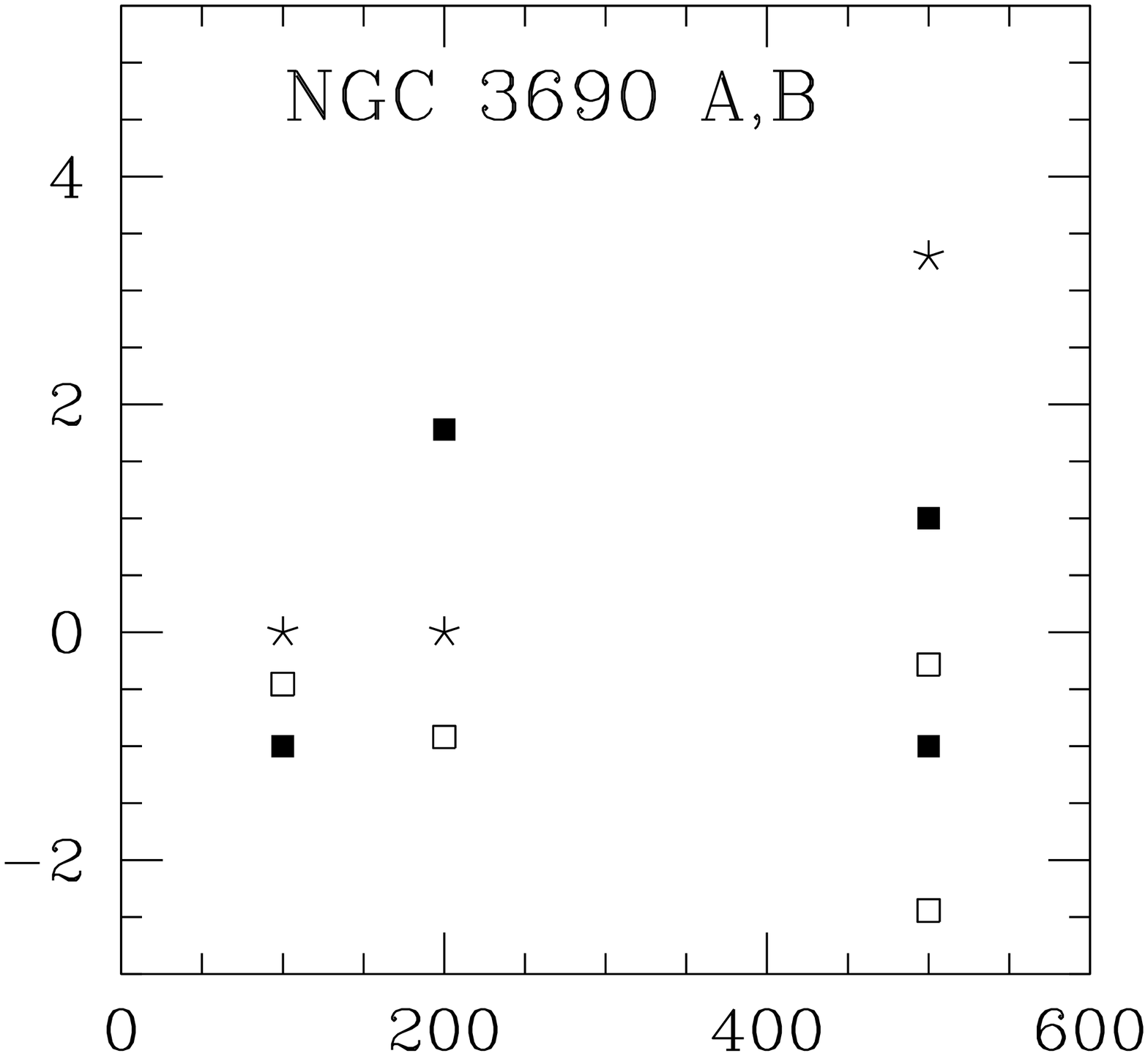}
\includegraphics[width=55mm]{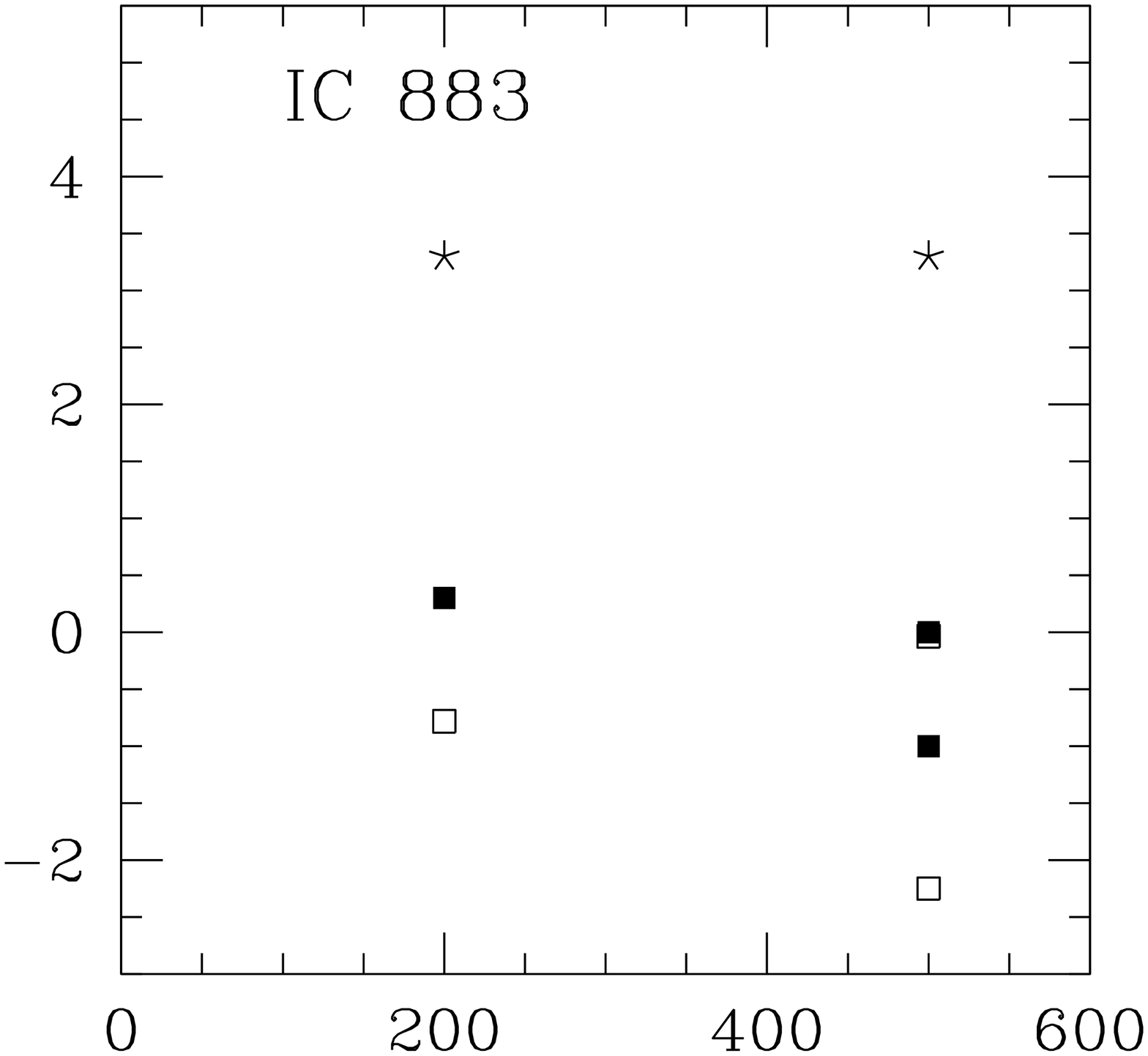}
\includegraphics[width=55mm]{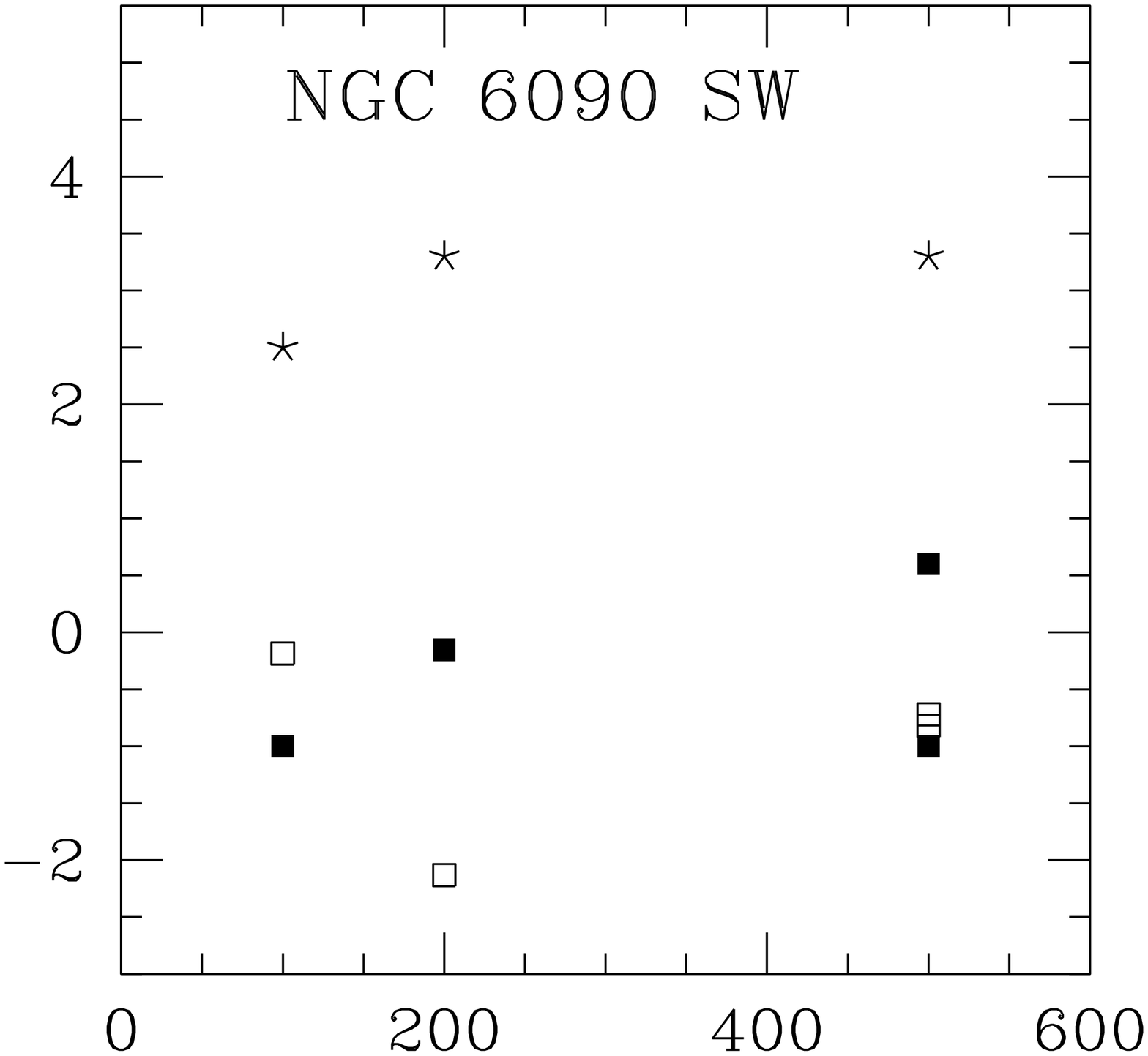}
\includegraphics[width=55mm]{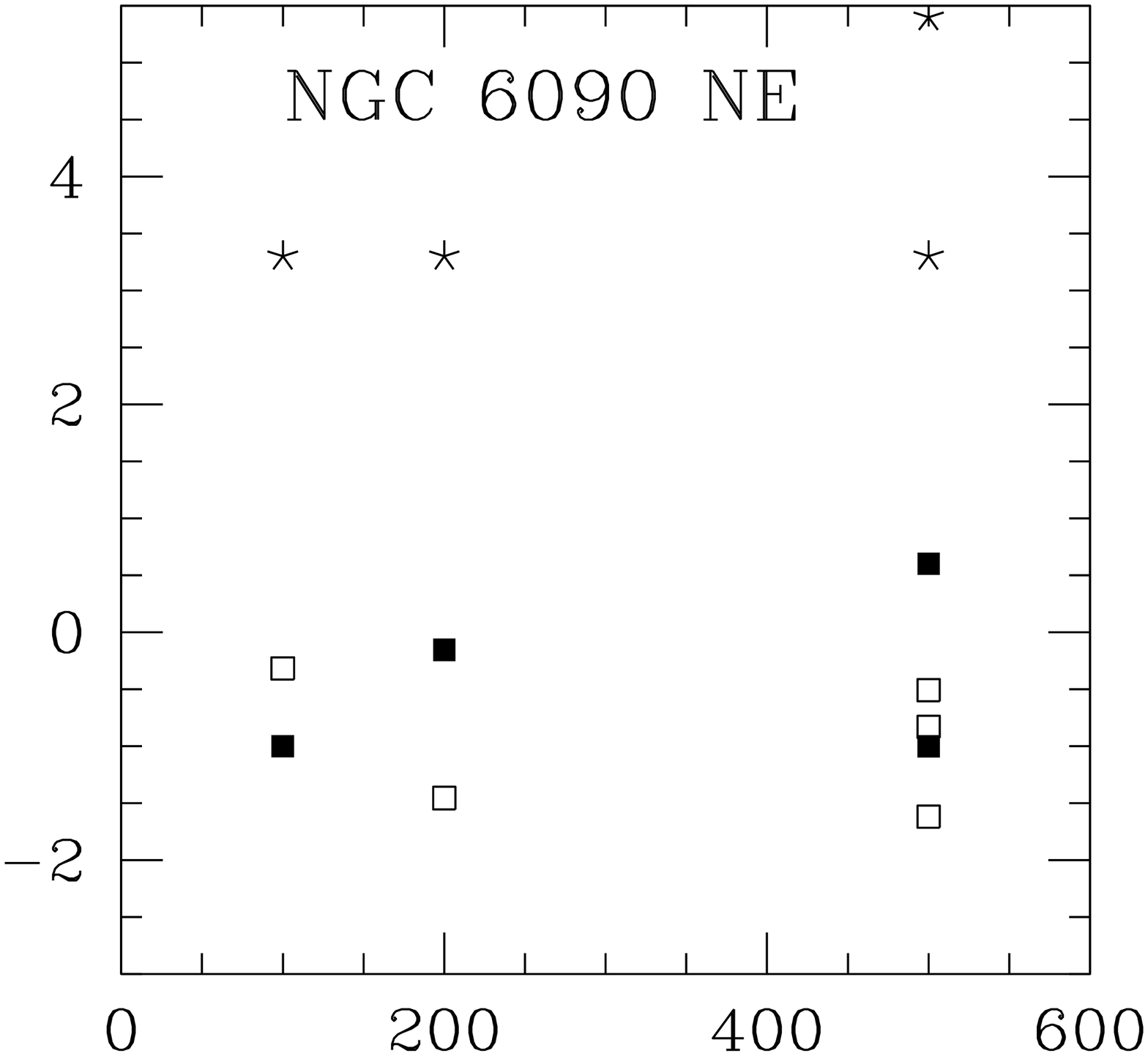}
\includegraphics[width=55mm]{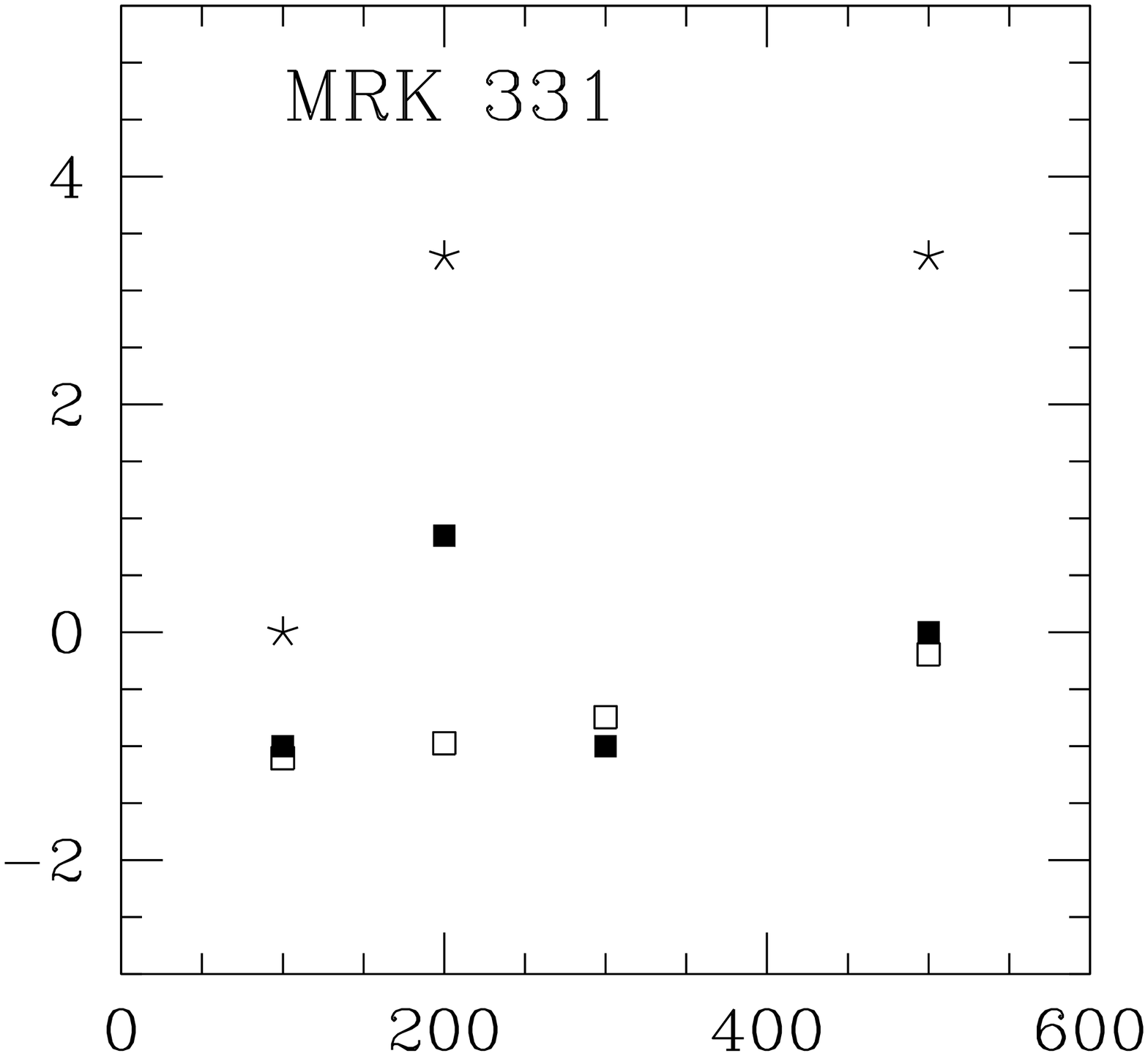}
\caption
{In each of the diagrams corresponding to different galaxies and to different regions,
log d/g (4 10$^{-4}$) (filled squares), log (P$_{[OIII]}$ (open squares),
and starburst ages in Myr (asterisks) are given as function of \Vs ~(in \kms)}
\end{figure*}

\section*{Aknowledgments}

We are very grateful to the referee for important comments.
We thank  Sueli M. Viegas for many helpful discussions.

\section*{Appendix A}

\begin{table}
\centerline{Table A1}
\centerline{Models corresponding to  CV01 ($D=10^{19}$ cm)}
\small{
\begin{tabular}{lcccllllll}
\hline
\ model  & \Vs\ & \n0\ &  $t$ & $U$ & d/g &  model \\
 &  \kms & \cm3 & Myr & & 4 $10^{-4}$ & (CV01) \\
\hline
\ M1 & 500  & 300 & 3.3 & 1. & 0.1 & 11(10A) \\
\ M2 & 500  & 300 & 3.3 & 3.5 & 80 & 11(10A) \\
\ M3 & 100  & 100 & 0.0 & 0.01& 0.1& 1(6A) \\
\ M4 & 100  & 100 & 0.0 & 0.02& 40 & 1(6A) \\
\ M5 & 200  & 100 & 0.0 & 0.01& 60 & 1(6A) \\
\ M6 & 100 & 100 &  2.5 & 0.1 & 0.1& 6(6A) \\
\ M7 & 100 & 100 & 2.5  & 1.  & 0.1& 7(6A) \\
\ M8 & 100 & 100 & 2.5  & 1. &  20 & 7(6A) \\
\ M9 & 500 & 300 & 3.3 & 0.1 & 0.1 & 10(10A) \\
\ M10& 500 & 300 & 3.3& 10. &  10.&  12(10A) \\
\ M11 & 500 & 300 & 3.3 & 1. & 1.& 11(10A) \\
\ M12 & 200 & 100 & 0.0 & 0.01 & 1. & 1(6A) \\
\ M13 & 500 & 300 & 3.3 & 0.01 & 10 & 9(10A) \\
\ M14 & 500 & 300 & 3.3 & 0.01 & 0.1& 9(10A) \\
\ M15 & 200 & 100 & 3.3 & 0.01 & 2. & 9(10A)  \\ 
\ M16 & 500 & 300 & 3.3 & 0.1 & 1. & 10(10A) \\
\ M17 & 500 & 300 & 3.3 & 0.01& 4.  & 9(10A) \\
\ M18 & 200 & 100 & 3.3 & 0.01 & 0.7 & 9(10A) \\
\ M19$^1$  & 100 & 100 & 2.5& 0.01 & 0.1 & 5(5A) \\
\ M20 & 500 & 300 &  5.4 & 1. & 0.1 & 19(10B) \\
\ M21 & 100 & 100 & 3.3 & 0.01 & 0.1 &  9(6A) \\
\ M22 & 500 & 300 & 3.3 & 0.01 & 1. & 9(10A) \\
\ M23 & 300 & 300 & BB$^2$ & 10. & 0.1& 4(3) \\
\hline
\end{tabular}}

$^1$ $D=10^{17}$ cm

$^2$ corresponding to $T=$ 10\,000 K

\label{A1}
\end{table}

In Table A1 the models are defined and referred to the models in CV01.
The input parameters appear in columns 2-6. In the last column
the number before the parenthesis refer to the number of the model
in the tables (which appears inside the parenthesis) of CV01.

\section*{Appendix B}

The data from NED come from:

Spinoglio et al. (1995), 
Moshir et al. (1990), 
Soifer et al. (1989),
Griffith et al. (1994),
Condon et al. (1998),
Fabbiano, Kim, \& Trinchieri (1992),
Kinney et al. (1993),
De Vaucouleurs et al. (1991),
Huchra (1977),
Doroshenko \& Terebezh (1979),
Balzano \& Weedman (1981),
Rieke \& Low (1972),
Lebofsky \& Rieke (1979),
Eales, Wynn-Williams, \& Duncan (1989),
Dunne et al. (2000),
Griffith et al. (1995),
Godwin et al. (1977),
Gregory \& Condon (1991),
Becker, White, \& Edwards (1991),
Joyce \& Simon (1976),
Allen (1976),
Carico et al. (1992),
Odewanhn \& Aldering (1995),
Becker, White, \& Helfand (1995).

%\bsp

\label{lastpage}

\end{document}